\documentclass[11pt]{article}

\usepackage{amsmath, amssymb, amsfonts, bm}
\usepackage{geometry}
\usepackage[T1]{fontenc}
\usepackage{lmodern}
\usepackage{array}
\newcolumntype{L}[1]{>{\raggedright\arraybackslash}p{#1}}
\usepackage{booktabs}
\usepackage{dsfont}

\usepackage{tabularx}
\usepackage{longtable}
\usepackage{pdflscape}
\usepackage[section]{placeins}
\newcolumntype{Y}{>{\centering\arraybackslash$\displaystyle}X<{$}}
\newcolumntype{Z}{>{\raggedright\arraybackslash}X}
\renewcommand{\arraystretch}{1.15}

\usepackage{url}
\usepackage{doi}
\usepackage[authoryear]{natbib}
\usepackage{enumitem}
\usepackage{graphicx}
\graphicspath{{figures/}}
\usepackage{float}
\usepackage[title,titletoc]{appendix}
\let\cleardoublepage\clearpage

\title{HODL Strategy or Fantasy?\\[0.75em]
\large 480 Million Crypto Market Simulations and the Macro-Sentiment Effect}

\author{Weikang Zhang \and Alison Watts}
\date{}

\begin{document}
\maketitle

\begin{abstract}
Crypto enthusiasts claim that buying and holding crypto assets yields high returns, 
often citing Bitcoin's past performance to promote other tokens and fuel fear of 
missing out. However, understanding 
the real risk–return trade-off and what factors affect future crypto returns is crucial 
as crypto becomes increasingly accessible to retail investors through major brokerages.
We examine the HODL strategy through two independent analyses.
\textbf{First}, we implement 480 million Monte Carlo simulations across 378 non-stablecoin crypto assets, 
net of trading fees and the opportunity cost of 1-month T-bills. 
We find strong evidence of survivorship bias and extreme downside concentration. 
At the 2--3 year horizon, the median excess return is $-28.4\%$ and 
$\mathrm{CVaR}_{1\%}\approx 1.08$, implying tail scenarios wipe out 
the principal after deducting all costs, while the top-quartile mean 
reaches $+1{,}326.7\%$. These results challenge the ``HODL'' narrative: 
across a broad set of crypto assets, simple buy-and-hold loads extreme downside risk onto most investors, and the ``miracles'' 
mostly belong to the luckiest quarter.
\textbf{Second}, using Bayesian multi-horizon local projection, we find that 
endogenous predictors (realized risk-return metrics) exhibit economically negligible 
population-level effects with limited cross-basket consistency. In contrast, macro-finance factors, particularly 
the 24-week exponential moving average of the Fear \& Greed Index, demonstrate persistent 
long-horizon effects and high cross-basket stability: where significant, a one-standard-deviation shock 
reduces forward top-quartile mean excess returns by 15--22 percentage points and median 
returns by 6--10 percentage points over 1--3 year horizons. These findings imply that 
realized risk-return distributions offer limited generalizable guidance for investment 
decisions; macro-sentiment conditions emerge as the dominant indicators of future outcomes. 

\end{abstract}

\section{Introduction}

``HODL'', a popular crypto slang term 
for buy-and-hold investing, has become a widely promoted 
strategy in cryptocurrency markets. But does this approach 
in fact deliver superior long-run outcomes, or does it instead 
expose typical retail investors to greater risk?

Understanding the risk-return characteristics of cryptocurrency 
buy-and-hold strategies has become increasingly important as the global 
cryptocurrency market cap has surpassed \$3 trillion\footnote{As of November 
2025; source: CoinGecko, \url{https://www.coingecko.com/en/charts}.} 
and major brokerages such as Robinhood and Fidelity now offer crypto 
trading on their platforms. Prior research has documented that cryptocurrencies 
exhibit extreme tail risk \citep{GKILLAS2018109, Osterrieder2017, NzokemMaposa2024} 
and respond to macroeconomic shocks \citep{MA2022101711, KARAU2023102880}, but two critical 
questions remain unanswered. First, what are the market-wide risk and return 
distributions of buy-and-hold outcomes when accounting for transaction costs, 
opportunity costs, and flexible holding horizons? Second, should investors rely on current realized risk-return distributions
to estimate their future outcomes? This paper addresses both questions 
through large-scale Monte Carlo simulation and Bayesian multi-horizon local 
projection analysis.

Crypto influencers often cite Bitcoin's success story to promote other 
tokens, but how far is the distance between the average investor and 
the luckier top-quartile group? We implement 480 million Monte Carlo 
simulations across eight baskets—seven major tokens (BTC, ETH, ADA, BNB, 
DOGE, LINK, XRP) and one market-wide basket randomly sampling from 378 
non-stablecoin cryptocurrencies. In the market-wide basket at the longest 
holding horizon (731--1095 days), the median excess return is $-28.4\%$, 
while the mean among the top 25\% reaches $+1{,}326.7\%$. Beyond 181 days, 
more than 50\% of scenarios result in losses exceeding 10\%, with tail risk 
(CVaR$_{1\%}$) approaching total capital loss after deducting opportunity 
costs and transaction fees. This disparity exposes a dangerous form of 
survivorship bias embedded in the cryptocurrency investment narrative.

To address whether investors can rely on current realized risk-return distributions 
to estimate future outcomes, we implement a Bayesian multi-horizon local projection 
framework that jointly models endogenous predictors (realized risk-return metrics) 
and stability-selected macro-finance factors with distinct hierarchical shrinkage 
structures across forecast horizons. Our analysis reveals a fundamental asymmetry: while some basket-specific endogenous 
predictors exhibit strong effects, a Bayesian random-effects meta-analysis shows 
their population-level influence remains economically negligible with limited 
cross-basket consistency. In contrast, macro-finance sentiment indicators demonstrate 
both persistent long-horizon effects and high cross-basket stability. Where significant, the 24-week 
exponential moving average of the Fear \& Greed Index emerges as the most stable 
predictor across baskets, reducing forward mean excess returns of the top 25\% 
by 15--22 percentage points and median returns by 6--10 percentage points per 
standard-deviation shock over 1--3 year horizons. This asymmetry implies that 
realized risk-return distributions offer limited generalizable guidance for long-term 
investment decisions. Sentiment effects, however, exhibit substantial heterogeneity: 
meme coins such as DOGE display approximately 3 times greater sensitivity than 
established cryptocurrencies.

Our findings challenge both pillars of the HODL narrative. First, the 
market-wide risk-return distribution does not justify buy-and-hold for 
most investors: median outcomes are deeply negative, tail risk approaches 
total loss, and the gap between typical and top-quartile returns reflects 
survivorship bias rather than attainable outcomes. Second, past realized distributions offer 
limited generalizable guidance for future outcomes: endogenous 
risk-return metrics exhibit economically negligible population-level 
effects, while macro-finance sentiment indicators—particularly the 
Fear \& Greed Index—demonstrate persistent long-horizon effects with 
high cross-basket stability. These results suggest investors should 
focus on macro-sentiment conditions and macroeconomic factors rather than 
extrapolate from historical performance or selective success stories.

\section{Literature Review}

Empirically testing "HODL" is challenging because retail investors have 
heterogeneous beliefs about market cycles, different holding periods, 
and different levels of tolerance for extreme tail outcomes.
To simulate cryptocurrency returns, two branches of simulation approaches are common. 
The first branch is parametric price-path Monte Carlo: in this type of simulation, researchers usually 
assume an explicit stochastic data-generating process (DGP), then simulate full price paths to evaluate risk and 
return distributions. For example, \citet{Likitratcharoen2021} implement a 
parametric price-path Monte Carlo by assuming Bitcoin follows a geometric Brownian motion 
and simulating 1,000,000 daily returns with pseudo-random normal shocks to compute VaR.
The second branch is nonparametric simulation: rather than specifying an 
explicit DGP, studies generate synthetic long-horizon paths by stationary block-bootstrap 
resampling of historical returns \citep{PolitisRomano1994} to preserve serial dependence and non-Gaussian features.

From the perspective of retail investors, there are a few questions often 
asked about cryptocurrency investments: What are the expected returns? 
What is the downside risk? Would a longer holding horizon decrease risks?
Regarding expected returns, empirical evidence suggests that cryptocurrency returns are systematically related to market capitalization.
\citet{Li15082020}, in research across more than 1{,}800 cryptocurrencies , find that returns are related to size : small-cap coins earn significantly higher subsequent returns than large-cap coins.

When turning to the question of downside risk, researchers have employed extreme value theory and related statistical methods to quantify the tail risk properties of cryptocurrencies.
\citet{GKILLAS2018109} apply extreme value theory to the return tails of five major cryptocurrencies and, using Value-at-Risk (VaR) and Expected Shortfall (ES) as tail risk measures, find that Bitcoin Cash is the riskiest, whereas Bitcoin and Litecoin exhibit the lowest tail risk.
Compared with major fiat (G10) currencies, Bitcoin’s returns exhibit heavier tails, much higher volatility, and substantially larger losses in extreme events \citep{Osterrieder2017}.
In another study on crypto tail risk relative to traditional markets, Aubain Nzokem and Daniel Maposa show that, compared with the S\&P 500, Bitcoin’s daily returns are markedly more heavy-tailed, making extreme moves more likely \citep{NzokemMaposa2024}.

The third question is whether longer holding horizons reduce risk. 
This has been examined in both traditional and cryptocurrency markets.
For example, using a stationary block bootstrap on 39 developed 
markets (1841–2019), \citet{Anarkulova2022} show that while average real payoffs 
rise with horizon, there is still about a 12\% chance of an inflation-adjusted loss over 30 years.
In the context of holding horizon and holding risks in the 
cryptocurrency market, \citet{conlon2024} find that, on average, Bitcoin behaves as a 
short-horizon hedge for USD exchange rates; however, at longer horizons it can co-move positively 
with large USD losses, so a longer holding horizon does not guarantee reduced risk in crypto holdings.
Another study analyzes the comovement between Bitcoin and the U.S. stock market and 
finds right-tail dependence that is stronger at long holding horizons; this co-move effect decreases significantly as the horizon shortens from yearly to monthly \citep{MAGHYEREH2021101355}.

After understanding the risk and return distribution and their dynamics over time, 
a natural question arises: What factors may affect these distributions? 
Although macroeconomic shocks are known to influence cryptocurrency returns 
\citep{MA2022101711, KARAU2023102880},
a comparative impulse-response framework 
that jointly quantifies the effect sizes of endogenous risk-return metrics versus 
macro-finance factors across multiple forecast horizons has not been explored.
A natural framework for studying such shock transmission is impulse response analysis.
Local Projections (LP; \citet{Jorda2005}) estimate impulse responses without fully specifying a 
multivariate dynamic system. At the population level, with unrestricted lag structures and the same shock 
identification, LPs and VARs target the same impulse responses; differences are primarily finite-sample bias–variance trade-offs.
Later research by \citet{KilianKim2011} shows that, in small samples, LP intervals are often wider and have less accurate coverage than bias-adjusted (bootstrap) VAR intervals. 
To improve LP estimation, \citet{BarnichonBrownlees2019} note that LP impulse responses can be highly 
variable and jagged in practice and propose Smooth Local Projections, penalized B-spline (ridge) smoothing across horizons,
to deliver smoother, more precise impulse responses without altering identification.
Starting from the local-projection setup, \citet{Ferreira2025} introduce Bayesian Local Projections (BLP), 
which shrink LP coefficients through hierarchical informative priors to handle the LP, VAR bias, variance trade-off while preserving LPs' 
robustness to misspecification. With informative priors, BLP delivers uncertainty comparable to a BVAR with standard macro priors, and its 
posterior mean can be viewed as an optimally weighted combination of LP and (iterated) VAR information at each horizon.

When applying impulse response methods to high-dimensional settings with many potential predictors, 
variable selection becomes crucial for two reasons. 
First, macroeconomic variables are often highly correlated, making it difficult to identify 
individual effects. 
Second, identifying a parsimonious set of predictors enhances interpretability
and improves the stability of coefficient estimates.
Two prominent approaches help identify the relevant predictors among many candidates: 
regularizing priors and stability selection.
To handle high-dimensional Bayesian models, the horseshoe global–local prior adapts to unknown sparsity by 
aggressively shrinking near-zero coefficients toward zero while preserving large signals \citep{CarvalhoPolsonScott2010Horseshoe}.
Empirically, horseshoe-type priors, with a calibrated global scale and the regularized 
horseshoe, lead to more reliable estimation and more stable posterior computation \citep{PiironenVehtari2017,PiironenVehtari2017hyperprior}.

A complementary, resampling-based approach to identifying robust predictors is stability selection, 
which addresses selection instability in high-dimensional, correlated settings.
In correlated, high-dimensional problems, CV with unstable selectors leads to noisy 
error curves and unstable selections \citep{Breiman1996}.
Following Breiman’s instability–stabilization view, stability selection repeatedly applies a sparse selector to 
many half-sample subsamples and retains variables with high selection frequency, 
and it provides error-control through an upper bound on the expected number of false selections \citep{MeinshausenBuehlmann2010,ShahSamworth2013}.
This idea extends naturally to joint-sparsity settings: recent studies pair multitask group-Lasso with stability selection, 
yielding more stable feature selection under LD-driven collinearity \citep{NouiraAzencott2022,nouira2025}.

Empirical studies suggest cryptocurrencies are not insulated from macroeconomic shocks.
On FOMC days, a 1-bp unexpected tightening (measured by the two-year Treasury yield) 
sends Bitcoin down about 0.25\% immediately, with a larger, persistent multi-day 
decline; the effect is strongest in bull markets \citep{MA2022101711}.
Since late 2020, unexpected U.S. monetary 
tightenings depress Bitcoin prices \citep{KARAU2023102880}.

Beyond macroeconomic shocks, sentiment-driven dynamics also play a significant role in cryptocurrency markets.
Investor sentiment—both rational and irrational—significantly and 
positively drives Bitcoin returns, especially after COVID-19, with 
patterns associated with FOMO (fear of missing out) \citep{Guler03072023}.
Relatedly, \citet{LI2021101829} find that a higher prior MAX—the largest daily return in the past 
month—predicts higher subsequent returns.

\subsection{Research Gap and Contributions}
The literature above identifies two simulation approaches for 
cryptocurrency returns (parametric price-path Monte Carlo and nonparametric 
block-bootstrap), demonstrates that cryptocurrencies respond to 
macro-finance shocks, and introduces Bayesian local projection methods 
for studying impulse responses.

However, two critical gaps remain. First, while prior simulation 
studies either assume stylized parametric processes or apply block-bootstrap 
to historical returns, existing work does not simultaneously combine 
flexible holding horizons (spanning 1 to 1,095 days), 
realistic transaction costs benchmarked against risk-free returns, 
and broad cryptocurrency coverage. Prior studies focus on Bitcoin or 
a few high-cap tokens, leaving survivorship bias and downside risk 
concentration across the broader market unquantified. 
We address this by simulating 480 million episodes across eight baskets, 
including an aggregate basket of 378 non-stablecoin cryptocurrencies, 
to capture both token-specific and market-wide risk distributions.

Second, while prior work documents crypto responses to macroeconomic and 
sentiment shocks \citep{MA2022101711, KARAU2023102880, Guler03072023}, to our knowledge there is no 
published framework that jointly (i) models multiple risk--return 
targets, (ii) contrasts basket-specific endogenous distribution measures 
with macro-finance predictors, and (iii) evaluates these relations 
across horizons. We develop a Bayesian multi-horizon 
local projection model to estimate impulse responses of four risk-return 
metrics at six forecast horizons (30, 90, 180, 365, 730, and 1,095 days), quantifying the relative strength and persistence of basket-specific 
risk-return responses to these two predictor types.

\section{Data}

We first compiled the top 800 crypto assets by market capitalization from CoinMarketCap\footnote{\url{https://coinmarketcap.com}, homepage pages 1–8; accessed April 2025.}. 
For each asset, we mapped the CoinMarketCap symbol to a Yahoo Finance ticker of the form SYMBOL-USD (e.g., BTC-USD), 
and, if unavailable, attempted the raw SYMBOL as a fallback. Using the yfinance Python package, we then downloaded the longest available daily data, specifically the High, Low, 
Close, and Volume, for each asset.

\paragraph{Data cleaning}
We removed assets whose first available date was on or after 2024-01-01.
We then removed stablecoins by computing each token’s average daily Close and the standard deviation of Close over the sample; tokens with an average daily Close within \([0.97, 1.03]\) USD and a standard deviation \(\le 0.03\) were classified as stablecoins and excluded.
We removed tokens whose average daily volume over the last 365 calendar days of the sample was \(<\$100{,}000\) and tokens whose latest observation was before 2025-04-26. For missing data, we trimmed each token’s history to begin on the first date on which High, Low, Close, and Volume are all observed. Tokens that still contained any missing values thereafter were dropped. As a final quality screen, we excluded any token with \(\ge 10\) days where \(\mathrm{High}=0\) or \(\mathrm{Volume}=0\), or \(\ge 10\) days with \(\mathrm{Volume}<\$500\).

Beyond the crypto asset panel, we also collect supplementary macro-financial indicators.

\noindent We collect the Crypto Fear \& Greed Index\footnote{A market–sentiment gauge for crypto 
(primarily Bitcoin), scaled 0–100 where 0 denotes ``Extreme Fear'' and 100 ``Extreme Greed''; 
source: Alternative.me, \url{https://alternative.me/crypto/fear-and-greed-index/}.} 
via Alternative.me’s API, aggregate the daily series to a weekly mean (Monday–Sunday), 
and assign each week’s timestamp to the Monday at the start of that week.

We collect macro-finance series from FRED, take the last available business day each week (Friday, or Thursday when Friday is missing) as the weekly value, stamp it on the Monday of that week, and list the associated codes and series names in Appendix A.2.1.

We also compute the log return of Bitcoin. Let \(P_t^{\mathrm{BTC}}\) be the closing price on the last day of week \(t\) (Sunday; if missing, use the last available trading day that week). The weekly log return is
\[
r_t^{\mathrm{BTC}}=\ln\!\left(\frac{P_t^{\mathrm{BTC}}}{P_{t-1}^{\mathrm{BTC}}}\right).
\]
For alignment, we reindex each \(r_t^{\mathrm{BTC}}\) to the Monday that starts week \(t\).

\section{Monte Carlo Simulation}

Our Monte Carlo simulation approximates the full distribution of one-time buy–hold–sell outcomes
that a crypto investor could experience without perfect foresight. We executed 10,000,000 simulations for each of six holding-period intervals per basket—480,000,000 in total across eight baskets.
\begin{table}[h]
\centering
\caption{Notation for the Simulation Section}
\label{tab:notation}
\begin{tabular}{ll}
\toprule
Symbol & Description \\
\midrule
\(\ell,h\)         & Lower/upper bound of allowed holding horizons (days) \\
\(\tau_i\)         & Holding period of episode \(i\) (days) \\
\(T\)              & Length of the selected basket-price sample (days) \\
\(s_i,e_i\)        & Buy and sell day indices for episode \(i\) \\
\(c_i\)            & Index of the selected coin in the basket \\
\(H_{t,c},L_{t,c}\)& Daily high and low prices of coin \(c\) on day \(t\) \\
\(P_i^{\text{buy}}\)& Simulated buy price for episode \(i\) \\
\(P_i^{\text{sell}}\)& Simulated sell price for episode \(i\) \\
\(r_t^{\text{rf}}\) & Daily simple risk-free rate \\
\(\gamma_t\)       & Cumulative log risk-free return up to day \(t\) \\
\(R_i^{\text{rf}}\) &{Risk-free return for episode \(i\)'s holding period} \\
\(\phi\)           & One-way proportional trading fee \\
\(G_i\) & Net return of episode \(i\) after fees \\
\(X_i\)            & Excess return of episode \(i\) over cash \\
\bottomrule
\end{tabular}
\end{table}

\subsection{Draw a Holding Horizon}

We pre-set six inclusive horizon intervals: 1--30, 31--90, 91--180, 181--365, 366--730, 731--1095.
For each simulated episode, the simulator randomly picks an integer from the lower
bound to the upper bound of the chosen interval. This design gives the model
flexible holding periods: the shortest interval \([1,30]\) captures investors who treat
crypto almost like a checking-account balance, ready to move in and out, whereas
the longest interval \([731,1095]\) represents holders who worry less about short-run
fluctuation and focus on multi-year returns.

Given inclusive horizon bounds \(\ell < h\),
\[
  \tau_i \sim \operatorname{DiscreteUniform}\{\ell,\ell+1,\dots,h\}
  \quad (i = 1,\dots,n),
\]
where \(\tau_i\) denotes the randomly drawn holding horizon for episode \(i\).

\subsection{Determine Buy \& Sell Dates}

For each holding horizon, the simulator determines the latest possible buying
and selling dates. Since Python indexing starts at 0, we subtract 1 to ensure the
latest buying date remains within the tokens' data range.

Let $T$ denote the total length of the dataset in trading days.
The buy date $s_i$ must be chosen such that it leaves enough days ($\tau_i$)
within the selected time horizon, ensuring that the sell date $e_i$ does not
exceed the dataset boundaries:

\[
0 \;\le\; s_i \;\le\; T-\tau_i-1,
\qquad
e_i \;=\; s_i + \tau_i \;<\; T.
\]

Accordingly, we randomly select the buy date from a discrete uniform
distribution:

\[
s_i \;\sim\; \operatorname{DiscreteUniform}\{0,1,\dots,T-\tau_i-1\},
\qquad
e_i = s_i + \tau_i.
\]

\paragraph{}%
In the situation of \(T-\tau_i-1 < 0\), the draw is discarded.

\subsection{Draw a Coin}

We construct eight baskets in total. Seven baskets are single-token baskets (\texttt{BTC}, \texttt{ADA}, \texttt{BNB}, \texttt{DOGE}, \texttt{ETH}, \texttt{LINK}, \texttt{XRP}), while the \texttt{ALL} basket contains the entire dataset comprising \( C = 378 \) tradable tokens.

For each simulated episode, we select a coin by uniformly sampling from pre-assigned consecutive integer indices:

\[
c_i \;\sim\; \operatorname{DiscreteUniform}\{0,\dots,C-1\},
\]

where \( c_i \) denotes the coin selected for episode \( i \). When the basket contains only a single token (\(C=1\)), this selection becomes deterministic, with \( c_i = 0 \).

\subsection{Data Validity Filter}

When we draw a coin from the \texttt{ALL} basket, some coins may not have values on certain dates, introducing null values into our model. Therefore, we apply a validity filter to ensure that all sampled price points are valid.

Let \(H_{t,c}\) and \(L_{t,c}\) denote the daily high and low prices, respectively, of coin \(c\) on day \(t\).  
An episode is accepted only if:
\[
H_{s_i,c_i} \ge L_{s_i,c_i} > 0, \quad H_{e_i,c_i} \ge L_{e_i,c_i} > 0.
\]

If these conditions are not met, the draw is rejected and resampled.

\subsection{Loop of Sampling}

To ensure exactly \(n\) valid episodes are collected per holding-period interval, especially after filtering out invalid data points due to varying token lengths and the potential lack of data for longer horizon bands, we iteratively sample and validate episodes until the target number is reached or until 50 consecutive failed attempts occur. After each sampling iteration, the remaining episodes needed are updated as:

\[
\text{needed episodes} = n - \text{(number of valid episodes collected)}
\]

\subsection{Intra-Day Trading}

As we defined the buy and sell dates in the previous section, the model ensures that a token's buying date will always be earlier than its selling date. The simulator then draws its buy and sell prices as follows:

\begin{align}
P_i^{\text{buy}}  &= L_{s_i,c_i} + U_i^{(1)}\!\bigl(H_{s_i,c_i}-L_{s_i,c_i}\bigr),\\
P_i^{\text{sell}} &= L_{e_i,c_i} + U_i^{(2)}\!\bigl(H_{e_i,c_i}-L_{e_i,c_i}\bigr),
\end{align}
where \(U_i^{(1)},U_i^{(2)}\stackrel{\text{i.i.d.}}{\sim}
\operatorname{Uniform}(0,1)\).


\subsection{Risk-Free Benchmark over the Same Holding Period}

To calculate the \emph{excess} return for each crypto investment episode, we benchmark it against the return an investor would have earned by continuously rolling over the 1-month U.S.\ Treasury bill across the exact same calendar days.

Let \(r_t^{\text{rf}}\) denote the \emph{daily} risk-free simple return. We first pre-compute the cumulative log returns up to each day \(t\):

\[
\gamma_t \;=\; \sum_{s=0}^{t} \ln\!\bigl(1 + r_s^{\text{rf}}\bigr).
\]

Then, for an episode \(i\) with buy date \(s_i\) and sell date \(e_i\), the holding-period-specific risk-free return is calculated as:

\[
R_i^{\text{rf}} \;=\; \exp\!\bigl(\gamma_{e_i} - \gamma_{s_i}\bigr) - 1.
\]


\subsection{Net Trade Return After Fees}

To bring the simulation closer to real-world trading, we apply a proportional fee to \emph{each side} of every spot trading.  
Although fee schedules vary across exchanges, users typically pay on the order of 0.10 \% per side.  
We therefore set the one-way fee at \(\phi = 0.001\)\footnote{At the default (VIP 0) tier, major crypto exchanges quote taker fees of approximately 0.10\%---Binance 0.10\%, Gate.io 0.10\%, and OKX 0.10\% (maker 0.08\%). Figures are taken from the exchanges' official fee schedules (accessed May 2025).}

Let \(\phi\) denote that one-way proportional fee.  
If a given episode buys at \(P_i^{\text{buy}}\) and later sells at \(P_i^{\text{sell}}\), the net (post-fee) return is

\[
G_i
\;=\;
(1-\phi)\,
\frac{P_i^{\text{sell}}}{P_i^{\text{buy}}}\,
(1-\phi) \;-\; 1.
\]

\subsection{Excess Return}

For every valid simulated episode we compute the \emph{excess} return as

\[
X_i \;=\; G_i \;-\; R_i^{\text{rf}} .
\]

Here \(G_i\) is the net trade return after fees, and \(R_i^{\text{rf}}\) is the corresponding risk-free benchmark.

\section{Monte Carlo Simulation Results}\label{sec:results}
After conducting \(480{,}000{,}000\) Monte--Carlo episodes, we calculate a series of statistical metrics that include central tendency (mean and median), 
standard deviation, downside tail risk (VaR and CVaR), non-annualized risk-adjusted return (Sharpe and Sortino ratios), 
the proportion of episodes with losses $>10\%$\footnote{\texttt{p\_sig\_loss} in the code, defined as excess return $<-10\%$.}, 
the 75-th-percentile excess return, and the mean excess return above that threshold. These statistical metrics are computed in both overall and weekly aggregations, 
with minor differences: the overall (non-weekly) aggregation\footnote{The overall perspective is structured at two complementary levels: (i) aggregated by holding horizon and basket, 
and (ii) aggregated by basket, irrespective of holding horizon.} employs a 1\% threshold for downside risk and includes additional statistics such as the inter-quartile range, skewness, 
and kurtosis. In contrast, the weekly aggregation sets the downside risk threshold at 10\%, omitting skewness and kurtosis due to imbalanced weekly sample sizes throughout the dataset. 
Comprehensive explanations and formulas are detailed in Appendix~A\footnote{Refer to Appendix~A, Section~A.1.3 (\S\ref{app:a1-metrics}) for the definitions of all statistics; see Sections~A.1.1--A.1.2 (\S\ref{app:a1-notation}--\S\ref{app:a1-thresholds}) for notation and tail-probability thresholds.}.

\FloatBarrier                    
\begin{figure}[!t]               
  \centering
  \includegraphics[width=0.75\linewidth]{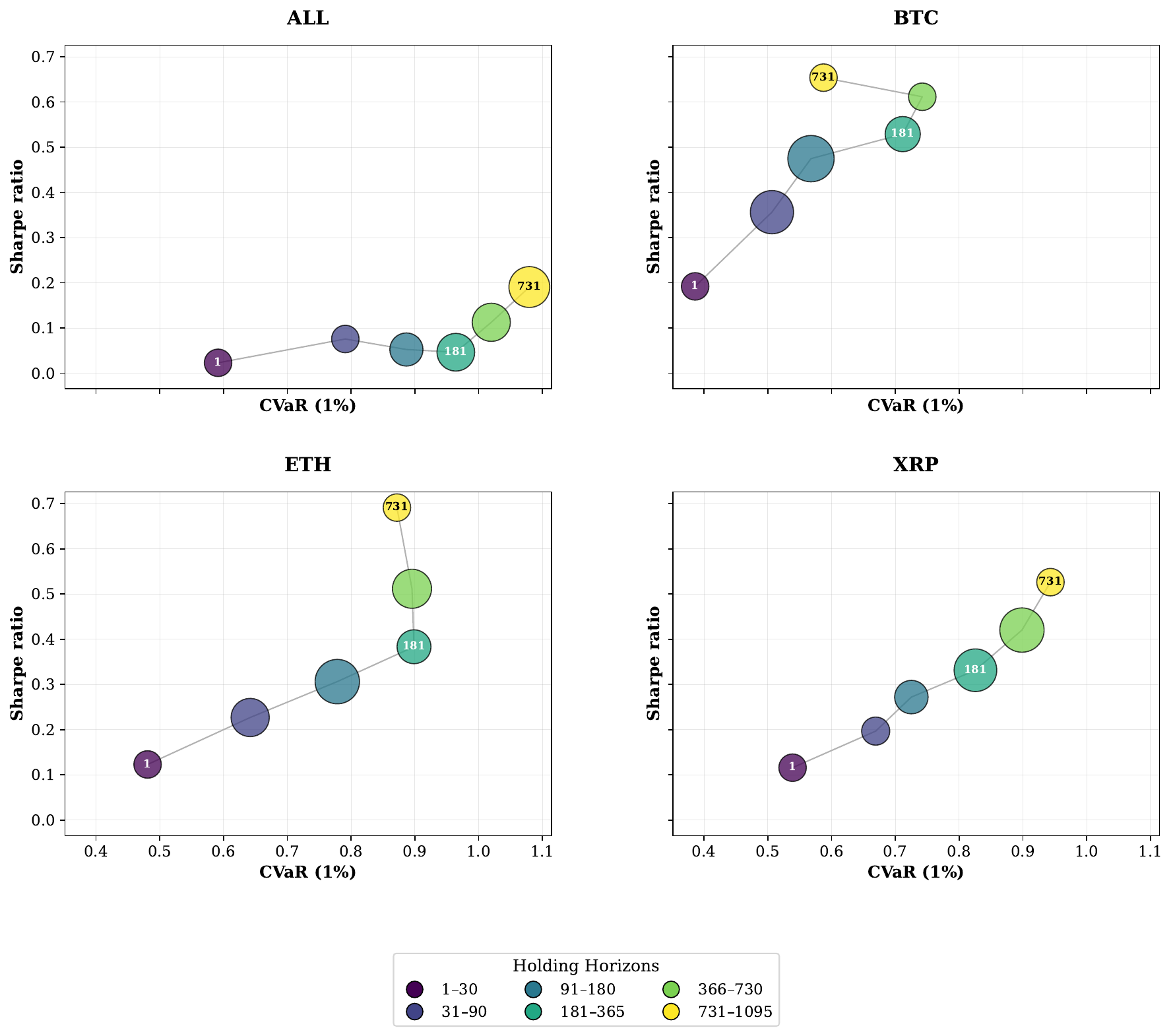}
  \caption{Risk–Return Trade-Off Across Holding Horizons}
  \label{fig:risk_return_horizon}
\end{figure}

Figure~1 illustrates the risk--return trade-off across different holding horizons for four baskets based on our simulation results, while the remaining four baskets are shown in Appendix~B, Figure~\ref{fig:B1}.
The \(x\)-axis represents the 1\% Conditional Value-at-Risk (CVaR), indicating the average loss experienced in the worst 1\% of scenarios, while the \(y\)-axis depicts the Sharpe ratio. Both metrics are computed separately for each basket and holding horizon. The size of each bubble represents the proportion of outcomes with losses exceeding 10\%. Bubble sizes are calculated for each basket–holding-horizon pair and then rescaled relative to the mean and dispersion of that basket’s other horizons; larger bubbles therefore indicate a  higher-than-average risk of losses within the same basket, while smaller bubbles indicate comparatively lower risks.\footnote{%
Bubble area \(A = \min\!\{\max[\,400\,(1+1.5\,z),\,400],\,2800\}\),
with \(z = (p_{\text{sig loss}} - \mu)/\sigma\).
This design ensures that the bubbles remain visible and emphasizes the risk associated with each holding horizon.} 

\noindent From Figure~\ref{fig:risk_return_horizon} we can observe that, as the length of the holding horizon 
increases, the $\text{CVaR}(1\%)$ climbs markedly for the \textsc{ALL} and \textsc{XRP} baskets. 
The \textsc{ETH} basket exhibits a similar upward trajectory through the 181--365-day horizon, 
but its tail risk subsequently moderates, declining from 0.899 to 0.872 in the longest period. Even though \textsc{BTC} 
performs relatively better than the other baskets in this simulation, its \(\text{CVaR}_{1\%}\) falls while the Sharpe ratio increases 
between the 181--365-day and 731--1095-day horizons; however, the longest horizon still shows a 58.7\,\% tail loss.

\noindent One notable point that grabs our attention is the \textsc{ALL} basket. As defined 
in our Monte-Carlo simulation, this basket randomly selects a single coin from 378 tokens for each episode.
This method captures the market-wide risk–return distribution rather than the experience of any 
single investor. The simulation highlights a critical insight: buy and hold is not a safe strategy for most 
cryptocurrencies. Specifically, the CVaR at $1\%$ approaches nearly 1 for holding horizons between 181 
and 365 days, and exceeds 1 for longer horizons, indicating that the worst 1\% of outcomes not only 
wipe out the principal but also forgo the alternative returns from Treasury bills. Given that each scenario is simulated independently millions of times, 
these findings reflect systemic market risk rather than a few outliers.

\FloatBarrier
\begin{figure}[!t] 
  \centering
  \includegraphics[width=0.55\linewidth]{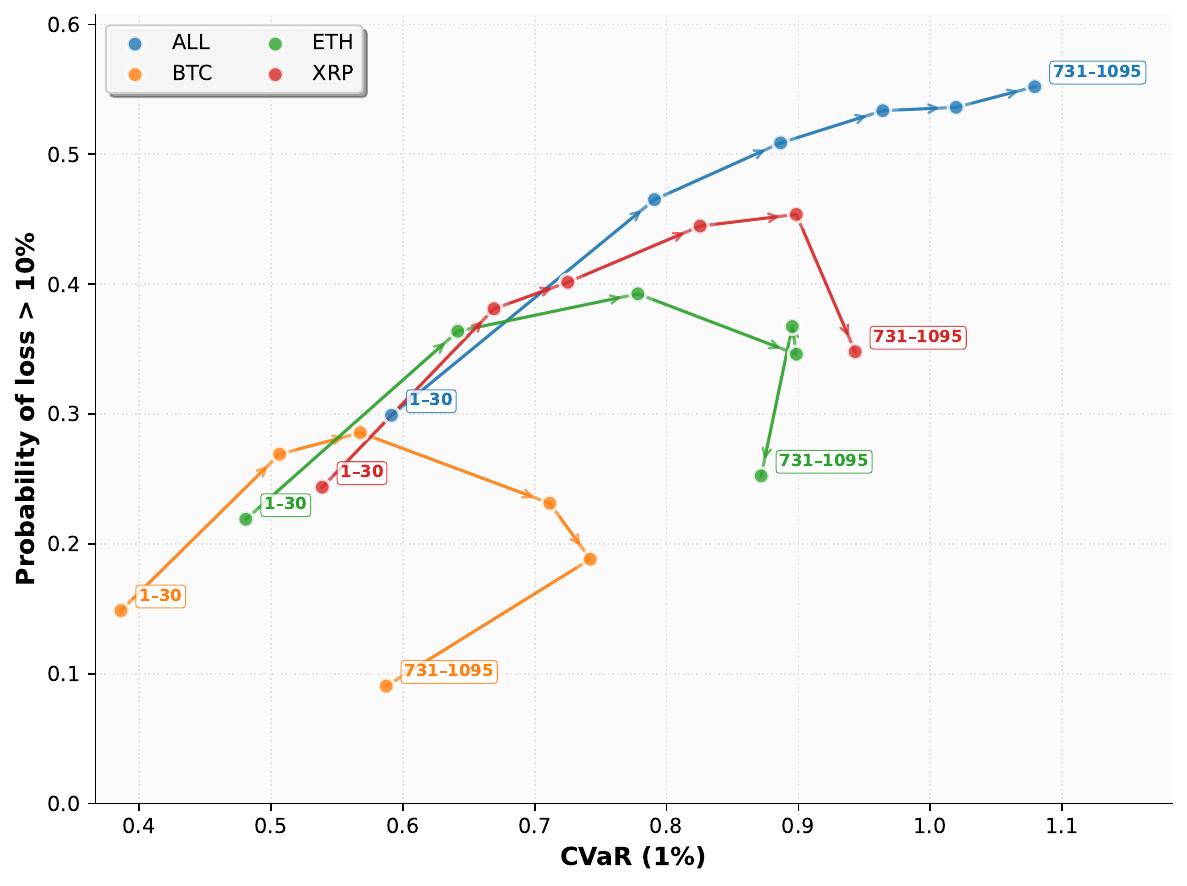}
  \caption{Trade-off between moderate and extreme risks across holding horizons}
  \label{fig:2}
\end{figure}

Figure~\ref{fig:2} reveals the trade-off between moderate risk (probability of losses greater than 10\%) and extreme risk (CVaR 1\%). The $x$-axis represents the CVaR (1\%), and the $y$-axis represents the probability of losses exceeding 10\%. Arrows indicate the progression from shorter to longer holding periods, with clearly marked starting and ending periods. We plot four baskets in Figure~\ref{fig:2} for clarity, and an additional four are provided in Figure~\ref{fig:B2}.

Some argue that purchasing cryptocurrencies, forgetting about them, 
and never selling (``HODL''\footnote{HODL is a term commonly used in the 
crypto community meaning to buy-and-hold indefinitely.}) is an ideal investment strategy.
Our analysis addresses this claim from two perspectives. First, we find that for certain baskets (tokens),
longer holding horizons, typically from 181--365 days onward, show the probability of moderate losses beginning
to decline, while extreme risk (CVaR) rises through mid-range horizons before declining at the longest 
holding periods. However, this pattern is not generalizable when considering
the crypto market as a whole. In fact, observations from the \textsc{ALL} basket show that, 
as the holding period lengthens, both the probability of experiencing moderate losses and the severity of
extreme losses increase significantly. Specifically, the worst 1\% group's losses approach 100\% of
their initial investment, and the probability of experiencing losses greater than 10\% exceeds 50\%
once the holding period extends beyond 181--365 days. Crucially, this scenario is not extreme but represents 
the overall distribution observed in our data. In summary, cryptocurrency investors should be cautious
and avoid relying solely on successful cases like Bitcoin and Ethereum to justify the 'HODL' strategy. 
In most situations, both moderate and extreme risks rise as the holding period lengthens.

\FloatBarrier
\begin{figure}[!t]  
  \centering
  \includegraphics[width=0.75\linewidth]{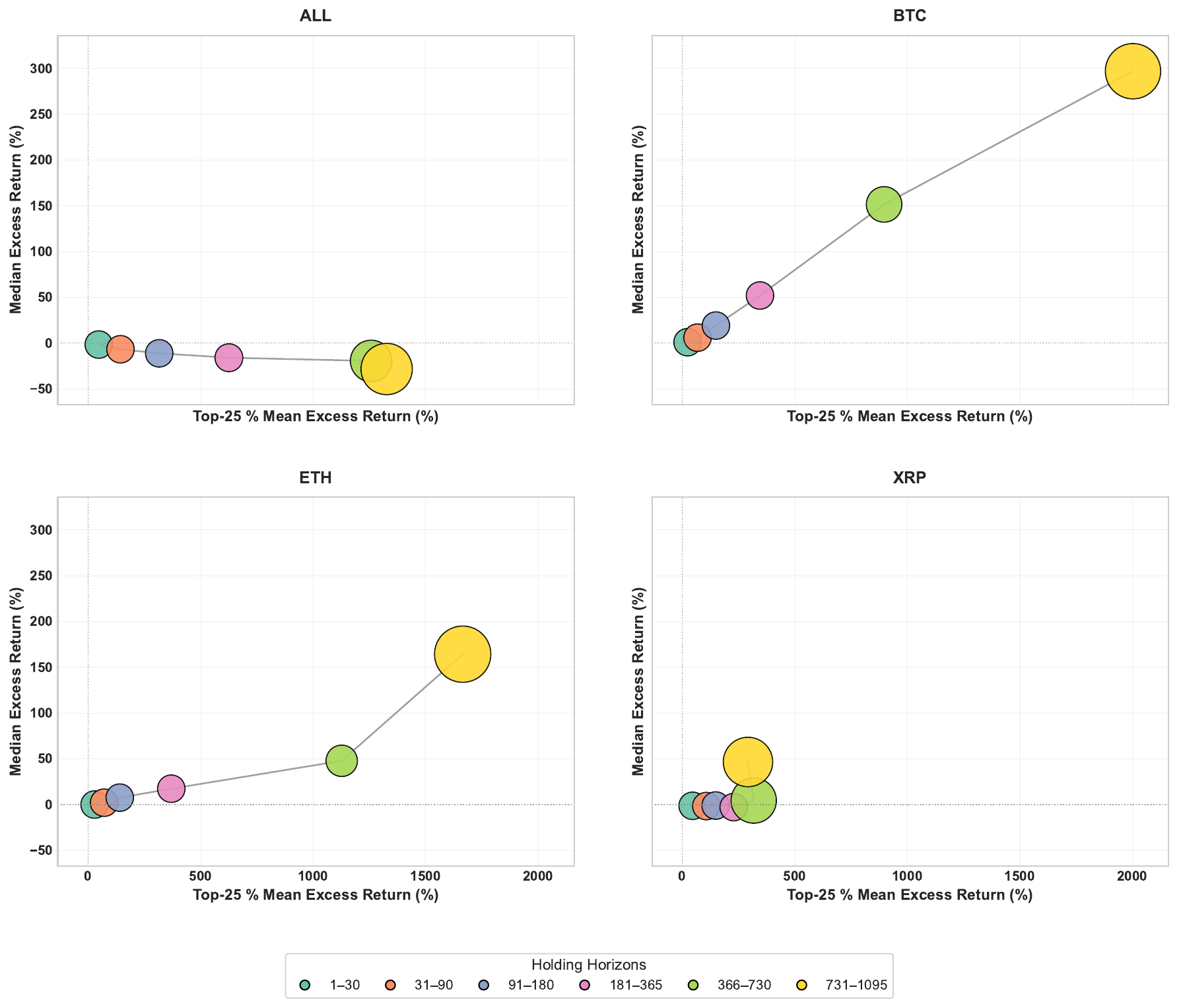}
  \caption{Median vs.\ top-25\% mean excess returns across holding horizons}
  \label{fig:3}
\end{figure}

Figure~\ref{fig:3} plots four baskets, showing median excess returns
against the mean excess return of the top-25\,\% observations; the
remaining four baskets are presented in Figure~\ref{fig:B3}.
 The \(x\)- and \(y\)-axes are
expressed in percent. Colours encode holding horizons, whereas bubble
area visualises dispersion: within each basket the area is proportional
to the inter-quartile range (IQR) of excess returns for that horizon.
A larger bubble therefore indicates that the central 50 \% of
observations are more widely spread (higher volatility), while a smaller
bubble indicates a tighter clustering (lower volatility) relative to
other horizons in the same basket.\footnote{%
Bubble area is computed as
\(A=\min\{\max[\,400\,(1+1.5\,z),\,400],\,2800\}\),
where
\(z=(\mathrm{IQR}-\mu_{\mathrm{IQR}})/\sigma_{\mathrm{IQR}}\)
standardises the IQR within each basket.}

From Figure~\ref{fig:3} we observe that, as the holding horizon
increases, the top-25\,\% mean excess return generally rises.  
XRP shows a slight backward step when moving from the 366–730-day window
to the 731–1095-day window, yet its overall trend remains upward.  
Dispersion (IQR) for each basket tends to
increase with holding length, especially across the 366–1095-day span.  
Viewed token-by-token, this pattern appears to support the ``HODL''
claim.  The ALL basket, however, tells a different story: while
its top-quartile mean at the longest horizon (731–1095 days) exceeds
1326.7\,\%, the median return is -28.4\,\%.  Its larger IQR confirms the wider spread of outcomes. 
Taken together, these findings suggest that many ``HODL missionaries''
rely on the performance of the fortunate upper quartile to overstate the
typical rewards available to the average investor across a broad
universe of cryptocurrencies.

\clearpage          
\cleardoublepage

\section{Methodology}
\subsection{Initial configuration}

The model takes one basket at a time (for example, BTC) for each run.
The endogenous block consists of four targets: the median excess return, the 10\% conditional value-at-risk (CVaR), the mean of the top quantile of returns (top 25\%), and the Sharpe ratio.
The macro-financial block includes a set of macro-financial variables. All variables were subjected to unit-root tests prior to model estimation and were pre-labeled according to their stationarity status for later use in the model.
Posterior sampling relied on the No-U-Turn Sampler (NUTS) \citep{JMLR:v15:hoffman14a} with four sequential chains. For all eight baskets, we fixed the sampler hyperparameters at \texttt{target\_accept}=0.95 and \texttt{max\_treedepth}=12.

\subsection{Load basket and horizon panels}

We load the merged dataset of sell-side statistics. For each basket, observations are grouped according to the week of selling. Specifically,
if a basket is sold within the same calendar week—defined from Monday through Sunday—the selling date is shifted  to the start of that week (Monday). 
This ensures a consistent weekly index across baskets.
The endogenous metrics are constructed directly from realized risk–return measures.
In other words, for each basket, holding horizon, and selling week, we compute the realized statistics as defined in the initial
configuration.The macro-finance variables are aligned to the weekly time index of the endogenous features.

We load each variable's stationarity test results and assess unit-root status with DF-GLS, KPSS, and Zivot–Andrews under two specifications: constant only (c)
and constant plus linear trend (ct) at the 5\% significance level \citep{elliott1996efficient,kwiatkowski1992kpss,zivotandrews1992}. Within each specification,
a series is labeled stationary if both DF-GLS and Zivot–Andrews reject the unit-root null (p < 0.05) 
 while KPSS fails to reject stationarity (p > 0.05). It is labeled unit root if the reverse pattern holds. 
 When the three tests point in different directions, the outcome is labeled ambiguous.

We classify the unit-root test outcomes into three categories: \textsc{Level}, \textsc{Trend}, and \textsc{RW}, using constant-only (\(c\)) versus constant-plus-trend (\(ct\)) specifications.
If the \(c\) specification is stationary, we keep the series in its original form (\textsc{Level}).
If only the \(ct\) specification is stationary, we remove a deterministic trend via a strictly causal rolling linear detrend and use the residuals (\textsc{Trend}).
If the outcome indicates a unit root or is ambiguous, we treat the series as a random walk (\textsc{RW}).

For macro variables, there is no horizon dimension, so the label is taken directly from the unit-root test outcome.
For endogenous variables, a single series can exhibit different unit-root outcomes across prediction horizons.
To ensure consistency, we aggregate per-horizon tags using a majority rule to obtain a single global tag for each variable.
This means the category with the highest count becomes the global tag, and we apply the same treatment across all horizons.
In the case of ties, the outcome is resolved conservatively with the ordering \(\textsc{RW} \succ \textsc{Trend} \succ \textsc{Level}\).
This design ensures horizon-consistent preprocessing for each variable and improves comparability of estimates and impulse responses across horizons.

\paragraph{Fractional differencing for {\normalfont\textsc{RW}} macro inputs.}
For macro–finance series labeled \textsc{RW}, we apply strictly causal fractional differencing \citep{granger1980long,hosking1981fractional} with fixed order \(d=0.5\)
and a truncation at \(K=200\) lags:
\[
y^{\mathrm{MF}}_{t}
= \sum_{k=0}^{K} w_k\, x^{\mathrm{MF}}_{t-k},
\qquad
w_k = (-1)^k \binom{d}{k},
\qquad
\binom{d}{k}=\frac{d(d-1)\cdots(d-k+1)}{k!},
\quad
d=0.5,\; K=200.
\]
Here, $x^{\mathrm{MF}}_{t}$ is the macro–finance series and
$y^{\mathrm{MF}}_{t}$ is its fractionally differenced value at time $t$; $w_k$
weights the $k$-period lag $x^{\mathrm{MF}}_{t-k}$. The implementation is
strictly causal with a finite truncation of the infinite sum.
To avoid numerical instability, we implement the weights through the stable
recurrence $w_0=1$ and $w_k=w_{k-1}\frac{-(d-(k-1))}{k}$ $(k\ge1)$\footnote{We truncate the recurrence earlier: once a weight’s absolute value falls below $10^{-10}$.},
which is equivalent to the binomial definition.

For endogenous variables labeled as \textsc{RW}, we transform the series into first differences:
\[
y_t^{(RW)} \;=\; y_t - y_{t-1}, \qquad t = 2,3,\dots,T.
\]
This ensures that variables classified as random walks are rendered stationary (I(0)) before entering the model.

The same detrending is applied to endogenous targets and macro finance  series when labeled \textsc{Trend}.
Let $L=52$ weeks be the rolling window length.
At each time $t\ (\ge L)$, define the past-only window $\mathcal W_t=\{t-L+1,\dots,t\}$ and index observations by $j=0,1,\dots,L-1$ so that the last point is $j=L-1$.
Within $\mathcal W_t$, fit an OLS line
\[
x_{t-L+1+j} \;=\; \alpha_t+\beta_t\,j+\varepsilon_{t,j},\qquad j=0,1,\dots,L-1,
\]
using only data in $\mathcal W_t$ (strictly causal).
Let $\hat\alpha_t,\hat\beta_t$ be the OLS coefficients.
We take the residual at the end of the window as the detrended value:
\[
x^{(\textsc{Trend})}_t \;=\; x_t - \big(\hat\alpha_t+\hat\beta_t\,(L-1)\big),\qquad t=L,L+1,\dots
\]
If the available history is shorter than $L$ for the entire series, we conservatively fall back to the first difference,
\[
x^{(\textsc{Trend})}_t \;=\; x_t - x_{t-1}.
\]

After transforming each macroeconomic series into a strictly causal $I(0)$ series by fractionally differencing the unit-root tags and rolling linear detrending the trend-stationary tags, we construct 
\emph{EMA}  and \emph{VOL}  features at 
multi-week horizons $w \in \{4,8,12,24\}$. This serves two purposes: macro factors 
typically have lagged effects, and including raw weekly lags would greatly expand 
the feature amount and lead to basket-specific lag heterogeneity under bootstrap resampling. 
By contrast, EMA and VOL compress lag structure, 
retaining long-memory while keeping the predictor set parsimonious.

For each transformed series $y^{\mathrm{MF}}_t$, the exponentially weighted moving average is
\[
\mathrm{EMA}^{(w)}_t \;=\; \alpha_w y^{\mathrm{MF}}_t + (1-\alpha_w)\,\mathrm{EMA}^{(w)}_{t-1},
\qquad \alpha_w = \frac{2}{w+1}.
\]
The recursion starts only once $w$ past observations are available.

We compute realized volatility from the trailing $w$-week window as
\[
\bar{y}^{\mathrm{MF}}_{t,w} = \frac{1}{w}\sum_{j=0}^{w-1} y^{\mathrm{MF}}_{t-j}, 
\qquad
\mathrm{VOL}^{(w)}_t \;=\;
\sqrt{\frac{1}{\,w-1\,}\sum_{j=0}^{w-1}\!\big(y^{\mathrm{MF}}_{t-j} - \bar{y}^{\mathrm{MF}}_{t,w}\big)^2}.
\]

Together, $\mathrm{EMA}^{(w)}_t$ captures lagged movements and 
$\mathrm{VOL}^{(w)}_t$ captures rolling-window volatility.

\paragraph{Tensor alignment.}
To build horizon-consistent panels, all horizons are harmonized to the same macro-finance feature set and column order.
For each horizon \(h\), the feature matrix is reindexed to a common column set \(\mathcal C\).
Next, we restrict the weekly index to the intersection of dates available across all horizons (after transforms such as differencing, fractional differencing, and rolling detrend/EMA/VOL), so that every observation uses the same feature set and the same time index\footnote{On the common time grid \(\mathcal T\) we assemble \(Y\in\mathbb R^{T\times H\times n_y}\) and \(X\in\mathbb R^{T\times H\times n_x}\), aligned along time \(T=|\mathcal T|\), horizons \(H\), and variables \((n_y,n_x)\).}.

\subsection{Causal standardisation}

After the stationarity transforms in §6.2, each horizon $h$ uses an aligned matrix $X_h$ with $T_h$ time points and one column per feature or target.
Denote the value in column $k$ at time $t$ by $x^{(k)}_{t,h}$.
We apply an expanding-window z-score transformation to avoid look-ahead bias:
\[
\mu^{(k)}_{t,h} = \frac{1}{t}\sum_{i=1}^{t} x^{(k)}_{i,h},
\qquad
\sigma^{(k)}_{t,h} = \sqrt{\frac{1}{t}\sum_{i=1}^{t}\!\bigl(x^{(k)}_{i,h}-\mu^{(k)}_{t,h}\bigr)^2}.
\]
To ensure numerical stability, the denominator is clipped at $\varepsilon=10^{-2}$:
\[
s^{(k)}_{t,h}=\max\{\sigma^{(k)}_{t,h},\varepsilon\},
\qquad
z^{(k)}_{t,h}=\frac{x^{(k)}_{t,h}-\mu^{(k)}_{t,h}}{s^{(k)}_{t,h}},\quad t=1,\dots,T_h.
\]

\subsection{Stability selection}

We adapt the stability selection framework of \citet{MeinshausenBuehlmann2010} to our grouped macro-finance predictor structure.
\paragraph{Purged time-series split}
Our horizons at time $t$ are the \emph{realised} risk–return distribution. 
Thus, the effective horizon must be fully realized by time $t$; otherwise trailing windows overlap across folds 
(for example, a 30-day horizon only becomes fully observed at $t{+}30$). To break this overlap we leave a per-fold gap following \citet{lopez2018advances}.
Let $\tau_k$ be the first index of the $k$-th validation block; for horizon $h$ we truncate the training indices to

\[
\mathcal I_{k}^{\text{train}}(h) \;=\; \{\,t : t \le \tau_k - g(h) - 1\,\}, 
\qquad
\mathcal I_{k}^{\text{test}} \;=\; \{\,t : \tau_k \le t \le \tau_k^{\max}\,\},
\]
with gap
\[
g(h) \;=\; \left\lceil \frac{h}{7} \right\rceil + 1,
\qquad h \in \mathcal H,
\]
where $\mathcal H$ is the set of horizons in days. Each $g(h)$ converts the daily horizon to weeks through the ceiling and adds one extra week as a buffer. 
This prevents overlapping trailing windows between train and validation.\footnote{Folds are generated with scikit-learn's \citep{scikit-learn} \texttt{TimeSeriesSplit}.}

\paragraph{Hyperparameter selection}
For each horizon $h\in\mathcal H$ we use the purged folds \(\mathcal I^{\mathrm{train}}_f(h),\mathcal I^{\mathrm{test}}_f(h)\) 
defined above.\footnote{We attempt $K=3$ purged folds for all horizons; 
if any horizon would yield fewer than two valid splits we downgrade to $K=2$. 
Horizons that still fail the two-fold requirement skip CV and fall back to the shared shrinkage level.} 
Let $\mathcal F_h$ denote the folds that remain after the purge gap.

Let $X_h$ denote the z-scored predictors at horizon $h$ (with $p$ screened variables), and let $Y_h$ be the corresponding standardized target matrix (with $m$ target series). 
The multitask Elastic Net with $\rho=1/2$ \citep{zou2005regularization} solves
\[
\widehat{B}_h(\alpha)
\;\in\;\arg\min_{B\in\mathbb R^{p\times m}}
\left\{
\frac{1}{2T}\,\bigl\|Y_h - X_h B\bigr\|_F^2
\;+\;
\alpha\Bigl(\tfrac{1}{4}\,\|B\|_F^2 + \tfrac{1}{2}\sum_{j=1}^p \bigl\|B_{j\cdot}\bigr\|_2\Bigr)
\right\},
\]
so no intercept is needed. For each horizon with $|\mathcal F_h|\ge 2$ we pick
\[
\widehat{\alpha}(h)\ \in\ \arg\min_{\alpha\in\mathcal A}\ 
\frac{1}{|\mathcal F_h|}\sum_{f\in\mathcal F_h}
\frac{1}{|I^{\mathrm{test}}_{f}(h)|}\,
\bigl\|Y_{h,I^{\mathrm{test}}_{f}(h)} - X_{h,I^{\mathrm{test}}_{f}(h)}\,\widehat{B}^{(f)}_h(\alpha)\bigr\|_F^2,
\]
where each candidate $\alpha$ is run through purged cross-validation: 
every fold re-fits the Elastic Net on its training slice to obtain $\widehat{B}^{(f)}_h(\alpha)$, 
applies those coefficients to the unseen observations, and records the mean-squared validation loss for that setting. 
The $\alpha$ that attains the smallest validation loss becomes $\widehat{\alpha}(h)$.
\footnote{\raggedright $\mathcal A$ denotes the candidate $\alpha$ sequence, which follows scikit-learn's default log-spaced path used by \texttt{MultiTaskElasticNetCV}.\par}
Horizons without two valid folds inherit a shared shrinkage level
\[
\alpha_{\mathrm{shared}} =
\begin{cases}
\mathrm{median}\,\{\widehat{\alpha}(h): |\mathcal F_h|\ge 2\}, & \text{if at least one horizon has two valid folds},\\[4pt]
0.1, & \text{otherwise},
\end{cases}
\]

\paragraph{Bootstrap-based feature selection}
To preserve short-run temporal dependence in the bootstrap resampling, we use a non-overlapping block bootstrap (NBB) 
for feature selection \citep{lahiri2003resampling}.
For each horizon $h$, each bootstrap sample is constructed from length-$b$ contiguous time blocks drawn with replacement from the aligned in-sample tensor described in §6.2. 
Block start positions are spaced by $b$ (non-overlapping) and sorted from earliest to latest.\footnote{More specifically, candidate block starts $\{0,b,2b,\ldots,n-b\}$ are drawn i.i.d.\ with replacement, 
then sorted and concatenated as length-$b$ intervals until the index reaches $n$ (truncating any excess).}

The block size follows the $n^{1/3}$ rule \citep{hall1995blocking} and is clipped to a weekly range:
\[
  b \;=\; \min\!\bigl\{20,\;\max\!\{4,\ \operatorname{round}(n^{1/3})\}\bigr\}.
\]
Inside each block the observations remain consecutive, so every resampled block respects the original time order.

Let $R$ denote the number of bootstrap draws.\footnote{We set $R=1000$ in our bootstrap model.}
For each bootstrap draw $r=1,\dots,R$ we sample an index sequence $I_r$ of length $T$ (as defined in §6.2) by concatenating the length-$b$ blocks above, and refit the multitask Elastic Net on $(X_{h,I_r}, Y_{h,I_r})$ with the fixed hyperparameter $\alpha_h^\star$ selected in the prior hyperparameter-selection stage, keeping the Elastic Net $\ell_1$ ratio at $\rho = 1/2$ as before:
\[
\widehat{B}_h^{(r)} \in 
\arg\min_{B\in\mathbb R^{p\times m}}
\Bigl\{
\frac{1}{2|I_r|}\,\|Y_{h,I_r}-X_{h,I_r}B\|_F^2
+\alpha_h^\star\bigl(\tfrac{1}{4}\|B\|_F^2+\tfrac{1}{2}\sum_{j=1}^p\|B_{j\cdot}\|_2\bigr)
\Bigr\}.
\]
We treat the untransformed macro–finance variables as the base variables, 
indexed by the set $\mathcal{V}$. For each base variable $v\in\mathcal{V}$, let $\mathcal{G}_v$ collect 
all EMA and VOL transforms constructed from $v$. In the bootstrap, for each draw $r$ we declare transform $j$ 
selected whenever $\big\|\widehat{B}_{h,j\cdot}^{(r)}\big\|_2>0$. These selections allow us to compute (i) the base–level stability and 
(ii) the within–base conditional stability for each transform $\gamma\in\mathcal{G}_v$.

\[
\hat\pi_{\mathrm{base}}(v)
=\frac{1}{R_{\mathrm{valid}}}\sum_{r=1}^{R}\mathbf 1\!\left\{\exists\,\gamma\in\mathcal G_v:\ \|( \widehat{B}_h^{(r)})_{[v,\gamma],\cdot}\|_2>0\right\},
\qquad
\hat\pi_{\mathrm{cond}}(v,\gamma)
=\frac{\sum_{r=1}^{R}\mathbf 1\!\left\{\|( \widehat{B}_h^{(r)})_{[v,\gamma],\cdot}\|_2>0\right\}}
{R_{\mathrm{valid}}\,\hat\pi_{\mathrm{base}}(v)},
\]
If a base variable is never selected, we set its conditional stability to zero.
For each horizon $h$, we retain a base variable $v$ if its base-level stability 
satisfies $\hat{\pi}_{\mathrm{base}}(v)\ge \tau_g$. For every base that passes 
this threshold, we then consider its transforms and keep a 
transform $\gamma$ only if its conditional stability meets 
$\hat{\pi}_{\mathrm{cond}}(v,\gamma)\ge \tau_c$.\footnote{In our implementation we use $\tau_g = 0.55$ and $\tau_c = 0.50$.} Within each of the EMA and VOL types, 
we retain the top-1 transform ranked by $\hat{\pi}_{\mathrm{cond}}(v,\gamma)$. 
The selected feature set at horizon $h$ is denoted $\mathcal{S}_h$; the 
overall selected feature set is their union $\mathcal{S}=\bigcup_{h\in\mathcal{H}}\mathcal{S}_h$, which preserves horizon-specific signals while avoiding multi-collinearity within transform families.

\subsection{Pre-Bayesian Preparation}

After the causal standardisation (§6.3) and the stability-selection filter (§6.4), each horizon $h\in\mathcal H$ retains the feature set $\mathcal S_h$ and their 
union $\mathcal S = \bigcup_h \mathcal S_h$. On the common weekly grid $\mathcal T = \{1,\dots,T\}$, we work with the standardized tensors
\[
\bigl\{(\widetilde{Y}^{(C)}_{t,h},\,\widetilde{X}^{\mathcal S}_{t,h}) : t\in\mathcal T,\ h\in\mathcal H\bigr\},\qquad \widetilde{Y}^{(C)}\in\mathbb R^{T\times H\times n_y},\; \widetilde{X}^{\mathcal S}\in\mathbb R^{T\times H\times |\mathcal S|},
\]
where $\widetilde{Y}^{(C)}_{t,h}$ and $\widetilde{X}^{\mathcal S}_{t,h}$ denote the post–expanding-$z$-score targets and screened features observed at time $t$.

\paragraph{Training window.}
Before fitting the Bayesian multi-horizon local projection, we reuse the same gap as the purged time-series split, to 
ensure every observation has a fully realised future outcome. 
For each horizon $h$, define the effective gap in weeks
\begin{equation}\label{eq:training-gap}
g(h) \;=\; \Bigl\lceil\frac{h}{7}\Bigr\rceil + 1,
\end{equation}
let $g_{\max}=\max_{h\in\mathcal H} g(h)$. The estimation sample ends at
\[
t^\star \;=\; T - g_{\max},
\]
so the training index becomes $\mathcal T_{\mathrm{train}} = \{1,\dots,t^\star\}$. 

\paragraph{Future-target tensor.}
For $t\in\mathcal T_{\mathrm{train}}$ and $h\in\mathcal H$, the dependent variable corresponds to the realized outcome at week $t+g(h)$. We therefore construct
\[
\widetilde{Y}^{(F)}_{t,h} \;\equiv\; \widetilde{Y}^{(C)}_{t+g(h),h},\qquad \widetilde{Y}^{(F)}\in\mathbb R^{t^\star\times H\times n_y},
\]
which contains only fully realized future outcomes. 

\paragraph{Reference scaling.}

To enable conversion of posterior draws from standardized units back to original scales, 
we record the expanding standard deviations at the end of the training period $t^\star$. These reference scales—denoted $\sigma_y(h)$ 
for each target and $\sigma_x^{\mathcal S}(h,j)$ for each selected feature—provide a consistent basis for converting z-scored coefficients 
to native units (see Appendix~\ref{app:a3} for details). 

\subsection{Bayesian multi-horizon local projection}
\label{sec:bayes-model}
Building on \citet{Ferreira2025} and related work on smooth Bayesian local 
projections \citep[e.g.,][]{tanaka2020bayesian,huber2024general}, we adopt a Bayesian local projection 
framework for multi-horizon impulse response estimation.
We model the standardized tensors from §6.5 using the notation established earlier.
We define the horizon-specific design vector $Z_{t,h}$ as
\[
  Z_{t,h}
  \;=\;
  \begin{bmatrix}
    \widetilde{Y}^{(C)}_{t,h} \\[3pt]
    \widetilde{X}^{(\mathcal{S})}_{t,h}
  \end{bmatrix}
  \in \mathbb{R}^{P},
  \qquad
  P \equiv P_y + P_x,\quad
  P_y = n_y,\quad
  P_x = |\mathcal{S}|.
\]
By construction, $Z_{t,h}$ only uses information observed at week $t$, while the aligned future outcome satisfies
\[
  \widetilde{Y}^{(F)}_{t,h} \;=\; \widetilde{Y}^{(C)}_{t+g(h),h},
\]
as defined in §6.5.

\paragraph{Likelihood.}
For each horizon $h$ we use a Student-$t$ likelihood \citep{Geweke1993Bayesian} with horizon-specific linear predictor:
\begin{align}
  \widetilde{Y}^{(F)}_{t,h} \;&\sim\; \mathrm{Student}\text{-}t_{\nu=6}\bigl(\mu_{t,h},\,\Sigma_h\bigr), 
  \label{eq:likelihood}\\
  \mu_{t,h} \;&=\; \alpha_h + Z_{t,h}\,\beta_h,\qquad
  \alpha_h\in\mathbb{R}^{n_y},\;\; \beta_h\in\mathbb{R}^{P\times n_y}.
  \label{eq:linear-predictor}
\end{align}

\paragraph{Residual covariance.}
For estimation stability, we assume independent Student-$t$ innovations across targets at each horizon,
\[
\Sigma_h=\mathrm{diag}\!\big(\sigma_{h,1}^2,\ldots,\sigma_{h,n_y}^2\big).
\]
Instead of capturing residual correlation across targets, our hierarchical shrinkage priors—with shared global scales 
$\tau_y$ and $\tau_x$ across all targets—already induce cross-target information pooling at the coefficient level.

\paragraph{Smoothing across unevenly-spaced horizons.}
Our six forecast horizons (30, 90, 180, 365, 730, 1095~days) are unevenly spaced in time.
If we smooth coefficients uniformly across horizon indices,
the prior would treat each horizon distance equally.
To impose constant smoothness based on the real time distance,
we map each horizon to weeks $w_h \equiv \lceil h/7\rceil$ and scale the RW1 innovations
by the square root of the weekly spacing, so each coefficient path evolves as
\begin{equation}
  \Delta_h \;=\; w_h - w_{h-1},\qquad
  b_h \;=\; b_{h-1} \;+\; \epsilon_h\,\tau\,\sqrt{\Delta_h},
  \quad \epsilon_h\sim\mathcal{N}(0,1).
  \label{eq:rw1-smoothing}
\end{equation}
This ensures $\mathrm{Var}(b_h - b_{h-1}) = \tau^2\,\Delta_h$,
so that the prior variance scales proportionally with time gap between horizons.

\subsubsection*{Priors}

\paragraph{Hierarchical intercept priors.}
For each target $y$, we assign a hierarchical prior on the intercepts across forecast horizons:
\begin{align}
  \alpha_{h,y} \;&=\; \mu_{\alpha,y} + \sigma_{\alpha,y}\,\tilde\alpha_{h,y}, \qquad
  \tilde\alpha_{h,y}\sim\mathcal{N}(0,1), \nonumber\\
  \mu_{\alpha,y} &\sim \mathcal{N}(0,1), \qquad
  \sigma_{\alpha,y} \sim \mathcal{HN}(0.5), \qquad y=1,\dots,n_y .
  \label{eq:alpha-prior}
\end{align}
This non-centered parameterization allows intercepts to vary moderately across horizons 
while pooling information at the target level through shared mean $\mu_{\alpha,y}$ and scale $\sigma_{\alpha,y}$.

\paragraph{Coefficient blocks.}
The slope coefficients consist of an endogenous block and a macro-finance block:
\[
\beta_h=\begin{bmatrix}\beta^{(y)}_h \\[2pt] \beta^{(x)}_h\end{bmatrix},
\quad
\beta^{(y)}_h\in\mathbb{R}^{P_y\times n_y},\;
\beta^{(x)}_h\in\mathbb{R}^{P_x\times n_y}.
\]
For both coefficient blocks we retain the RW1 design from \eqref{eq:rw1-smoothing}: 
for $h\ge 2$ each step is multiplied by $\sqrt{\Delta_h}$. The endogenous coefficients use $p=1,\dots,P_y$ 
while the macro-finance block uses $p=1,\dots,P_x$. The hierarchical shrinkage factors $s^{(y)}_{p,y}$ and $s^{(x)}_{p,y}$ 
control the innovation variance for each coefficient path.

We adopt a global--local shrinkage hierarchy \citep[e.g.,][]{CarvalhoPolsonScott2010Horseshoe,PiironenVehtari2017}:
\[
s^{(y)}_{p,y} \;=\; \tau_y\,\lambda^{(y)}_{p,y},
\qquad
s^{(x)}_{p,y} \;=\; \tau_x\,\lambda^{(g)}_{p}\,\lambda^{(\ell)}_{p,y},
\]
where $\tau_y,\tau_x$ are global scales and 
$\lambda^{(y)}_{p,y}$, $\lambda^{(g)}_{p}$, $\lambda^{(\ell)}_{p,y}$ 
are local (and group-level) multipliers.
The key difference between the endogenous and macro-finance shrinkage factors 
is that macro-finance predictors employ a three-level hierarchy:
the group factor $\lambda^{(g)}_{p}$ is shared across all targets, 
so when $\lambda^{(g)}_{p}\approx 0$, feature $p$ is jointly shrunk to zero 
across all target variables, reflecting that this macro-finance predictor provides 
limited explanatory power for the outcomes.
The local factor $\lambda^{(\ell)}_{p,y}$ then controls the strength of feature $p$'s 
effect on each individual target variable.
This design reflects that macro shocks jointly affect risk and return outcomes,
while allowing target-specific heterogeneity in both magnitude and direction.
In contrast, endogenous coefficients use a two-level hierarchy 
without cross-target group sharing.

The hyperpriors are:
\begin{align*}
  \tau_y &\sim \mathcal{HN}(\tau_{0,y}),\quad
  \tau_x \sim \mathcal{HN}(\tau_{0,x}),\qquad
  \lambda^{(y)}_{p,y},\,\lambda^{(g)}_{p},\,\lambda^{(\ell)}_{p,y} \stackrel{\text{i.i.d.}}{\sim} \mathcal{H}t_{\nu=4}(1), \\
  \intertext{where}
  \tau_{0,y} &= \frac{p_{0,y}}{\max(P_y-p_{0,y},1)}\cdot T_{\mathrm{train}}^{-1/2},
  \quad p_{0,y} \equiv \min\{\max(0.25\,P_y,1),\,5\}, \\
  \tau_{0,x} &= \frac{p_{0,x}}{\max(P_x-p_{0,x},1)}\cdot T_{\mathrm{train}}^{-1/2},
  \quad p_{0,x} \equiv \min\{\max(0.10\,P_x,1),\,5\}.
\end{align*}
Here, $\mathcal{HN}(\sigma)$ and $\mathcal{H}t_{\nu}(\sigma)$ denote half-normal and half-Student-$t$ 
distributions with mode at zero. We use $\nu=4$ for moderately heavy tails, 
allowing important features to escape shrinkage while pushing most coefficients toward zero.

\medskip
\noindent\emph{(i) Endogenous block \(\beta^{(y)}_h\): global--local shrinkage RW1.}
For $p=1,\dots,P_y$ and $y=1,\dots,n_y$,
\begin{align}
\beta^{(y)}_{1,p,y} &\sim \mathcal{N}(0,\,0.3^2), \\
\beta^{(y)}_{h,p,y} &= \beta^{(y)}_{h-1,p,y} \;+\; s^{(y)}_{p,y}\,\sqrt{\Delta_h}\,\varepsilon^{(y)}_{h-1,p,y},
\qquad \varepsilon^{(y)}_{h-1,p,y}\sim\mathcal{N}(0,1),\quad h\ge 2,
\end{align}
with $s^{(y)}_{p,y}=\tau_y\,\lambda^{(y)}_{p,y}$ and hyperpriors defined above.

\medskip
\noindent\emph{(ii) Macro-finance block \(\beta^{(x)}_h\): group–local shrinkage RW1.}
For $p=1,\dots,P_x$ and $y=1,\dots,n_y$ we retain the same RW1 recursion, but with a tighter initial variance:
\begin{align}
\beta^{(x)}_{1,p,y} &\sim \mathcal{N}(0,\,0.2^2), \\
\beta^{(x)}_{h,p,y} &= \beta^{(x)}_{h-1,p,y}
  + s^{(x)}_{p,y}\,\sqrt{\Delta_h}\,\varepsilon^{(x)}_{h-1,p,y},
  \qquad \varepsilon^{(x)}_{h-1,p,y}\sim\mathcal{N}(0,1),\quad h\ge 2,
  \label{eq:beta-x}
\end{align}
where the innovation scale is the three-level factor $s^{(x)}_{p,y} = \tau_x\,\lambda^{(g)}_{p}\,\lambda^{(\ell)}_{p,y}$ introduced above.

\section{Results}
\subsection{Convergence}
Table~\ref{tab:convergence} reports convergence diagnostics for all eight baskets. All baskets achieve convergence ($\hat{R}=1.00$, ESS $> 900$, divergences $< 0.05\%$), and no sampler runs into the tree-depth limit of 12. Detailed trace plots and energy diagnostics are available in Appendix~B.

\begin{table}[htbp]
    \centering
    \caption{MCMC convergence diagnostics by basket.}
    \label{tab:convergence}
    \resizebox{\linewidth}{!}{%
        \begin{tabular}{lrrrrrrr}
\toprule
Basket & Max $\hat{R}$ & Min ESS (bulk) & Min ESS (tail) & Max MCSE/SD & Min BFMI & Divergence rate & Treedepth saturation \\
\midrule
ADA & 1.00 & 1559 & 1462 & 0.118 & 0.76 & 0.0000 & 0.00 \\
ALL & 1.00 & 907 & 1450 & 0.069 & 0.71 & 0.0005 & 0.00 \\
BNB & 1.00 & 2441 & 2244 & 0.111 & 0.85 & 0.0000 & 0.00 \\
BTC & 1.00 & 1637 & 1484 & 0.041 & 0.68 & 0.0000 & 0.00 \\
DOGE & 1.00 & 1614 & 1570 & 0.143 & 0.87 & 0.0000 & 0.00 \\
ETH & 1.00 & 2501 & 2601 & 0.043 & 0.80 & 0.0000 & 0.00 \\
LINK & 1.00 & 1388 & 1278 & 0.182 & 0.81 & 0.0001 & 0.00 \\
XRP & 1.00 & 2243 & 2279 & 0.060 & 0.89 & 0.0001 & 0.00 \\
\bottomrule
\end{tabular}

    }
\end{table}

\subsection{Cross-Basket Stability of Predictors}

\begin{figure}[t]
    \centering
    \includegraphics[width=\linewidth]{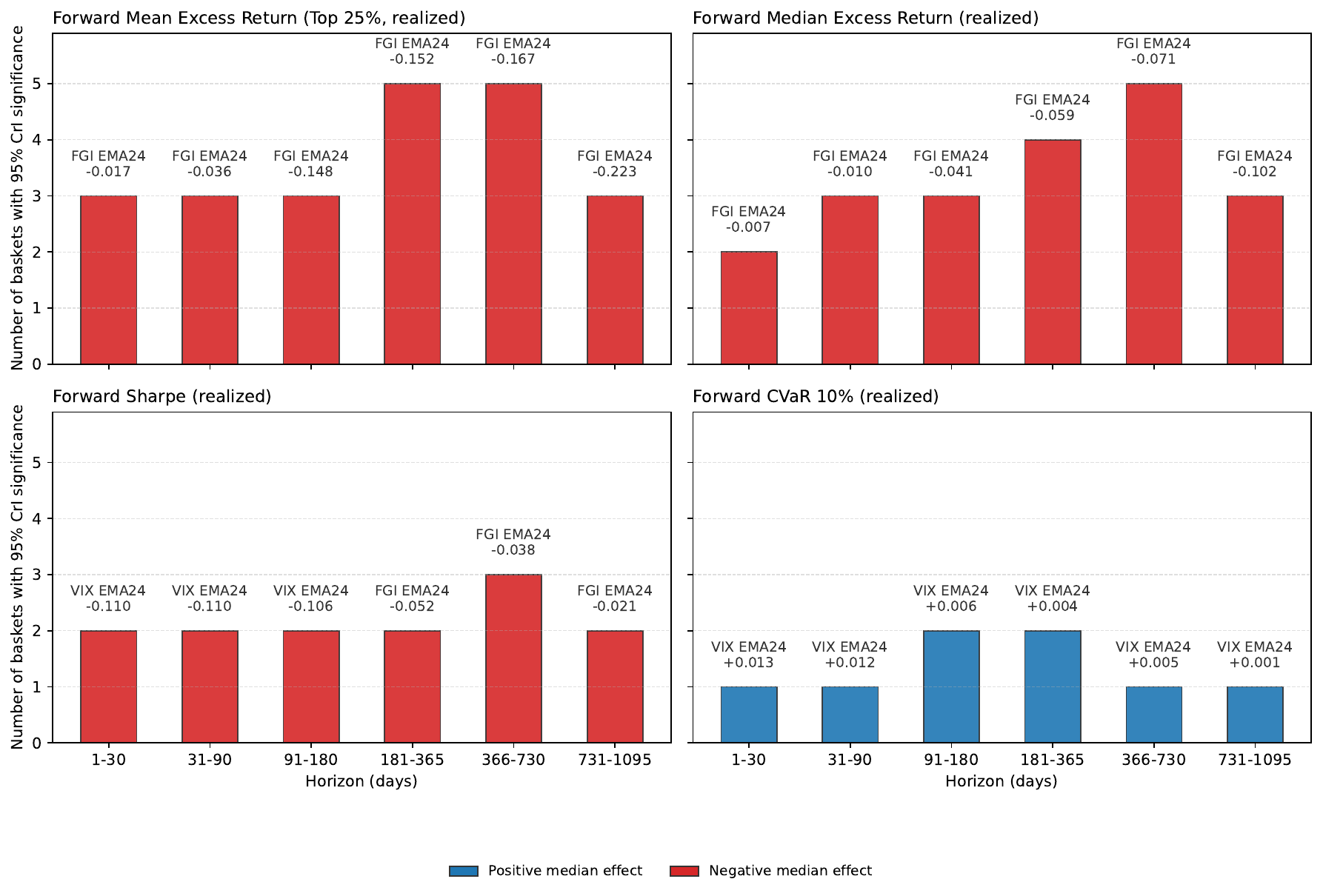}
    \caption{Cross-basket stability of top predictors by target and horizon}
    \label{fig:target_horizon_panel}
\end{figure}

Figure \ref{fig:target_horizon_panel} summarizes the cross-basket stability of top predictors for 
each target-horizon combination. For each target and horizon, we display the single most 
stable predictor—defined as the predictor significant in the most 
baskets.\footnote{In the case of ties, we sort by mean absolute median effect, 
then by maximum absolute median effect. See Appendix \ref{app:defs} for detailed ranking procedures.}
Each bar indicates the number of baskets (out of eight in total) in which the predictor achieves $95\%$ credible-interval 
significance for the corresponding target and horizon. Bar colors denote the sign of the mean 
median effect\footnote{``Median effect'' denotes the posterior median impulse response produced by the BLP model.} across baskets where the predictor is $95\%$ significant for that target--horizon pair: blue for positive effects and red for negative effects. 
Annotations display the predictor label and the mean median effect across baskets, 
measured in the target's native units per one-standard-deviation shock in the predictor.

Figure \ref{fig:target_horizon_panel} reveals several key findings. 
First, the predictors with the highest cross-basket significance are exclusively macrofinance 
factors. Second, no single predictor exhibits universal significance: even FGI EMA24, which is the most stable 
predictor for the forward top-25\% and median excess-return targets, does not achieve $95\%$ credible-interval significance across 
all eight baskets (predictor abbreviations follow Appendix \ref{app:a3-labels}).

A one-standard-deviation increase in FGI EMA24 lowers the forward mean excess return of the top 25\% quantile
by roughly 15--22 percentage points across the 1--3 year horizons we examine.
For the forward median excess return at 1--3 year horizons, a one-standard-deviation 
shock in FGI EMA24 is associated with reductions of 6--10 percentage points.

For the forward Sharpe ratio, VIX EMA24 delivers strong negative effects around $-0.1$ in two baskets across the 1--180 day horizons, while FGI EMA24 becomes the most stable predictor once horizons extend to 181--365 days and beyond. 
Tail-risk CVaR$_{10\%}$ shows little evidence of a systematic response: VIX EMA24 and T10Y2Y RVol24 
each achieve significance in at most two baskets per horizon, with economically negligible 
effect sizes.

\begin{figure}[t]
    \centering
    \includegraphics[width=\linewidth]{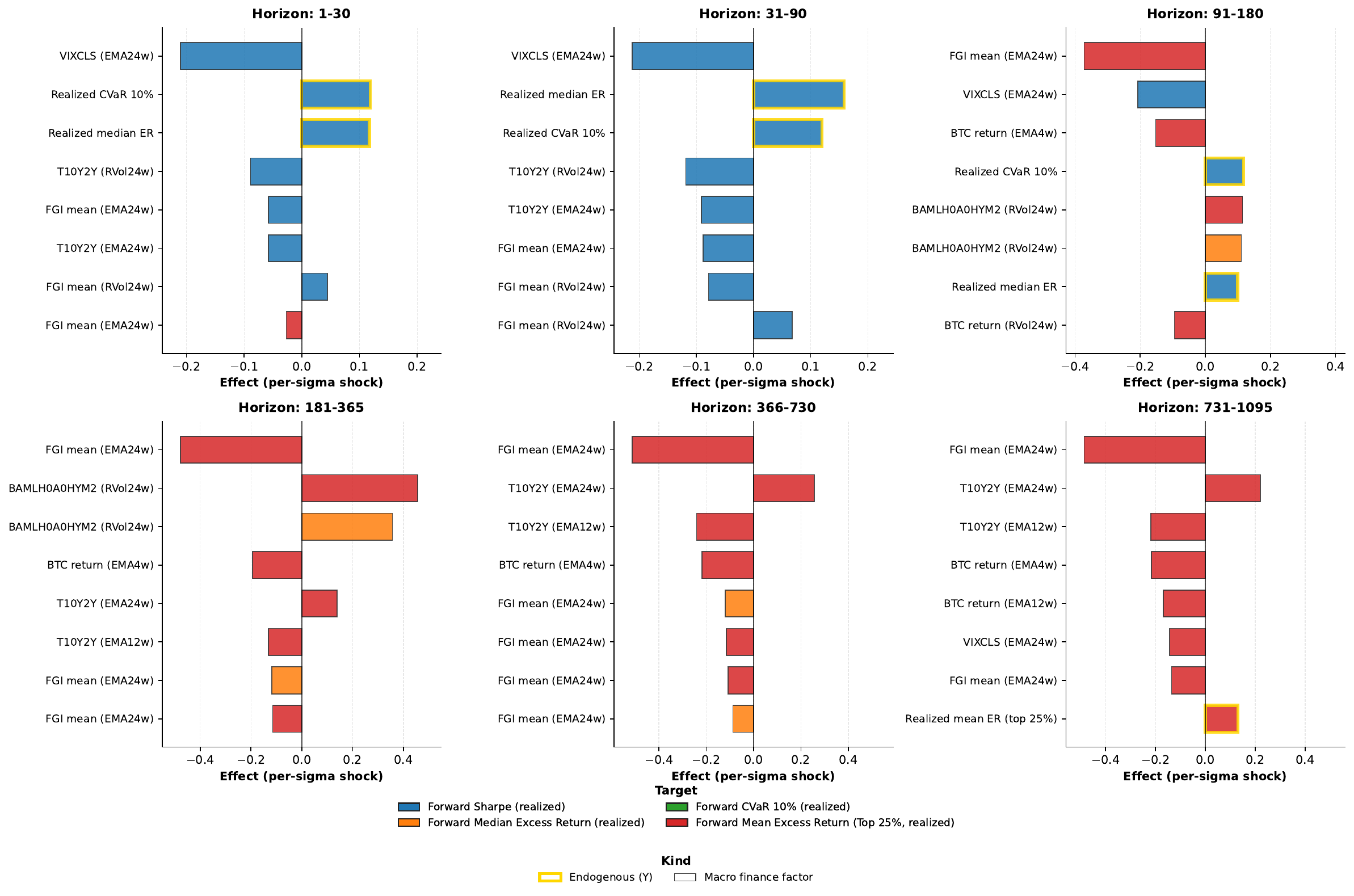}
    \caption{Largest predictor effects by horizon}
    \label{fig:top8_horizon}
\end{figure}

Figure \ref{fig:top8_horizon} lists the eight largest median effects at each horizon, ordered by absolute median impact.
Unlike Figure \ref{fig:target_horizon_panel}, which fixes the target to study cross-basket stability,
this panel surfaces the predictor--target combinations that deliver the strongest per-sigma effects regardless of target choice.
Each horizontal bar represents the median posterior effect in the target's native units per one-standard-deviation shock to the predictor,
with colors indicating the target variable (see the legend); gold borders denote endogenous predictors while dark borders denote macrofinance factors.
Each displayed effect is 95\% credible-interval significant.

Endogenous predictors, realized risk and return metrics, exhibit large effects primarily at shorter forecast 
horizons. Among the 24 top-8 entries across horizons spanning 1–180 days, 6 are endogenous predictors. 
Realized median excess return and realized CVaR$_{10\%}$ show positive effects on forward Sharpe ratio, 
while Realized mean ER (top 25\%) exhibits negative effects. After 180 days, however, endogenous predictors nearly vanish from the top-8 list---only a single entry (Realized mean ER (top 25\%) at the 731--1095 day horizon) remains.
At short-to-medium horizons (1–90 days), VIX EMA24w exhibits the largest negative effects on forward Sharpe ratio, 
while FGI EMA24w also shows negative effects but with smaller magnitudes. Starting at 180 days, FGI EMA24w's largest effects 
shift to forward mean excess return for the top 25\% quantile, where it repeatedly emerges as the single most influential predictor across all 
three long horizons, aligning with Figure \ref{fig:target_horizon_panel}.

T10Y2Y exhibits transformation-dependent relationships with forward mean
excess return for the top 25\% quantile. T10Y2Y EMA24w shows persistent positive effects, 
while T10Y2Y EMA12w exhibits negative effects.
This divergence suggests that different smoothing lengths of the 10-year minus 2-year Treasury spread capture distinct phases of yield-curve dynamics with 
contrasting implications for forward cryptocurrency returns.

\subsection{Endogenous Decay and Cross-Basket Heterogeneity}
The findings in the previous section reveal a clear pattern: the most stable predictors across baskets are 
exclusively macro-finance factors (Figure~\ref{fig:target_horizon_panel}). 
Endogenous predictors, realized risk and return metrics, appear among the largest effects at horizons 
spanning 1–180 days (Figure~\ref{fig:top8_horizon}), yet only one of them remains 
in the top-8 effects beyond 180 days. This raises a natural question: do current 
realized risk and return distributions retain any influence on forward 
risk and return outcomes at longer horizons, or does their effect decay entirely?

Unlike the macro-finance features, which must pass the stability filter to enter the BLP multi-horizon model, 
the four endogenous variables enter the model directly across all baskets and horizons.
This setup gives us the opportunity to analyze cross-basket effects of the endogenous variables.
We conduct a Bayesian random-effects meta-analysis to pool information across baskets and horizons, 
following the framework detailed in Appendix~\ref{app:a3-endogenous-decay}. 
To ensure comparability with the previous section, we scale the posterior coefficients by the 
target's reference standard deviation, so that effects are measured in the target's native units 
per one-standard-deviation increase in the predictor.

Tables~\ref{tab:endogenous-decay-top25} and~\ref{tab:endogenous-decay-sharpe} report 
basket-level effects for predictor-target-horizon combinations exhibiting 95\%-credible 
significance. Out of the four target variables, only forward Mean ER (Top 25\%) and 
forward Sharpe exhibit significant impulse responses for at least one 
basket-predictor-horizon combination. Forward Median ER and forward CVaR 10\% show no 
significant responses across all baskets and horizons.

For forward Sharpe, only two basket-predictor pairs, LINK's realized CVaR 10\% and 
ETH's realized median ER, exhibit significant positive effects across horizons. 
Both show responses that decay substantially with horizon: a 
one-standard-deviation shock increases forward Sharpe by more than 0.11 in Sharpe units at short-to-medium horizons, but decays to approximately 0.01--0.04 in Sharpe units 
at the longest horizon.

Despite these substantial basket-specific effects, Figure~\ref{fig:endogenous-decay-summary} reveals that 
population-level effects $\mu_h$ remain economically negligible. The pooled estimates 
show realized CVaR 10\% generating the strongest population-average response to forward 
Sharpe, around 0.0027--0.0033 in Sharpe units across horizons. This contrast with the basket-level effects underscores 
substantial cross-basket heterogeneity: while specific pairs exhibit economically meaningful 
impulse responses, these patterns are basket-specific rather than generalizable.

This pattern stands in contrast to the behavior of macro-finance predictors. 
While endogenous risk-return metrics exhibit decaying effects with limited 
cross-basket consistency at the population level, macro-finance factors, particularly 
market sentiment indicators, demonstrate more persistent effects across longer horizons. 
Figure~\ref{fig:target_horizon_panel} shows that FGI EMA24w emerges as the most stable 
predictor at all long-run horizons (181--1095 days) for both forward mean excess return 
(Top 25\%) and forward median excess return targets. 
This raises a natural question: does this long-horizon influence manifest similarly 
across cryptocurrency baskets, or does it exhibit significant heterogeneity?

\begin{figure}[t]
    \centering
    \includegraphics[width=\textwidth]{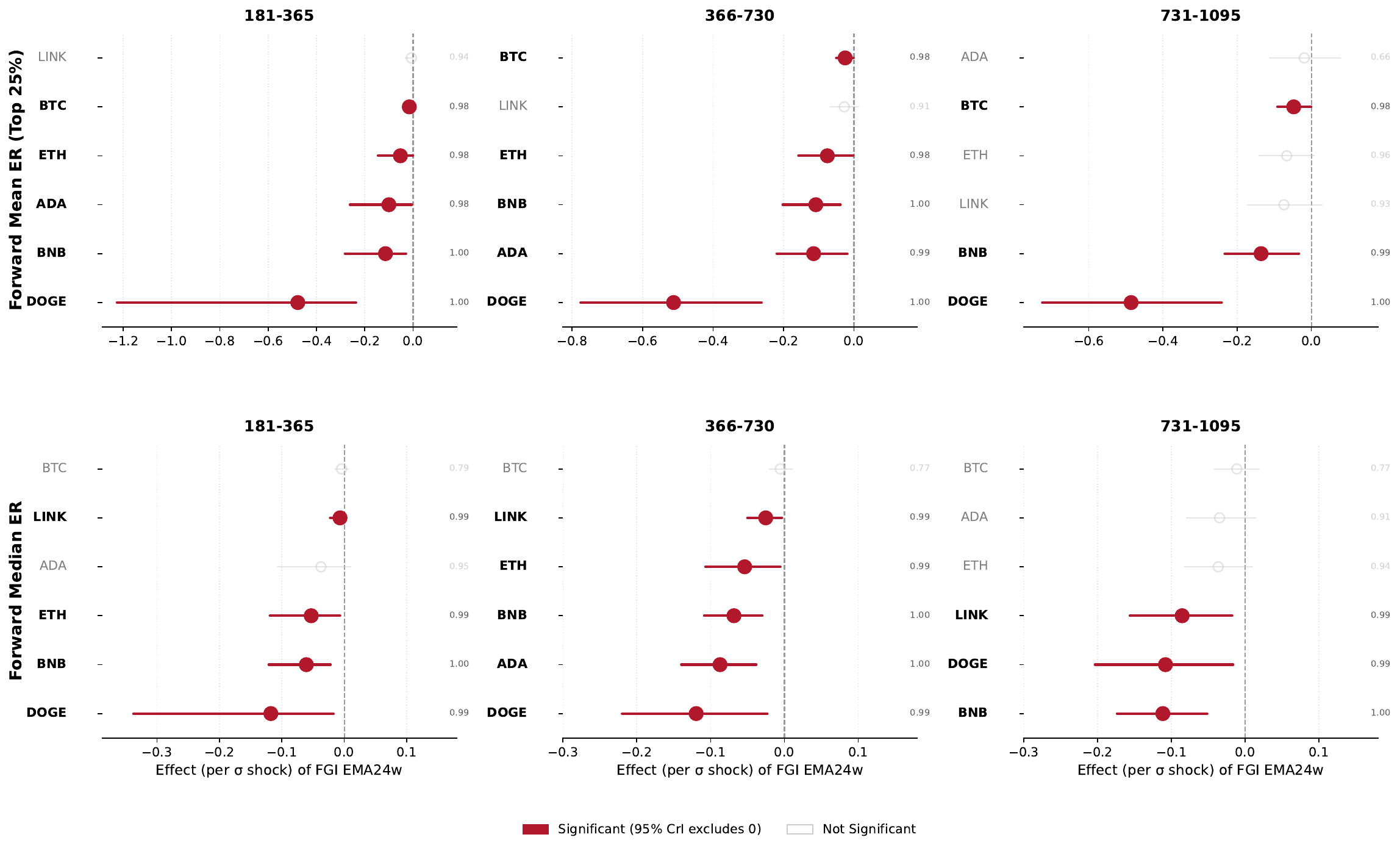}
    \caption{Cross-basket impulse responses to FGI sentiment shocks at long forecast horizons.}
    \label{fig:forest-long-horizon}
\end{figure}

Figures~\ref{fig:target_horizon_panel} and \ref{fig:top8_horizon} 
highlighted the consistent negative relationship between FGI EMA24w 
and both forward mean excess return for the top 25\% quantile and forward median 
excess return at long horizons. Figure~\ref{fig:forest-long-horizon} 
decomposes these responses across six baskets\footnote{XRP and ALL are excluded because stability selection never 
retains \texttt{FGI mean EMA24w}. In XRP the bootstrap selects \texttt{FGI mean RVol24w} instead, while in ALL only \texttt{FGI mean RVol12w} exceeds the conditional-stability threshold.} and three 
long-run forecast horizons (181--365, 366--730, and 731--1095 days). 
Each panel displays posterior median impulse responses to a 
one-standard-deviation shock in FGI EMA24w. Red markers indicate 
responses that pass a 95\% credible interval excludes zero; 
gray markers denote non-significant responses. Numbers to the right 
report the posterior directional probability.

Figure~\ref{fig:forest-long-horizon} reveals substantial cross-basket heterogeneity in 
sentiment effects. For forward mean excess return (Top 25\%), DOGE exhibits 
significantly stronger responses than other baskets: at the 365-day horizon, 
a one-standard-deviation increase in FGI EMA24w reduces forward mean excess return (Top 25\%) 
by approximately 48 percentage points for DOGE, an effect more than 3 times larger than 
estimates for BNB (-11 percentage points), ADA (-10 percentage points), and other 
baskets. For forward median excess return, DOGE's response of approximately 12 
percentage points at the same horizon is less drastic but still roughly double 
the effect size of other significant baskets. DOGE's amplified sensitivity persists 
across all three long horizons with posterior directional probabilities consistently 
exceeding 0.98, suggesting that meme coins like DOGE display heightened responsiveness 
to market sentiment across the entire forward return distribution, with particularly 
strong amplification in the upper tail.

Taken together, these findings answer the question posed at the outset: while 
endogenous risk-return metrics exhibit limited and decaying influence on forward 
outcomes at longer horizons, macro-finance sentiment indicators demonstrate more persistent 
and economically substantial effects. FGI EMA24w emerges as the most stable long-run 
predictor for forward return metrics across all tested macro-finance factors. 
However, the effects of sentiment shocks exhibit considerable cross-basket heterogeneity, 
with certain assets, particularly meme coins like DOGE, displaying disproportionate 
sensitivity to sentiment shifts.

\section{Robustness Checks}

\subsection{Local Projection Model}\label{sec:lp}

To assess whether our main findings are robust to the choice of statistical framework, 
we implement a classical local projection estimator \citep{Jorda2005} as an alternative 
to the Bayesian multi-horizon local projection model (§6.6). This provides a robustness 
check that tests whether the impulse response patterns remain qualitatively consistent 
under different estimation approaches.

We retain the identical data preparation from §6.5, including the same 
stability-selected features through moving block bootstrap (§6.4). The only change 
is replacing the Bayesian multi-horizon local projection model (§6.6) with a 
classical local projection approach. This ensures differences in results reflect 
the estimator choice rather than data preprocessing.

We use the same standardised objects as defined in §6.5:
\[
\bigl\{(\widetilde{Y}^{(F)}_{t,h},\,\widetilde{Y}^{(C)}_{t,h},\,\widetilde{X}^{\mathcal S}_{t,h})
  : t\in\mathcal T_{\mathrm{train}},\ h\in\mathcal H\bigr\}.
\]

For each horizon $h\in\mathcal H$ and target $k\in\{1,\dots,n_y\}$, 
we define the design vector
\[
\widetilde Z_{t,h}
\;=\;
\begin{bmatrix}
\widetilde{Y}^{(C)}_{t,h} \\
\widetilde{X}^{\mathcal S}_{t,h}
\end{bmatrix}
\in\mathbb R^{P},\qquad
P = n_y + |\mathcal S|,
\]
and estimate the horizon-specific regression
\begin{equation}\label{eq:lp-ols}
\widetilde{Y}^{(F)}_{t,h,k}
\;=\;
\alpha_{h,k} + \widetilde Z_{t,h}^\top \beta_{h,k} + \varepsilon_{t,h,k},
\qquad
t\in\mathcal T_{\mathrm{train}}.
\end{equation}

We estimate this through ordinary least squares separately for each $(h,k)$ pair, computing 
heteroskedasticity-robust (HC1) standard errors \citep{MacKinnonWhite1985}. This yields raw estimates 
$\hat\beta_{h,p,k}$ and $\hat s_{h,p,k}$ for each predictor $p\in\{1,\dots,P\}$.

\paragraph{Cross-horizon RW1 smoothing.}
As in the Bayesian model (§6.6), we smooth coefficient paths across horizons 
using a first-order random-walk structure, but here implemented as a 
quadratic penalty in GLS rather than as an RW1 prior.
For a given predictor $p\in\{1,\dots,P\}$ and target $k$, define the 
horizon-specific coefficient path
\[
\hat\beta_{p,k}
=
\bigl(\hat\beta_{h_1,p,k},\dots,\hat\beta_{h_H,p,k}\bigr)^\top
\in\mathbb R^{H},
\]
and let $\hat s_{h_j,p,k}$ denote the corresponding HC1 standard errors.  
We map horizons $h_j$ (in days) to weekly indices
$\omega_j = \lceil h_j/7\rceil$ with spacings
$\Delta_j = \omega_{j+1}-\omega_j>0$.
The RW1-penalized coefficient path
$\hat\beta^{\mathrm{rw1}}_{p,k}$ solves the optimization problem
\begin{equation}\label{eq:rw1-penalized}
\hat\beta^{\mathrm{rw1}}_{p,k}
=
\arg\min_{b\in\mathbb R^{H}} \sum_{j=1}^{H} \frac{\bigl(b_j - \hat\beta_{h_j,p,k}\bigr)^2}{\hat s_{h_j,p,k}^2} + \lambda_{\mathrm{rw1}} \sum_{j=1}^{H-1} \frac{\bigl(b_{j+1} - b_j\bigr)^2}{\Delta_j},
\end{equation}
with tuning parameter $\lambda_{\mathrm{rw1}}=1$.
We apply this smoothing separately for each predictor--target pair $(p,k)$. 
Intercepts $\alpha_{h,k}$ are not smoothed across horizons.
\paragraph{Stationary bootstrap and simultaneous bands.}
The RW1-penalized paths $\hat\beta^{\mathrm{rw1}}_{p,k}$ from \eqref{eq:rw1-penalized}
require standard errors that account for both serial dependence
and the smoothing across horizons induced by the RW1 penalization.
We construct confidence bands using a stationary bootstrap
\citep{PolitisRomano1994} combined with studentized $k$-max simultaneous bands \citep{RomanoWolf2007}.

Let $T = |\mathcal T_{\mathrm{train}}|$. For each bootstrap replication $b=1,\dots,B$, 
we resample time indices $\{t^{(b)}_1,\dots,t^{(b)}_{T}\}\subset\mathcal T_{\mathrm{train}}$
using the stationary bootstrap with mean block length
\[
\bar L = \max\bigl\{2,\, \min\bigl(T-1,\, 
  \max\bigl(1.75\,T^{1/3},\, L_{\mathrm{med}}\bigr)
\bigr)\bigr\},
\]
where $L_{\mathrm{med}}$ is the median horizon length in weeks.
For each bootstrap sample, we estimate the OLS regressions \eqref{eq:lp-ols} 
and apply the RW1 penalization \eqref{eq:rw1-penalized},
obtaining $\hat\beta^{\mathrm{rw1},(b)}_{h_j,p,k}$ and 
$\hat s^{\mathrm{rw1},(b)}_{h_j,p,k}$.

We construct studentized $t$-statistics
\[
t^{(b)}_{h_j,p,k}
=
\frac{
  \hat\beta^{\mathrm{rw1},(b)}_{h_j,p,k} - \hat\beta^{\mathrm{rw1}}_{h_j,p,k}
}{
  \hat s^{\mathrm{rw1},(b)}_{h_j,p,k}
}.
\]
To construct simultaneous bands, for each predictor--target pair $(p,k)$ 
we compute $M^{(b)}_{p,k}$, the $k_{\max}$-th largest value among 
$\{|t^{(b)}_{h_1,p,k}|,\dots,|t^{(b)}_{h_H,p,k}|\}$ with $k_{\max}=\min\{2,H\}$,
and let $\hat c_{p,k}$ be the 95\% quantile of $\{M^{(b)}_{p,k}\}_{b=1}^B$.
The resulting simultaneous 95\% confidence band is
\begin{equation}\label{eq:lp-simul-band}
\hat\beta^{\mathrm{rw1}}_{h_j,p,k}
\;\pm\;
\hat c_{p,k}\,\hat s^{\mathrm{rw1}}_{h_j,p,k},
\qquad
j=1,\dots,H,
\end{equation}
which controls the family-wise error rate across horizons at 5\% 
for each predictor--target pair $(p,k)$.
To match the Bayesian results, we scale by the target-specific reference scale 
$\sigma_{h_j,k}$ (defined in §6.5) to obtain impulse responses
$\widehat{\mathrm{IRF}}^{\mathrm{LP}}_{h_j,p,k}
= \hat\beta^{\mathrm{rw1}}_{h_j,p,k}\,\sigma_{h_j,k}$
with confidence bands
$[\hat\beta^{\mathrm{rw1}}_{h_j,p,k} \pm \hat c_{p,k}\,\hat s^{\mathrm{rw1}}_{h_j,p,k}]\,\sigma_{h_j,k}$.

\subsection{Local Projection Robustness Results}\label{sec:lp_robustness_results}
Table~\ref{tab:blp_lp_summary_overall} compares the Bayesian multi-horizon local projection (BLP)
and classical local projection (LP) estimates across all $N=2{,}256$ 
predictor-target-horizon combinations spanning eight baskets and six horizons.
We assess consistency using three metrics: sign match (whether the two 
models agree on direction), 95\% interval overlap (whether credible and 
confidence intervals intersect), and significance rates (proportion of 
effects significant at 95\% level).

\begin{table}[htbp]
    \centering
    \caption{BLP vs LP Robustness Summary}
    \label{tab:blp_lp_summary_overall}
    \begin{tabular}{lc}
    \toprule
    Metric & Value \\
    \midrule
    Total Samples & 2,256 \\
    Sign Match Rate (\%) & 64.3 \\
    95\% Interval Overlap (\%) & 98.7 \\
    BLP Significance (\%) & 9.1 \\
    LP Significance (\%) & 6.2 \\
    Sign Match | Both Sig (\%) & 100.0 \\
    Sign Match | BLP Sig (\%) & 79.1 \\
    \bottomrule
    \end{tabular}
    \end{table}

The results demonstrate strong consistency between the two approaches. The 95\% 
intervals overlap in 98.7\% of cases, indicating quantitatively similar uncertainty 
estimates across both frameworks. The BLP model exhibits a higher significance rate 
(9.1\%) compared to the LP model (6.2\%). We attribute this difference primarily 
to the Bayesian framework's hierarchical shrinkage structure. In contrast, the LP 
approach estimates each horizon-target pair separately through OLS.

When examining directional consistency, the two models show strong 
agreement: when BLP signals significance, 79.1\% share the 
same sign with the LP estimate, rising to 100\% when both methods 
flag an effect as significant.

\begin{figure}[htbp]
    \centering
    \includegraphics[width=\linewidth]{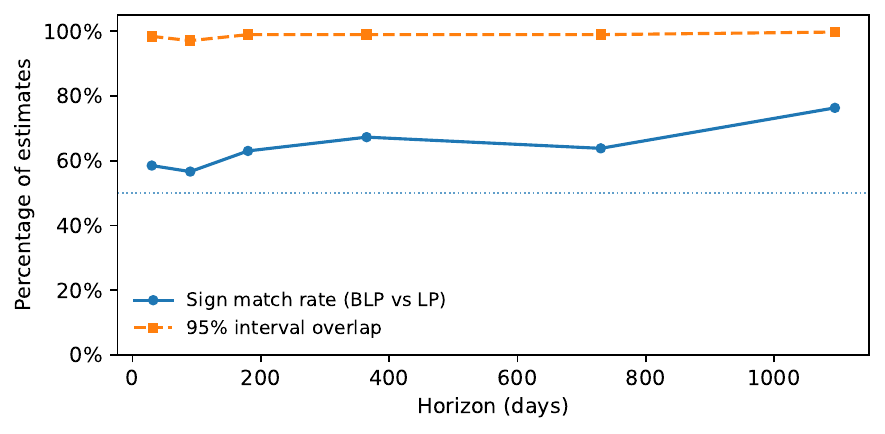}
    \caption{BLP--LP Agreement Across Projection Horizons}
    \label{fig:blp_lp_comparison}
\end{figure}

Figure~\ref{fig:blp_lp_comparison} shows model agreement varies across 
forecast horizons, with stronger consistency at longer horizons.
The 95\% interval overlap remains consistently high
across all horizons, while the sign match rate rises from 58.5\% at 30 days to 
76.3\% at 1,095 days. This pattern indicates the two approaches diverge more on 
short-term dynamics but converge on longer-term directional effects, confirming 
our main findings are robust to the choice of estimation framework.

\section{Conclusion}

This study set out to answer a question of growing importance for retail 
crypto investors: does the popular ``HODL'' strategy, buying and holding 
cryptocurrency, truly deliver on its promises? Through a combination of 
large-scale Monte Carlo simulation (7 individual tokens plus one basket containing 
378 non-stablecoin crypto assets, totaling 480 million scenarios) and Bayesian 
multi-horizon local projection analysis, we tested two related claims central to the 
crypto investment narrative. First, whether the risk and return distribution 
across the broad crypto market justifies the HODL strategy when accounting for 
trading costs, opportunity costs, and six holding-period intervals 
ranging from 1 to 1095 days. Second, whether past realized risk and return 
distributions are the dominant factor affecting future outcomes.

Our Monte Carlo simulation reveals two critical insights that challenge the ``HODL'' narrative. 
First, we quantify substantial survivorship bias: while people often cite Bitcoin or Ethereum as proof of the HODL strategy's success, our ALL basket, which randomly samples from 378 non-stablecoin 
tokens, tells a starkly different story. At the longest holding horizon (731–1095 days), the median excess 
return is \(-28.4\%\), yet the top-quartile mean reaches \(+1326.7\%\). This dramatic dispersion exposes how ``HODL missionaries'' 
exploit upper-quartile performance to overstate rewards available to the average investors. Moreover, for holding periods 
beyond 181–365 days, the probability of experiencing losses greater than 10\% 
exceeds 50\%, and the CVaR$_{1\%}$ approaches near-total loss (100\%).

Second, we reveal a complex trade-off: as holding periods lengthen, moderate and extreme risks often move in opposite directions. 
For successful tokens such as BTC and ETH, longer horizons show declining moderate loss 
probabilities (losses $>10\%$), dropping from peaks around 90--180 days to much lower levels at the longest horizons, while 
extreme tail risk (CVaR) rises through mid-range horizons before declining at 2--3 year holding periods.
Crucially, however, this pattern is not generalizable when we examine the market-level distribution. 
The ALL basket demonstrates that across the broader cryptocurrency market, both moderate and extreme 
risks increase simultaneously with holding length, suggesting no miraculous recovery but rather escalating 
risk with longer holding periods.

Our Bayesian multi-horizon local projection analysis directly addresses the second claim underlying 
the HODL narrative: that past realized risk and return distributions are the dominant factor 
affecting future outcomes. We find the opposite. While endogenous 
predictors (realized risk and return metrics) generate some of the largest short-horizon impulse 
responses, they exhibit limited cross-basket stability compared with macro-finance predictors. 
Meta-analysis across all baskets reveals economically negligible population-level effects for 
endogenous predictors. Instead, macrofinance sentiment indicators, particularly FGI EMA24w, emerge as 
both the most stable and most influential predictors of long-horizon (181–1095 days) return metrics. 
Where significant, a one-standard-deviation increase in FGI EMA24w lowers top-quartile returns by 
roughly 15–22 percentage points and median returns by 6–10 percentage points across these horizons. 
This dominance of sentiment over historical fundamentals undermines 
a core premise of the HODL narrative: that past realized distributions 
drive future outcomes. Moreover, we document striking cross-basket heterogeneity: DOGE exhibits 
sentiment sensitivity more than three times larger than BNB and 
ADA (48 versus 10–11 percentage points for top-quartile returns), 
underscoring the heightened speculative nature of meme coins.

Taken together, our findings fundamentally challenge the viability of the HODL strategy 
for retail investors. The evidence is unequivocal: relying on top-quartile success stories 
to forecast personal returns is not optimistic, it is dangerously misleading. 
The gap between median outcomes (\(-28.4\%\)) and top-quartile performance (\(+1326.7\%\)) 
in the ALL basket represents not a modest discount but a chasm between financial ruin and 
extraordinary gain. Moreover, the belief that historical risk-return distributions can 
guide future investment decisions finds no empirical support in our data. Instead, 
macroeconomic conditions and sentiment shifts, captured most powerfully by FGI EMA24w, emerge 
as the primary drivers of cryptocurrency returns across holding horizons. Retail investors 
should prioritize monitoring macro-sentiment indicators rather than extrapolating their 
expected returns from past return patterns.

\begin{appendices}
\section{Additional Tables and Figures}

\subsection{Monte Carlo Metrics and Definitions}\label{app:a1-mc}

For each \emph{basket $\times$ horizon}
\[
  \mathcal X=\{X_1,\dots ,X_N\},\qquad
  N=\lvert\mathcal X\rvert ,
\]
be the sample of excess-return observations.
Let $\mu$ be the sample mean, $\sigma$ the unbiased standard deviation, and
$q_p\equiv\operatorname{Quantile}_{p}(\mathcal X)$.

\subsubsection{Additional notation}\label{app:a1-notation}
\begin{itemize}\setlength\itemsep{0pt}
  \item $n_{\text{neg}}=\lvert\{\,i:X_i<0\,\}\rvert$ --- number of negative returns
  \item $\displaystyle
        \mu_{\text{neg}}=
        \frac{1}{n_{\text{neg}}}\sum_{X_i<0} X_i$
        --- mean of negative returns
\end{itemize}

\subsubsection{Tail-probability thresholds used in the paper}\label{app:a1-thresholds}
\[
  \alpha=10\%\quad\text{for weekly statistics},\qquad
  \alpha=1\%\quad\text{for  overall statistics}.
\]

\subsubsection{Metrics and definitions}\label{app:a1-metrics}

\begin{table}[H]
  \centering
  \renewcommand{\arraystretch}{1.25}
  \begin{tabular}{@{}L{0.27\textwidth} L{0.63\textwidth}@{}}
    \toprule
    Statistic & Formula \\ \midrule
    Sample size & $N$ \\
    Mean excess return & $\displaystyle \mu=\frac{1}{N}\sum_{i=1}^{N}X_i$ \\
    Median excess return & $\displaystyle \tilde X=q_{0.50}$ \\
    Unbiased standard deviation & $\displaystyle \sigma=\sqrt{\frac{1}{N-1}\sum_{i=1}^{N}(X_i-\mu)^2}$ \\
    Inter-quartile range & $\displaystyle \text{IQR}=q_{0.75}-q_{0.25}$ \\
    Sharpe ratio & $\displaystyle \text{Sharpe}=\frac{\mu}{\sigma}$ \\
    Sortino ratio & $\displaystyle \text{Sortino}= \frac{\mu}{\sigma_{\text{neg}}},\quad \sigma_{\text{neg}}^{\,2}=\frac{1}{n_{\text{neg}}-1}\sum_{X_i<0}\bigl(X_i-\mu_{\text{neg}}\bigr)^2$ \\
    Value-at-Risk & $\displaystyle \text{VaR}_{\alpha}= \max\!\bigl(0,\; -q_{\alpha}\bigr)$ \\
    Conditional VaR & $\displaystyle \text{CVaR}_{\alpha}= \max\!\Bigl(0,\; - \frac{1}{\lvert\{i:X_i\le q_{\alpha}\}\rvert}\sum_{X_i\le q_{\alpha}} X_i\Bigr)$ \\
    Probability of profit & $\displaystyle p_{\text{profit}}=\frac{\lvert\{i:X_i>0\}\rvert}{N}$ \\
    Probability of $>10\%$ loss & $\displaystyle p_{\text{sig loss}}=\frac{\lvert\{i:X_i<-0.10\}\rvert}{N}$ \\
    75-th percentile & $q_{0.75}$ \\
    Mean of top-quartile returns & $\displaystyle \bar X_{\text{top 25}}=\frac{1}{\lvert\{i:X_i\ge q_{0.75}\}\rvert}\sum_{X_i\ge q_{0.75}} X_i$ \\
    Top-quartile proportion & $\displaystyle p_{\text{top 25}}=\frac{\lvert\{i:X_i\ge q_{0.75}\}\rvert}{N}$ \\
    Skewness & $\displaystyle \gamma_{1}^{\text{(G1)}}=\frac{N\sqrt{N-1}}{N-2}\cdot\frac{\sum_{i=1}^{N}(X_i-\mu)^{3}}{\bigl(\sum_{i=1}^{N}(X_i-\mu)^{2}\bigr)^{3/2}}$ \\
    Excess kurtosis & $\displaystyle \kappa^{\text{(G2)}}=\frac{N(N+1)(N-1)\sum_{i=1}^{N}(X_i-\mu)^{4}}{(N-2)(N-3)\bigl(\sum_{i=1}^{N}(X_i-\mu)^{2}\bigr)^{2}} - \frac{3(N-1)^{2}}{(N-2)(N-3)}$ \\
    \bottomrule
  \end{tabular}
\end{table}

\subsubsection{Core aggregated statistics by basket and horizon}\label{app:a1-core}
\begin{landscape}
    \thispagestyle{plain} 
    \centering
    \footnotesize                
    \setlength{\tabcolsep}{5pt}

    \begin{longtable}{@{}ll*{10}{r}@{}}
    \caption{Key Monte-Carlo statistics by basket and holding horizon}\label{tab:agg-stats-core}\\
    \toprule
    Basket & Horizon (days) & $\mu$ & $\tilde X$ & $\sigma$ & Sharpe & Sortino & VaR$_{1\%}$ & CVaR$_{1\%}$ & $p_{\text{sig\,loss}}$ & $q_{0.75}$ & $\overline{X}_{\text{top25}}$\\
    \midrule
    \endfirsthead
    \toprule
    Basket & Horizon (days) & $\mu$ & $\tilde X$ & $\sigma$ & Sharpe & Sortino & VaR$_{1\%}$ & CVaR$_{1\%}$ & $p_{\text{sig\,loss}}$ & $q_{0.75}$ & $\overline{X}_{\text{top25}}$\\
    \midrule
    \endhead
    \midrule
    \multicolumn{12}{r}{\textit{Continued on next page}}\\
    \endfoot
    \bottomrule
    \endlastfoot
    ALL & 1--30 & 0.045 & -0.017 & 1.948 & 0.023 & 0.351 & 0.506 & 0.591 & 0.299 & 0.094 & 0.473 \\
    ALL & 31--90 & 0.216 & -0.068 & 2.847 & 0.076 & 1.143 & 0.725 & 0.791 & 0.465 & 0.248 & 1.438 \\
    ALL & 91--180 & 0.597 & -0.112 & 11.376 & 0.052 & 2.710 & 0.834 & 0.887 & 0.509 & 0.389 & 3.159 \\
    ALL & 181--365 & 1.324 & -0.160 & 28.489 & 0.046 & 5.128 & 0.937 & 0.964 & 0.534 & 0.558 & 6.256 \\
    ALL & 366--730 & 2.876 & -0.196 & 25.525 & 0.113 & 10.218 & 1.003 & 1.020 & 0.536 & 0.960 & 12.578 \\
    ALL & 731--1095 & 3.027 & -0.284 & 15.869 & 0.191 & 10.659 & 1.061 & 1.079 & 0.552 & 1.395 & 13.267 \\
    BTC & 1--30 & 0.031 & 0.009 & 0.160 & 0.192 & 0.371 & 0.328 & 0.386 & 0.149 & 0.088 & 0.234 \\
    BTC & 31--90 & 0.144 & 0.058 & 0.403 & 0.356 & 1.133 & 0.468 & 0.506 & 0.269 & 0.314 & 0.676 \\
    BTC & 91--180 & 0.396 & 0.191 & 0.833 & 0.475 & 2.537 & 0.536 & 0.568 & 0.286 & 0.624 & 1.494 \\
    BTC & 181--365 & 1.043 & 0.519 & 1.971 & 0.529 & 5.189 & 0.676 & 0.711 & 0.231 & 1.274 & 3.451 \\
    BTC & 366--730 & 2.949 & 1.514 & 4.822 & 0.612 & 14.555 & 0.700 & 0.742 & 0.188 & 3.595 & 8.967 \\
    BTC & 731--1095 & 6.743 & 2.969 & 10.311 & 0.654 & 38.751 & 0.544 & 0.587 & 0.091 & 8.309 & 20.028 \\
    ETH & 1--30 & 0.025 & 0.001 & 0.204 & 0.123 & 0.233 & 0.423 & 0.481 & 0.219 & 0.104 & 0.289 \\
    ETH & 31--90 & 0.104 & 0.021 & 0.458 & 0.227 & 0.643 & 0.605 & 0.642 & 0.364 & 0.316 & 0.710 \\
    ETH & 91--180 & 0.274 & 0.076 & 0.893 & 0.306 & 1.389 & 0.735 & 0.778 & 0.393 & 0.506 & 1.406 \\
    ETH & 181--365 & 0.866 & 0.174 & 2.258 & 0.383 & 3.366 & 0.873 & 0.899 & 0.346 & 0.808 & 3.696 \\
    ETH & 366--730 & 2.936 & 0.479 & 5.737 & 0.512 & 12.786 & 0.870 & 0.895 & 0.367 & 2.475 & 11.269 \\
    ETH & 731--1095 & 5.308 & 1.643 & 7.678 & 0.691 & 22.996 & 0.829 & 0.872 & 0.252 & 8.206 & 16.644 \\
    XRP & 1--30 & 0.057 & -0.013 & 0.488 & 0.116 & 0.518 & 0.446 & 0.539 & 0.244 & 0.078 & 0.459 \\
    XRP & 31--90 & 0.175 & -0.016 & 0.889 & 0.197 & 1.114 & 0.600 & 0.669 & 0.381 & 0.217 & 1.067 \\
    XRP & 91--180 & 0.265 & -0.010 & 0.976 & 0.272 & 1.481 & 0.653 & 0.725 & 0.401 & 0.278 & 1.483 \\
    XRP & 181--365 & 0.436 & -0.027 & 1.315 & 0.331 & 2.078 & 0.756 & 0.826 & 0.445 & 0.525 & 2.283 \\
    XRP & 366--730 & 0.684 & 0.045 & 1.628 & 0.420 & 3.186 & 0.849 & 0.898 & 0.454 & 1.131 & 3.172 \\
    XRP & 731--1095 & 0.805 & 0.466 & 1.529 & 0.526 & 3.437 & 0.903 & 0.943 & 0.348 & 1.411 & 2.918 \\
    \end{longtable}

\end{landscape}
\subsection{Methodology}\label{app:a2}
\subsubsection{Macro unit-root tests}\label{app:a2-1}
\begin{table}[H]
  \centering
  \renewcommand{\arraystretch}{1.15}
  \caption{Unit-root tests for macro series}
  \label{tab:macro-unitroot-pvalues}
  \resizebox{\textwidth}{!}{%
  \begin{tabular}{lrrrrrrll}
  \toprule
  Series & DF-GLS $p$ (c) & KPSS $p$ (c) & ZA $p$ (c) & DF-GLS $p$ (ct) & KPSS $p$ (ct) & ZA $p$ (ct) & Decision (c) & Decision (ct) \\
  \midrule
  BTC log return & 0.000 & 0.100 & 0.000 & 0.000 & 0.100 & 0.001 & STATIONARY & STATIONARY \\
  Crypto Fear \& Greed Index & 0.001 & 0.078 & 0.002 & 0.000 & 0.098 & 0.005 & STATIONARY & STATIONARY \\
  HY OAS spread (BAMLH0A0HYM2) & 0.014 & 0.076 & 0.584 & 0.092 & 0.023 & 0.556 & AMBIGUOUS & UNIT ROOT \\
  Fed funds rate (DFF) & 0.260 & 0.010 & 0.024 & 0.422 & 0.010 & 0.697 & AMBIGUOUS & UNIT ROOT \\
  US 10Y Treasury yield (DGS10) & 0.469 & 0.010 & 0.691 & 0.859 & 0.010 & 0.877 & UNIT ROOT & UNIT ROOT \\
  Trade-weighted USD (broad) (DTWEXBGS) & 0.625 & 0.010 & 0.323 & 0.232 & 0.022 & 0.421 & UNIT ROOT & UNIT ROOT \\
  Nasdaq Composite (NASDAQCOM) & 0.828 & 0.010 & 0.232 & 0.240 & 0.011 & 0.327 & UNIT ROOT & UNIT ROOT \\
  10Y-2Y Treasury spread (T10Y2Y) & 0.266 & 0.010 & 0.404 & 0.742 & 0.010 & 0.893 & UNIT ROOT & UNIT ROOT \\
  VIX index (VIXCLS) & 0.000 & 0.100 & 0.000 & 0.000 & 0.010 & 0.001 & STATIONARY & AMBIGUOUS \\
  \bottomrule
  \end{tabular}
  }
\end{table}

\subsection{Posterior Results}\label{app:a3}
\subsubsection{Posterior effect metrics}\label{app:a3-metrics}
\label{app:defs}

Let $\beta_{h,p,y}$ denote the local projection coefficient (see equation \eqref{eq:linear-predictor}). 
These coefficients represent standardized impulse responses: $\beta_{h,p,y}$ measures 
how target $y$ changes in standardized units at horizon $h$ when predictor $p$ increases 
by one standard deviation at time $t$, conditional on all other predictors in $Z_{t,h}$.
We denote the posterior samples by index $s=1,\ldots,S$, where $S$ is the total number of MCMC draws.

To analyze effects in target $y$'s native units, we scale the standardized 
coefficients by $\sigma_{h,y}$, the reference standard deviation of target $y$ 
at horizon $h$:
\begin{equation*}
    \beta^*_{h,p,y} = \beta_{h,p,y} \cdot \sigma_{h,y},
    \end{equation*}

    \begin{table}[H]
\centering
\small
\caption{Posterior Effect Metric Definitions}
\label{tab:defs}
\begin{tabularx}{\textwidth}{@{} l X @{}} \toprule
\textbf{Symbol} & \textbf{Definition / Computation} \\
\midrule
$\displaystyle \beta^*_{h,p,y}$
& Effect on target $y$ (native units) per one-standard-deviation shock to predictor $p$:
$\beta^*_{h,p,y}=\beta_{h,p,y}\,\sigma_{h,y}$. \\[3pt]

$\displaystyle \operatorname{med}_{h,p,y}$
& Posterior median:
$\operatorname{med}_{h,p,y}=Q_{0.50}\{\beta^*_{h,p,y,s}\}_{s=1}^{S}$. \\[3pt]

$\displaystyle \mathrm{CrI}_{0.95}(h,p,y)=[\ell_{h,p,y},u_{h,p,y}]$
& $95\%$ credible interval:
$\ell_{h,p,y}=Q_{0.025}\{\beta^*_{h,p,y,s}\}$,\;
$u_{h,p,y}=Q_{0.975}\{\beta^*_{h,p,y,s}\}$. \\[3pt]

$\displaystyle \pi_{+}(h,p,y)$
& Posterior probability of positive effect:
$\displaystyle \pi_{+}(h,p,y)=\frac{1}{S}\sum_{s=1}^{S}\mathds{1}\{\beta^*_{h,p,y,s}>0\}$. \\[4pt]

$\displaystyle \mathcal{S}_{0.95}(h,p,y)\in\{0,1\}$
& Significance indicator (two-sided):
$\displaystyle \mathcal{S}_{0.95}(h,p,y)=
\mathds{1}\big\{\ell_{h,p,y}>0 \;\lor\; u_{h,p,y}<0 \;\lor\; \pi_{+}(h,p,y)\ge 0.95 \;\lor\; \pi_{+}(h,p,y)\le 0.05\big\}$. \\[4pt]
\bottomrule
\end{tabularx}
\end{table}
        
\paragraph{Ranking and aggregation procedures.}
To identify the most important effects, we apply two complementary rankings:

\begin{enumerate}
\item \textbf{Per-horizon ranking.} At each horizon $h$, we select all significant predictor-target pairs (\,$\mathcal{S}_{0.95}(h,p,y)=1$\,) and keep the eight with the largest absolute median effects $|\operatorname{med}_{h,p,y}|$.

\item \textbf{Cross-basket stability ranking.} For each horizon-target pair $(h,y)$, we aggregate results across all baskets and rank predictors by the number of baskets where $\mathcal{S}_{0.95}(h,p,y)=1$; in the case of ties, we then sort by their mean and maximum absolute median effects. We report the four predictors with the highest basket counts for each $(h,y)$ combination.
\end{enumerate}

\subsubsection{Posterior label definitions}\label{app:a3-labels}
\begin{table}[H]
  \centering
  \footnotesize
  \renewcommand{\arraystretch}{1.2}
\begin{tabular}{@{}>{\raggedright\arraybackslash}p{0.30\textwidth}
                >{\raggedright\arraybackslash}p{0.30\textwidth}
                >{\raggedright\arraybackslash}p{0.30\textwidth}@{}}
    \toprule
    Label & Series & Transform \\
    \midrule
    VIX EMA24 / EMA24w & VIX index (VIXCLS) & 24-week exponential moving average. \\
    FGI EMA24 / EMA24w & Crypto Fear \& Greed Index & 24-week exponential moving average. \\
    BAMLH0A0HYM2 RVol24w & HY OAS spread (BAMLH0A0HYM2) & 24-week rolling volatility. \\
    T10Y2Y EMA24 / EMA24w & 10Y-2Y Treasury spread (T10Y2Y) & 24-week exponential moving average. \\
    T10Y2Y EMA12 / EMA12w & 10Y-2Y Treasury spread (T10Y2Y) & 12-week exponential moving average. \\
    T10Y2Y RVol24w & 10Y-2Y Treasury spread (T10Y2Y) & 24-week rolling volatility. \\
    BTC RVol12w & BTC log return & 12-week rolling volatility. \\
    BTC EMA4 / EMA4w & BTC log return & 4-week exponential moving average. \\
    BTC EMA8 / EMA8w & BTC log return & 8-week exponential moving average. \\
    BTC EMA12 / EMA12w & BTC log return & 12-week exponential moving average. \\
    NASDAQCOM EMA4 / EMA4w & Nasdaq Composite (NASDAQCOM) & 4-week exponential moving average. \\
    \bottomrule
  \end{tabular}
  \caption{Posterior label definitions referenced in Figures \ref{fig:target_horizon_panel} and \ref{fig:top8_horizon}.}
\end{table}

\subsubsection{A Two-Stage Bayesian Meta Framework}\label{app:a3-endogenous-decay}
We examine how endogenous predictors' effects evolve across forecast horizons by pooling information 
across baskets through a Bayesian meta-analysis \citep{sutton2001bayesian,Rover2020bayesmeta}. For each basket $b$, horizon $h$, predictor 
$p$, and target $y$, the local projection model (§\ref{sec:bayes-model}) yields posterior 
samples $\{\beta^{(s)}_{b,h,p,y}\}_{s=1}^S$. Following Appendix~\ref{app:defs}, we scale 
these standardized coefficients by the target's reference standard deviation to obtain effects 
in native units:
\[
\theta^{(s)}_{b,h,p,y}\;=\;\sigma_{h,y}\,\beta^{(s)}_{b,h,p,y},
\]
where $\theta$ measures the effect in target-native units per one-standard-deviation increase in predictor $p$.

\paragraph{Stage 1: Basket-level posterior summaries.}
For each $(b,h,p,y)$ we compute the posterior mean $\bar{\theta}_{b,h,p,y}$, standard deviation 
$s_{b,h,p,y}$, equal-tailed credible intervals at $68\%$ and $95\%$, and the posterior 
probability of positive effects $\text{ppos}_{b,h,p,y}=\Pr(\theta_{b,h,p,y}>0)$. 
We mark an observation as 95\%-significant if the credible interval excludes zero.

\paragraph{Stage 2: Meta-analysis across horizons.}
For each predictor-target pair $(p,y)$, we pool the Stage-1 summaries across 
basket-horizon combinations. Let $i=1,\ldots,N$ index these combinations, with 
$\bar{\theta}_i$ and $s_i$ denoting the posterior mean and standard deviation, 
and $h(i)\in\{30,\allowbreak 60,\allowbreak 90,\allowbreak 180,\allowbreak 365,\allowbreak 730,\allowbreak 1095\}$ the horizon in days. We fit a 
horizon-specific random-effects model
\[
\bar{\theta}_i \mid \mu_{h(i)},\tau \;\sim\; \mathcal{N}\!\left(\mu_{h(i)},\ s_i^2+\tau^2\right),
\]
where $\mu_h$ represents the population mean effect at horizon $h$ and $\tau>0$ 
quantifies between-basket heterogeneity (assumed constant across horizons).
To smooth temporal variation in effects, we impose a first-order random walk 
prior over horizons measured in days:
\[
\mu_{h_1}\sim\mathcal{N}(0,\sigma_\mu^2),\qquad 
\mu_{h_{j+1}}=\mu_{h_j}+\delta_j,\qquad 
\delta_j \sim \mathcal{N}\!\bigl(0,\ \sigma_{\mathrm{rw}}^2\,\Delta_j\bigr),
\]
with $\Delta_j=(h_{j+1}-h_j)/30$ to handle unequally spaced horizons, 
$\sigma_\mu=1.0$, $\sigma_{\mathrm{rw}}\sim\mathrm{HalfNormal}(0.05)$, and 
$\tau\sim\mathrm{HalfStudentT}(\nu=3,\sigma=0.3)$.

We perform inference using PyMC’s NUTS with 4 chains, 3,000 tuning iterations and 2,000 
posterior draws per chain (target acceptance 0.99).

\paragraph{Panel A: pooled effect $\mu(h)$.}
For each horizon, we plot the posterior median of $\mu_h$ with $68\%$ and $95\%$ credible bands.

\paragraph{Panel B: retention and half-life.}
We plot the retention ratio $R(h) = |\mu_h|/|\mu_{30}|$ (posterior median with $68\%$, 
$95\%$ credible bands). The half-life is the horizon where $R(h)$ falls to $0.5$. 
If fewer than $50\%$ of draws cross this threshold by 1095 days, we report the half-life as right-censored.

\paragraph{Panel C: cross-basket significance.}
For each horizon, we count how many baskets exhibit 95\%-significant effects 
(credible interval excludes zero).

\subsubsection{Meta-Analysis Model Convergence}\label{app:a3-endogenous-decay-diagnostics}
\begin{table}[H]
  \centering
  \renewcommand{\arraystretch}{1.2}
  \caption{Convergence diagnostics for horizon-specific random-effects meta-analysis models}
  \label{tab:endogenous-decay-convergence}
  \resizebox{\textwidth}{!}{%
    \begin{tabular}{llrrrrrr}
\toprule
Basket & Predictor & Max $\hat{R}$ & Min ESS (bulk) & Min ESS (tail) & Max MCSE/SD & Min BFMI & Divergence rate \\
\midrule
Forward CVaR 10\% & Realized CVaR 10\% & 1.00 & 2713 & 2895 & 0.015 & 0.79 & 0.00025 \\
Forward CVaR 10\% & Realized mean ER (top 25\%) & 1.00 & 2595 & 2885 & 0.015 & 0.80 & 0.00025 \\
Forward CVaR 10\% & Realized median ER & 1.00 & 2368 & 1925 & 0.013 & 0.82 & 0.00038 \\
Forward CVaR 10\% & Realized Sharpe & 1.00 & 2273 & 1779 & 0.015 & 0.74 & 0.00038 \\
Forward Mean ER (Top 25\%) & Realized CVaR 10\% & 1.00 & 3026 & 3129 & 0.013 & 0.78 & 0.00013 \\
Forward Mean ER (Top 25\%) & Realized mean ER (top 25\%) & 1.00 & 3036 & 2672 & 0.013 & 0.82 & 0.00000 \\
Forward Mean ER (Top 25\%) & Realized median ER & 1.00 & 3111 & 1689 & 0.014 & 0.79 & 0.00013 \\
Forward Mean ER (Top 25\%) & Realized Sharpe & 1.00 & 3513 & 2702 & 0.013 & 0.86 & 0.00013 \\
Forward Median ER & Realized CVaR 10\% & 1.00 & 2906 & 3054 & 0.013 & 0.78 & 0.00000 \\
Forward Median ER & Realized mean ER (top 25\%) & 1.00 & 3219 & 2858 & 0.013 & 0.75 & 0.00013 \\
Forward Median ER & Realized median ER & 1.00 & 2371 & 2984 & 0.013 & 0.79 & 0.00000 \\
Forward Median ER & Realized Sharpe & 1.00 & 3722 & 2857 & 0.014 & 0.83 & 0.00025 \\
Forward Sharpe & Realized CVaR 10\% & 1.00 & 2295 & 2276 & 0.014 & 0.80 & 0.00075 \\
Forward Sharpe & Realized mean ER (top 25\%) & 1.00 & 2373 & 2032 & 0.015 & 0.74 & 0.00063 \\
Forward Sharpe & Realized median ER & 1.00 & 2964 & 2450 & 0.015 & 0.78 & 0.00038 \\
Forward Sharpe & Realized Sharpe & 1.00 & 2340 & 2557 & 0.016 & 0.84 & 0.00100 \\
\bottomrule
\end{tabular}

  }
\end{table}

Table~\ref{tab:endogenous-decay-convergence} reports convergence diagnostics for all 16 target-predictor pairs. 
All models achieve convergence with $\hat{R}\leq 1.00$, ESS $\geq 1689$, and 
divergence rates $\leq 0.1\%$.

\begin{figure}[H]
  \centering
  \setlength{\tabcolsep}{4pt}
  \begin{tabular}{cc}
    \includegraphics[width=0.48\textwidth]{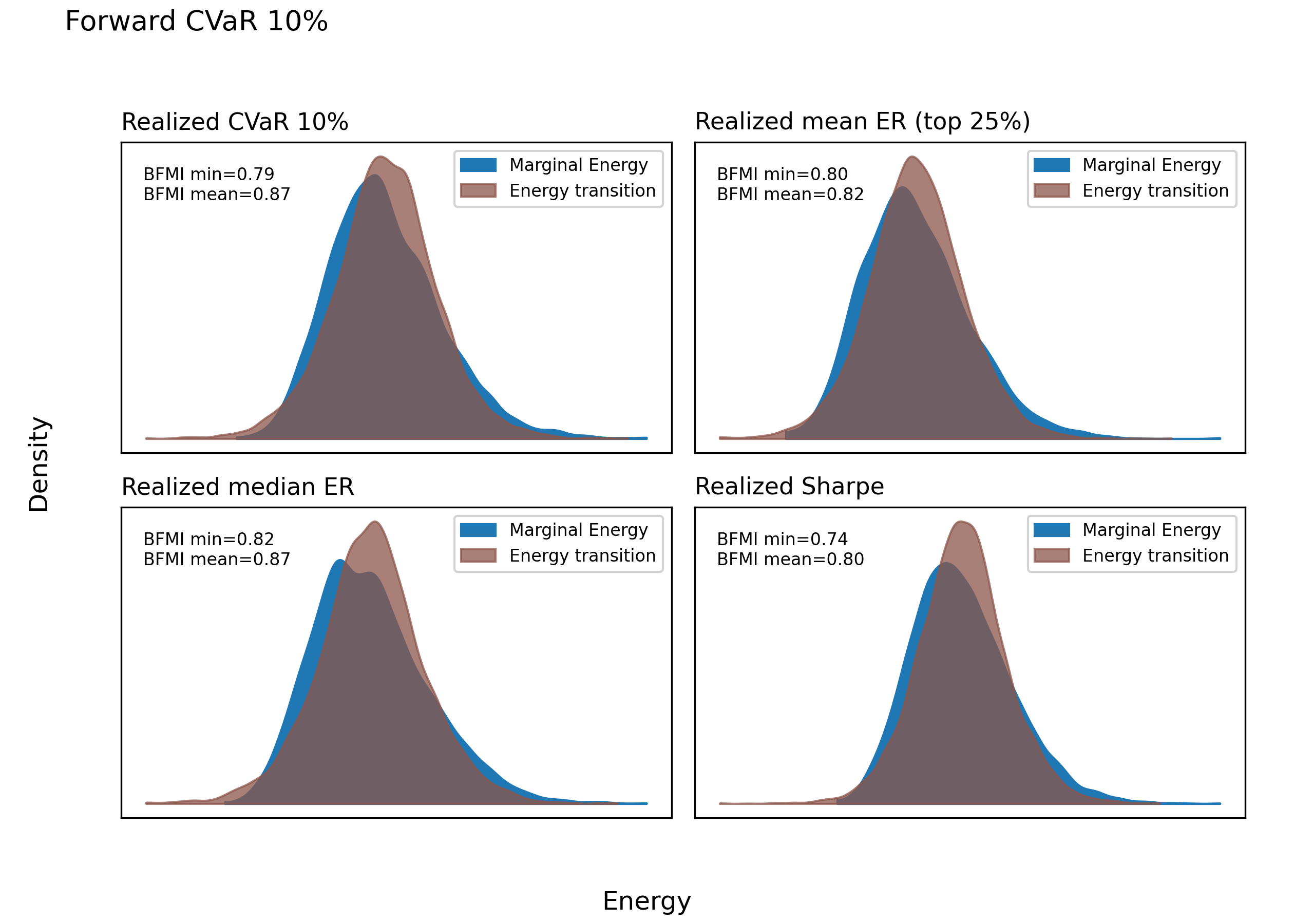} &
    \includegraphics[width=0.48\textwidth]{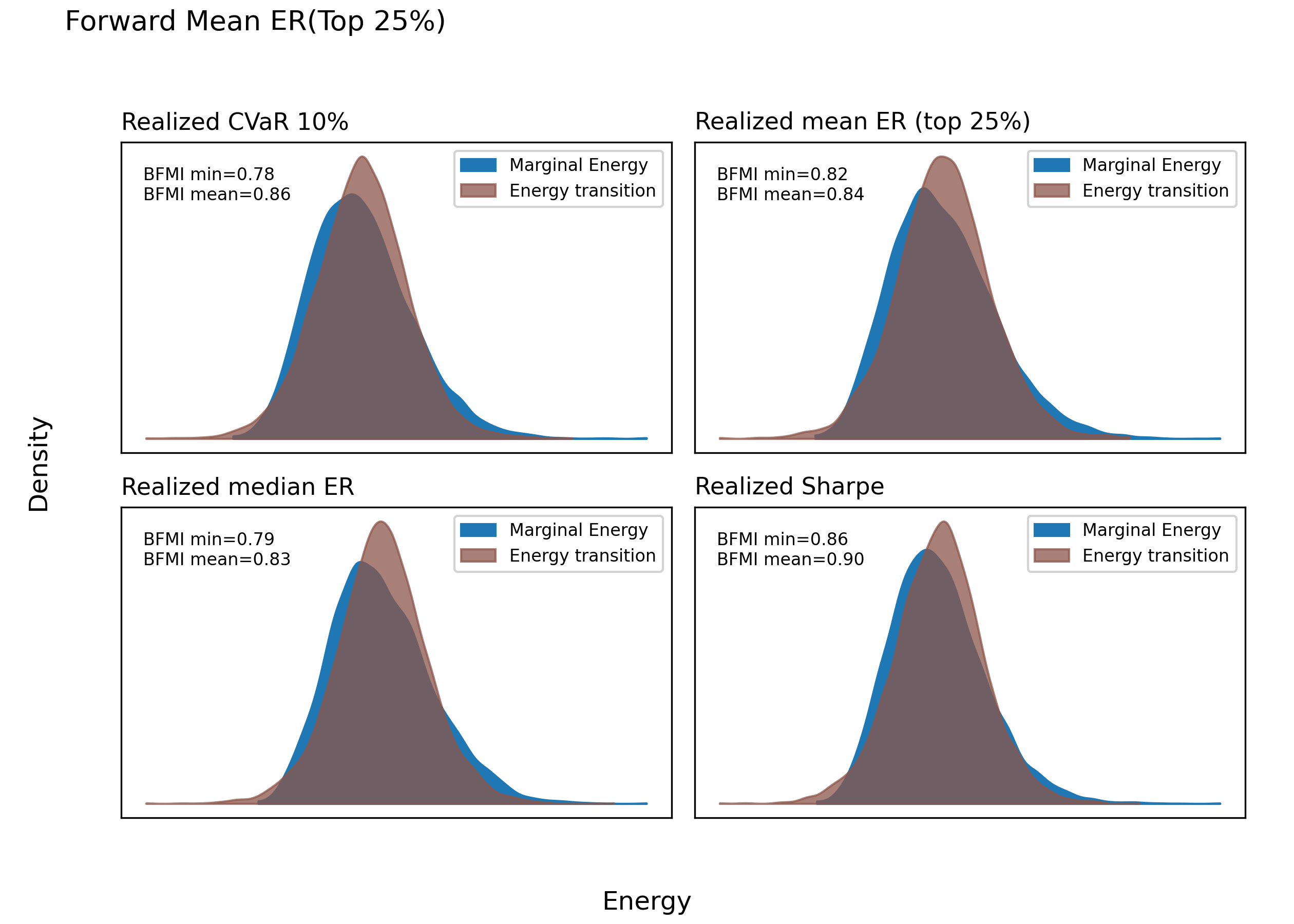} \\
    (a) Forward CVaR 10\% & (b) Forward Mean ER (Top 25\%) \\
    \includegraphics[width=0.48\textwidth]{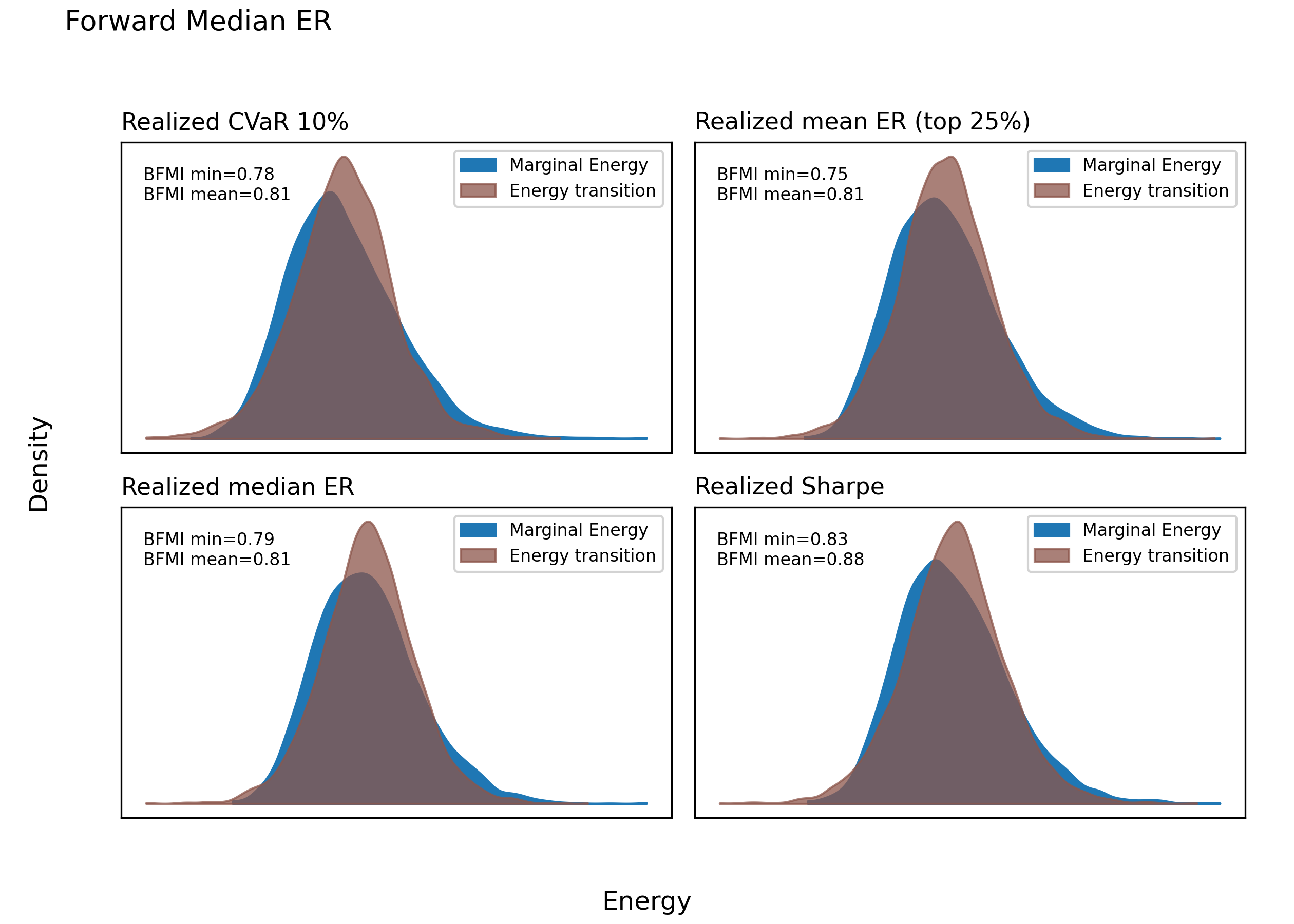} &
    \includegraphics[width=0.48\textwidth]{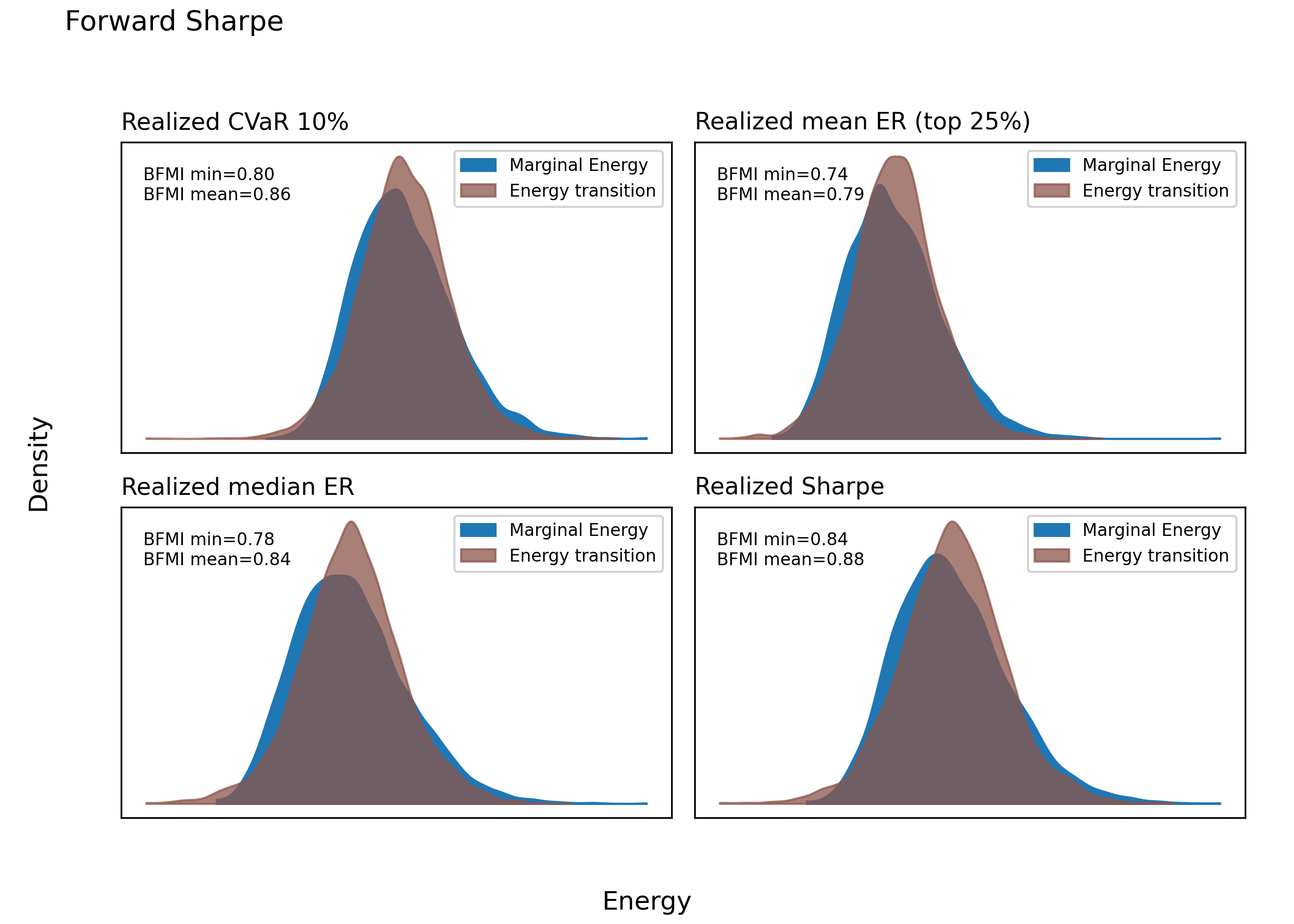} \\
    (c) Forward Median ER & (d) Forward Sharpe \\
  \end{tabular}
  \caption{Energy diagnostics for all target--predictor pairs.}
  \label{fig:endogenous-decay-energy}
\end{figure}

Figure~\ref{fig:endogenous-decay-energy} shows energy diagnostics for each target--predictor pair. 
With minimum BFMI of 0.74, the overlapping energy transition and 
marginal energy distributions indicate the models did not encounter geometric pathologies.

\subsubsection{Meta-Analysis Results}\label{app:a3-endogenous-decay-results}
\begin{table}[H]
  \centering
  \caption{Meta-analysis results: forward mean ER (Top 25\%) significant horizons}
  \label{tab:endogenous-decay-top25}
  \resizebox{\textwidth}{!}{%
    \begin{tabular}{lrllrrrrrr}
      \toprule
      Basket & Horizon (days) & Predictor & Target & Mean & SD & Pr(>0) & 95\% Credible Interval (lower) & 95\% Credible Interval (upper) & 95\% Credible Interval\\
      \midrule
      BNB & 366--730 & Realized mean ER (top 25\%) & Forward Mean ER (Top 25\%) & 0.0557 & 0.0267 & 0.9824 & 0.0041 & 0.1087 & \checkmark \\
      BNB & 731--1095 & Realized mean ER (top 25\%) & Forward Mean ER (Top 25\%) & 0.1302 & 0.0606 & 0.9841 & 0.0133 & 0.2480 & \checkmark \\
      ETH & 1--30 & Realized median ER & Forward Mean ER (Top 25\%) & 0.0057 & 0.0028 & 0.9764 & 0.0001 & 0.0112 & \checkmark \\
      ETH & 31--90 & Realized median ER & Forward Mean ER (Top 25\%) & 0.0080 & 0.0040 & 0.9775 & 0.0002 & 0.0159 & \checkmark \\
      ETH & 91--180 & Realized median ER & Forward Mean ER (Top 25\%) & 0.0150 & 0.0072 & 0.9800 & 0.0009 & 0.0294 & \checkmark \\
      ETH & 181--365 & Realized median ER & Forward Mean ER (Top 25\%) & 0.0358 & 0.0168 & 0.9828 & 0.0029 & 0.0693 & \checkmark \\
      ETH & 366--730 & Realized median ER & Forward Mean ER (Top 25\%) & 0.0640 & 0.0288 & 0.9862 & 0.0072 & 0.1218 & \checkmark \\
      ETH & 731--1095 & Realized median ER & Forward Mean ER (Top 25\%) & 0.0777 & 0.0343 & 0.9879 & 0.0101 & 0.1461 & \checkmark \\
      \bottomrule
    \end{tabular}
  }
\end{table}

\begin{table}[H]
  \centering
  \caption{Meta-analysis results: forward Sharpe significant horizons}
  \label{tab:endogenous-decay-sharpe}
  \resizebox{\textwidth}{!}{%
    \begin{tabular}{lrllrrrrrr}
      \toprule
      Basket & Horizon (days) & Predictor & Target & Mean & SD & Pr(>0) & 95\% Credible Interval (lower) & 95\% Credible Interval (upper) & 95\% Credible Interval\\
      \midrule
      LINK & 1--30 & Realized CVaR 10\% & Forward Sharpe & 0.1183 & 0.0520 & 0.9908 & 0.0167 & 0.2227 & \checkmark \\
      LINK & 31--90 & Realized CVaR 10\% & Forward Sharpe & 0.1194 & 0.0526 & 0.9908 & 0.0165 & 0.2248 & \checkmark \\
      LINK & 91--180 & Realized CVaR 10\% & Forward Sharpe & 0.1167 & 0.0513 & 0.9908 & 0.0167 & 0.2189 & \checkmark \\
      LINK & 181--365 & Realized CVaR 10\% & Forward Sharpe & 0.0430 & 0.0190 & 0.9902 & 0.0059 & 0.0809 & \checkmark \\
      LINK & 366--730 & Realized CVaR 10\% & Forward Sharpe & 0.0138 & 0.0061 & 0.9900 & 0.0018 & 0.0260 & \checkmark \\
      LINK & 731--1095 & Realized CVaR 10\% & Forward Sharpe & 0.0134 & 0.0060 & 0.9904 & 0.0017 & 0.0252 & \checkmark \\
      ETH & 1--30 & Realized median ER & Forward Sharpe & 0.1163 & 0.0584 & 0.9760 & 0.0009 & 0.2320 & \checkmark \\
      ETH & 31--90 & Realized median ER & Forward Sharpe & 0.1575 & 0.0789 & 0.9759 & 0.0013 & 0.3128 & \checkmark \\
      ETH & 91--180 & Realized median ER & Forward Sharpe & 0.0986 & 0.0494 & 0.9756 & 0.0006 & 0.1955 & \checkmark \\
      ETH & 181--365 & Realized median ER & Forward Sharpe & 0.0374 & 0.0187 & 0.9755 & 0.0003 & 0.0747 & \checkmark \\
      ETH & 366--730 & Realized median ER & Forward Sharpe & 0.0250 & 0.0124 & 0.9776 & 0.0008 & 0.0496 & \checkmark \\
      ETH & 731--1095 & Realized median ER & Forward Sharpe & 0.0366 & 0.0179 & 0.9781 & 0.0012 & 0.0718 & \checkmark \\
      \bottomrule
    \end{tabular}
  }
\end{table}

Tables~\ref{tab:endogenous-decay-top25} and~\ref{tab:endogenous-decay-sharpe} 
report basket-level posterior summaries $\bar{\theta}_{b,h,p,y}$ for combinations 
exhibiting 95\%-significant effects, as defined in Appendix~\ref{app:a3-endogenous-decay}.
Out of the four target variables, only forward Mean ER (Top 25\%) and forward Sharpe 
exhibit significant effects for at least one basket-predictor-horizon combination.
For forward Sharpe, only two basket-predictor pairs exhibit significant effects. LINK's 
realized CVaR 10\% shows positive effects that remain substantial over horizons 
between 1--180 days; one standard deviation would increase forward Sharpe by more 
than 0.11 in Sharpe units, but later decay to approximately 0.013 in Sharpe units at 
longer horizons. For ETH, realized median ER exhibits strong impact over horizons 
between 1--90 days (one standard deviation of realized median ER would increase 
forward Sharpe by more than 0.11 in Sharpe units), and then later decay with 
horizon; for the horizon 731--1095 days the effects decay to 0.0366 in Sharpe units.

For forward Mean ER (Top 25\%), a one-standard-deviation shock from realized median 
ER produces positive and significant effects on ETH's forward Mean ER (Top 25\%) 
across all horizons, with the effect at 731--1095 days reaching 7.77 percentage 
points. BNB's realized mean ER (top 25\%) also shows positive and significant 
effects on forward Mean ER (Top 25\%) at longer horizons (366--730 and 731--1095 days), 
with the longest-horizon effect reaching 13.02 percentage points.

\begin{figure}[H]
  \centering
  \setlength{\tabcolsep}{6pt}
  \begin{tabular}{cc}
    \includegraphics[width=0.48\textwidth]{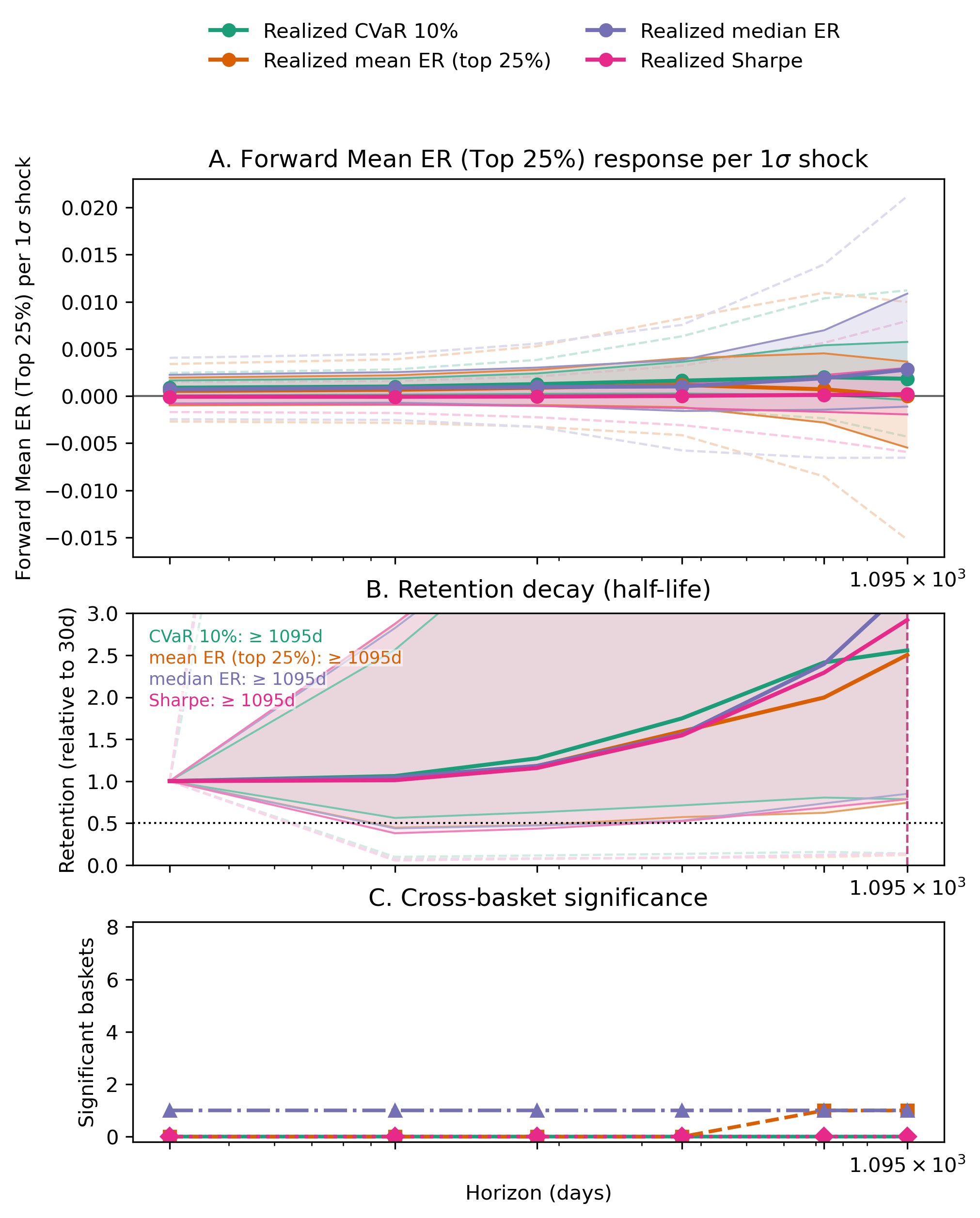} &
    \includegraphics[width=0.48\textwidth]{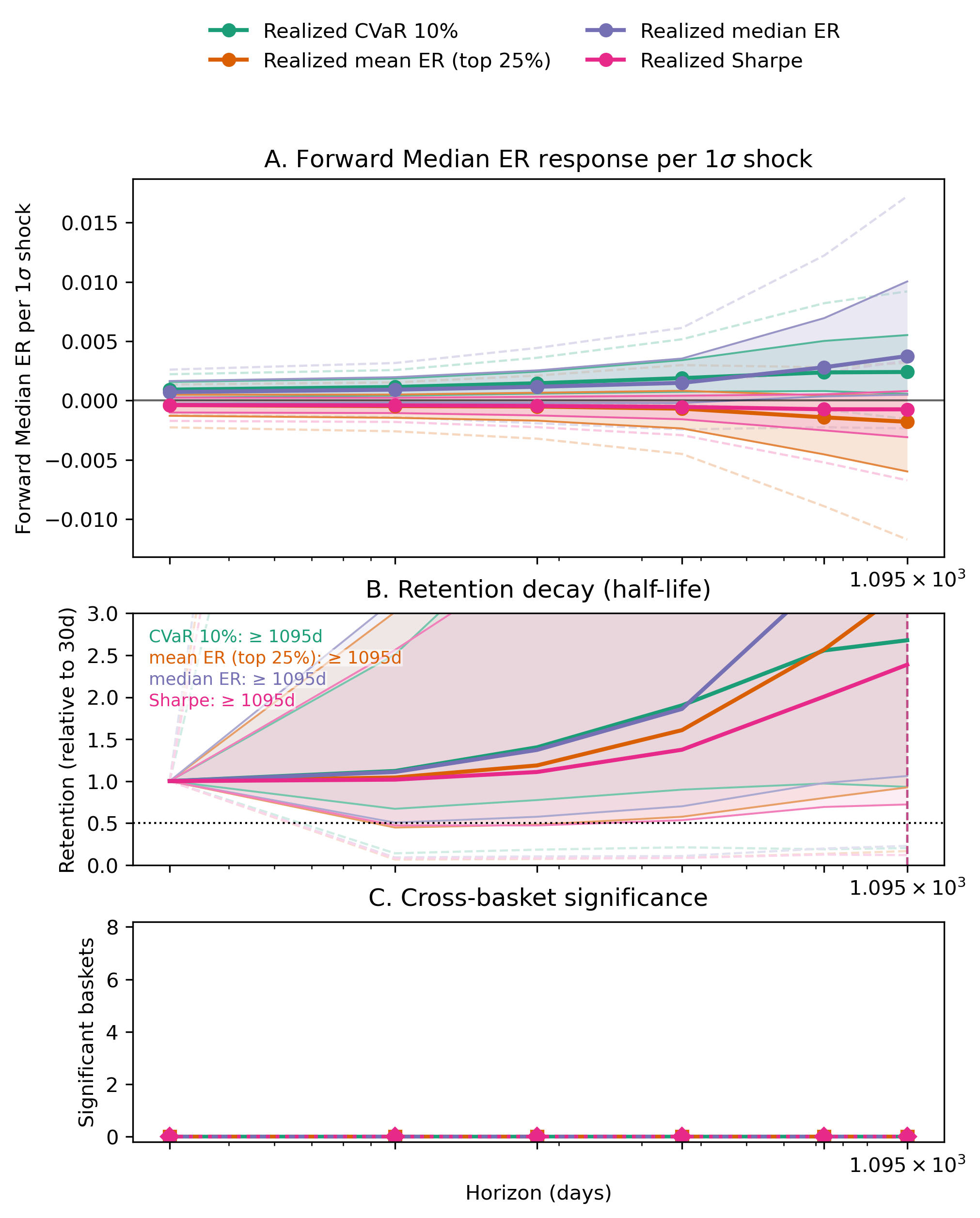} \\
    (a) Forward Mean ER (Top 25\%) & (b) Forward Median ER \\
    \includegraphics[width=0.48\textwidth]{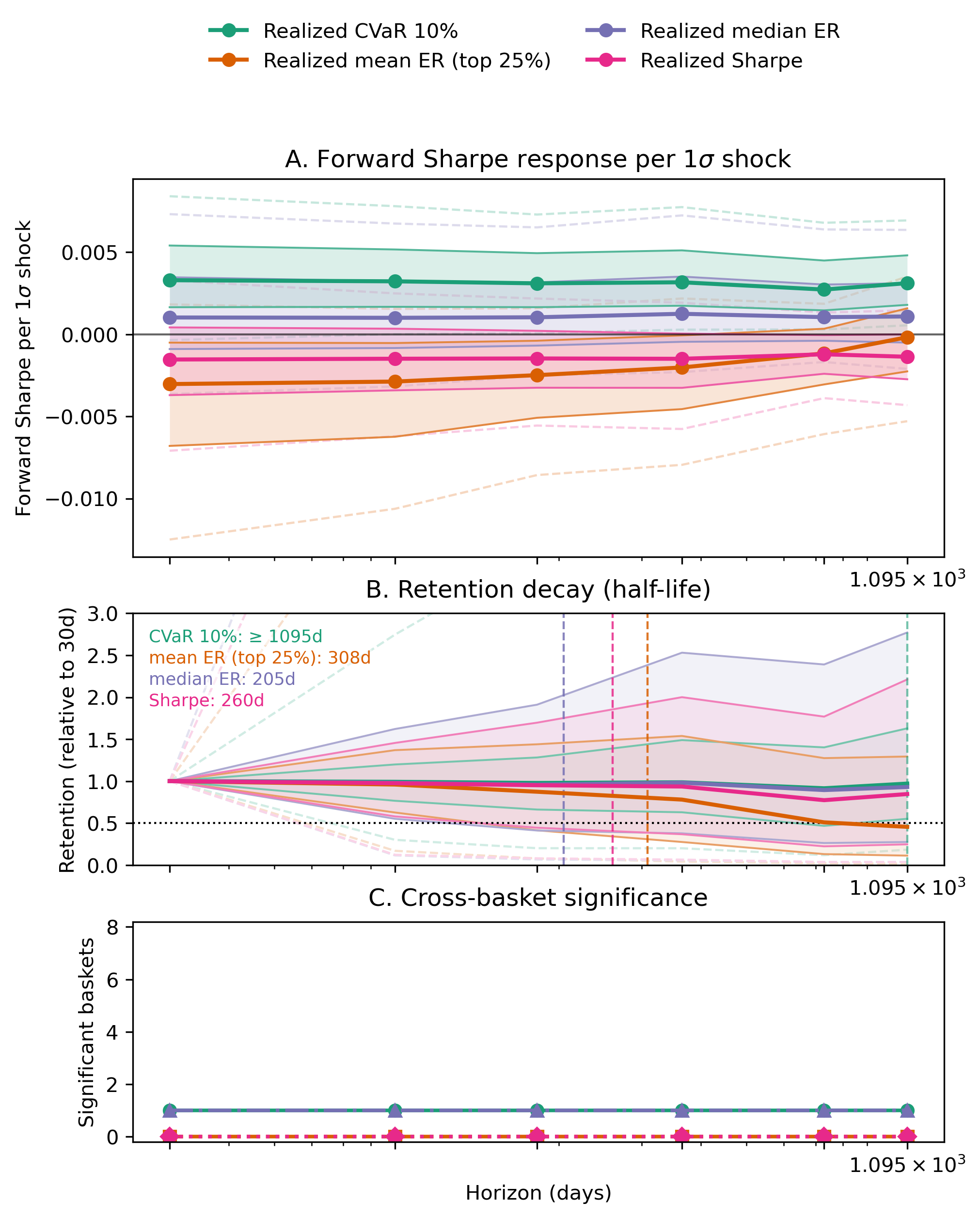} &
    \includegraphics[width=0.48\textwidth]{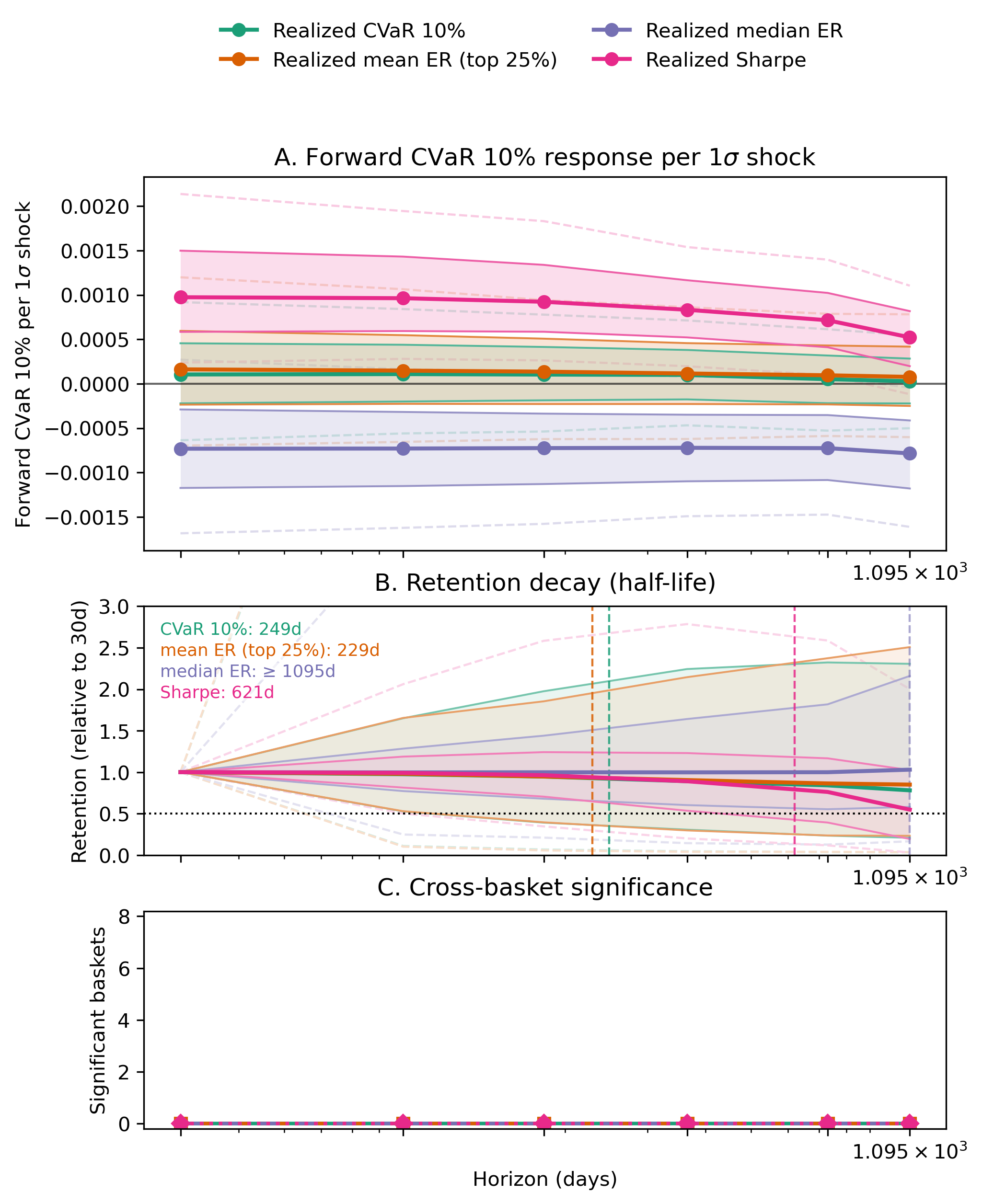} \\
    (c) Forward Sharpe & (d) Forward CVaR 10\% \\
  \end{tabular}
  \caption{Endogenous predictor effects across forecast horizons by target.}
  \label{fig:endogenous-decay-summary}
\end{figure}

Figure~\ref{fig:endogenous-decay-summary} reports the pooled (population-average) 
effect $\mu(h)$ under the random-effects model as defined in 
Appendix~\ref{app:a3-endogenous-decay}. At each horizon $h$, we plot the 
posterior median (solid line with markers), the 68\% 
credible band (shaded region), and the 95\% credible interval (outer dashed lines), 
with each predictor distinguished by color.
The pooled estimates reveal generally weak population-level effects across all targets. 
For forward Sharpe, realized CVaR 10\% shows the strongest pooled effect 
(one standard deviation would increase forward Sharpe by roughly 0.0027--0.0033 
in Sharpe units across horizons); however, most endogenous predictor effects 
remain economically negligible.
Overall, the section reveals substantial cross-basket heterogeneity for the endogenous predictors. 
While specific pairs exhibit large effects (e.g., LINK's CVaR 10\% driving more than 
11 percentage points increases in forward Sharpe), these patterns are basket-specific 
rather than generalizable.

\subsubsection{Long-horizon top-quartile impulse responses to FGI EMA24w shocks (Figure~\ref{fig:forest-long-horizon} data)}\label{app:a3-fgi-top25}
\begin{table}[H]
  \centering
  \footnotesize
\begin{tabular}{l l l r r r r c}
  \hline
  Basket & Target & Horizon & Median & Lower & Upper & $p_{pos}$ & Sig. (95\%) \\
  \hline
  DOGE & Forward mean ER (Top 25\%) & 181--365 & -0.477 & -1.226 & -0.236 & 0.000 & \checkmark \\
  BNB & Forward mean ER (Top 25\%) & 181--365 & -0.114 & -0.282 & -0.030 & 0.001 & \checkmark \\
  ADA & Forward mean ER (Top 25\%) & 181--365 & -0.100 & -0.261 & -0.006 & 0.018 & \checkmark \\
  ETH & Forward mean ER (Top 25\%) & 181--365 & -0.053 & -0.146 & -0.001 & 0.022 & \checkmark \\
  BTC & Forward mean ER (Top 25\%) & 181--365 & -0.016 & -0.033 & -0.001 & 0.018 & \checkmark \\
  LINK & Forward mean ER (Top 25\%) & 181--365 & -0.007 & -0.033 & 0.002 & 0.065 &  \\
  DOGE & Forward mean ER (Top 25\%) & 366--730 & -0.512 & -0.776 & -0.262 & 0.000 & \checkmark \\
  ADA & Forward mean ER (Top 25\%) & 366--730 & -0.115 & -0.219 & -0.019 & 0.009 & \checkmark \\
  BNB & Forward mean ER (Top 25\%) & 366--730 & -0.108 & -0.202 & -0.038 & 0.000 & \checkmark \\
  ETH & Forward mean ER (Top 25\%) & 366--730 & -0.076 & -0.158 & -0.002 & 0.021 & \checkmark \\
  LINK & Forward mean ER (Top 25\%) & 366--730 & -0.027 & -0.068 & 0.013 & 0.087 &  \\
  BTC & Forward mean ER (Top 25\%) & 366--730 & -0.025 & -0.050 & -0.001 & 0.021 & \checkmark \\
  DOGE & Forward mean ER (Top 25\%) & 731--1095 & -0.486 & -0.726 & -0.242 & 0.000 & \checkmark \\
  BNB & Forward mean ER (Top 25\%) & 731--1095 & -0.135 & -0.234 & -0.033 & 0.005 & \checkmark \\
  LINK & Forward mean ER (Top 25\%) & 731--1095 & -0.074 & -0.173 & 0.029 & 0.073 &  \\
  ETH & Forward mean ER (Top 25\%) & 731--1095 & -0.066 & -0.141 & 0.008 & 0.041 &  \\
  BTC & Forward mean ER (Top 25\%) & 731--1095 & -0.047 & -0.091 & -0.001 & 0.025 & \checkmark \\
  ADA & Forward mean ER (Top 25\%) & 731--1095 & -0.019 & -0.112 & 0.079 & 0.343 &  \\
  \hline
  \end{tabular}
\end{table}

\subsubsection{Long-horizon median impulse responses to FGI EMA24w shocks (Figure~\ref{fig:forest-long-horizon} data)}\label{app:a3-fgi-median}
\begin{table}[H]
  \centering
  \footnotesize
\begin{tabular}{l l l r r r r c}
    \hline
    Basket & Target & Horizon & Median & Lower & Upper & $p_{pos}$ & Sig. (95\%) \\
    \hline
    DOGE & Forward median ER & 181--365 & -0.118 & -0.338 & -0.017 & 0.013 & \checkmark \\
    BNB & Forward median ER & 181--365 & -0.061 & -0.121 & -0.022 & 0.004 & \checkmark \\
    ETH & Forward median ER & 181--365 & -0.053 & -0.119 & -0.007 & 0.007 & \checkmark \\
    ADA & Forward median ER & 181--365 & -0.037 & -0.107 & 0.010 & 0.051 &  \\
    LINK & Forward median ER & 181--365 & -0.007 & -0.023 & -0.001 & 0.012 & \checkmark \\
    BTC & Forward median ER & 181--365 & -0.004 & -0.015 & 0.007 & 0.211 &  \\
    DOGE & Forward median ER & 366--730 & -0.120 & -0.220 & -0.023 & 0.007 & \checkmark \\
    ADA & Forward median ER & 366--730 & -0.087 & -0.140 & -0.038 & 0.000 & \checkmark \\
    BNB & Forward median ER & 366--730 & -0.068 & -0.109 & -0.030 & 0.000 & \checkmark \\
    ETH & Forward median ER & 366--730 & -0.054 & -0.107 & -0.006 & 0.014 & \checkmark \\
    LINK & Forward median ER & 366--730 & -0.025 & -0.050 & -0.003 & 0.013 & \checkmark \\
    BTC & Forward median ER & 366--730 & -0.006 & -0.021 & 0.011 & 0.228 &  \\
    BNB & Forward median ER & 731--1095 & -0.112 & -0.174 & -0.051 & 0.000 & \checkmark \\
    DOGE & Forward median ER & 731--1095 & -0.108 & -0.204 & -0.016 & 0.011 & \checkmark \\
    LINK & Forward median ER & 731--1095 & -0.085 & -0.156 & -0.018 & 0.006 & \checkmark \\
    ETH & Forward median ER & 731--1095 & -0.037 & -0.082 & 0.009 & 0.061 &  \\
    ADA & Forward median ER & 731--1095 & -0.035 & -0.079 & 0.015 & 0.085 &  \\
    BTC & Forward median ER & 731--1095 & -0.011 & -0.042 & 0.019 & 0.233 &  \\
    \hline
    \end{tabular}
\end{table}

\section{Additional Figures}\label{app:fig}

\subsection{Additional figures for Monte Carlo simulation results}

\begin{figure}[H]          
  \centering
  \includegraphics[width=0.75\linewidth]{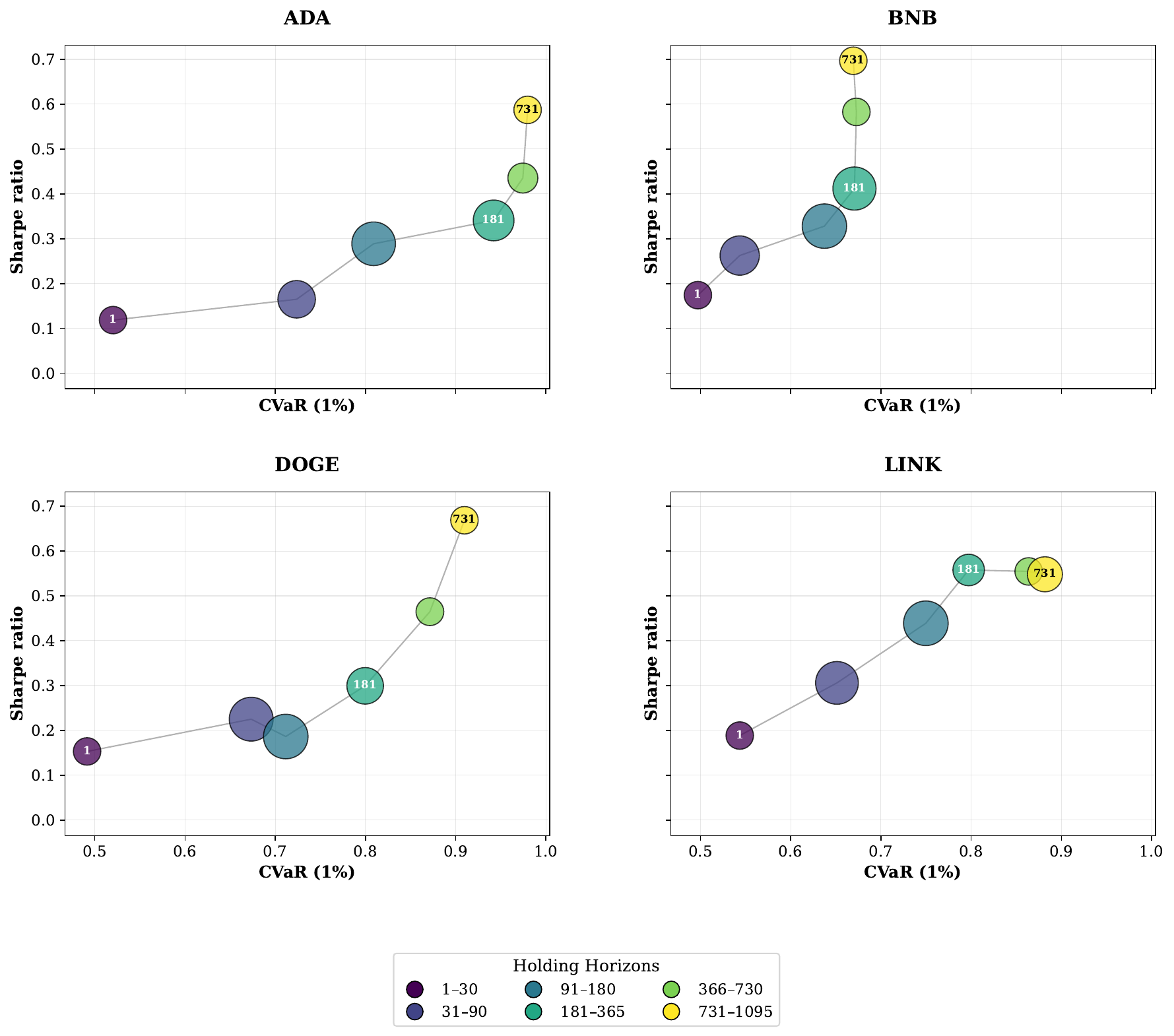}
  \caption{Risk–return trade-off across holding horizons for the remaining four baskets}
  \label{fig:B1}
\end{figure}

\begin{figure}[H]                     
  \centering
  \includegraphics[width=0.55\linewidth]{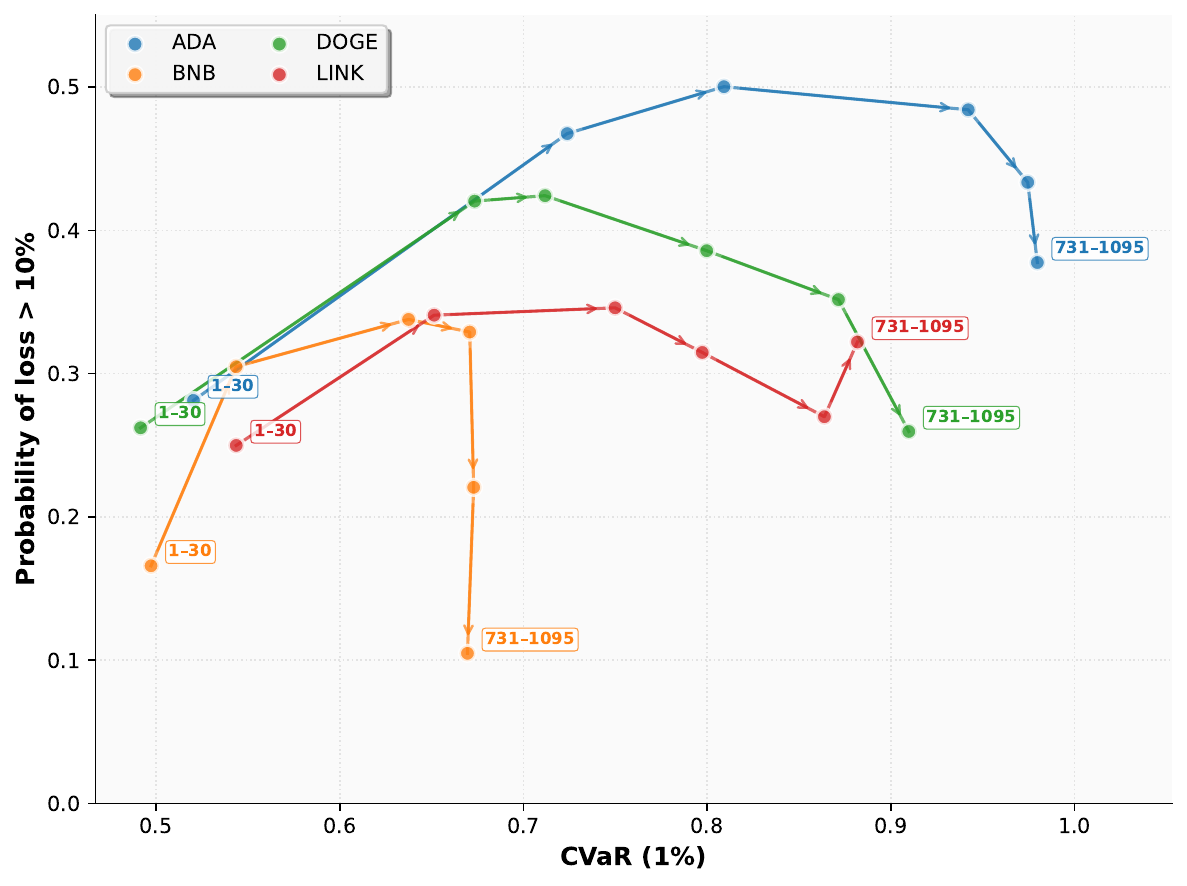}
  \caption{Trade-off between moderate and extreme risks across holding horizons for the remaining four baskets}
  \label{fig:B2}
\end{figure}

\begin{figure}[H]                           
  \centering
  \includegraphics[width=0.75\linewidth]{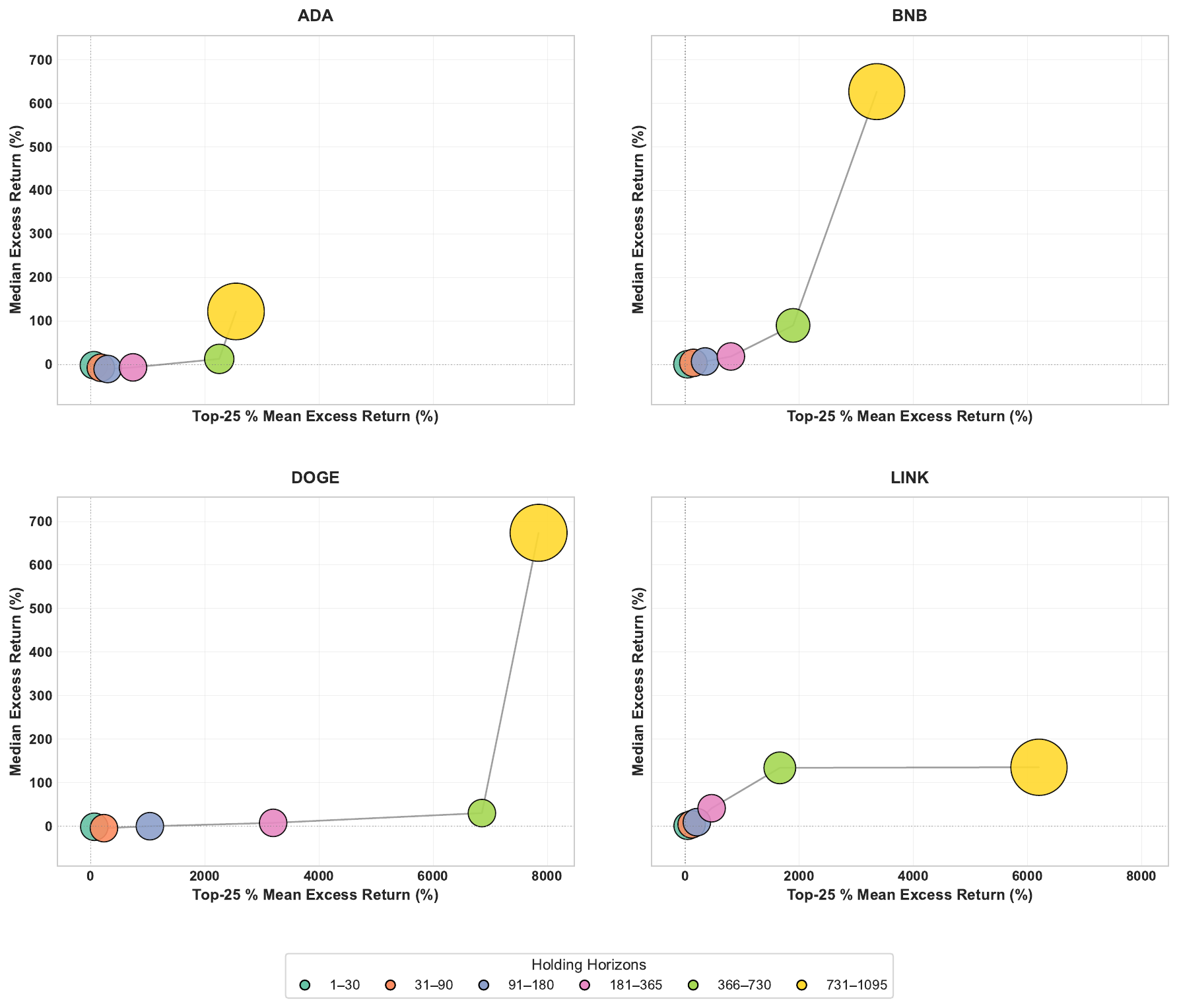}
  \caption{Median vs.\ top-25\% mean excess returns across holding horizons for the remaining four baskets}
  \label{fig:B3}
\end{figure}

\subsection{Convergence diagnostics}
\subsubsection{Trace and autocorrelation diagnostics}
\begin{figure}[H]
  \centering
  \begin{minipage}{0.47\linewidth}
    \centering
    \includegraphics[width=\linewidth]{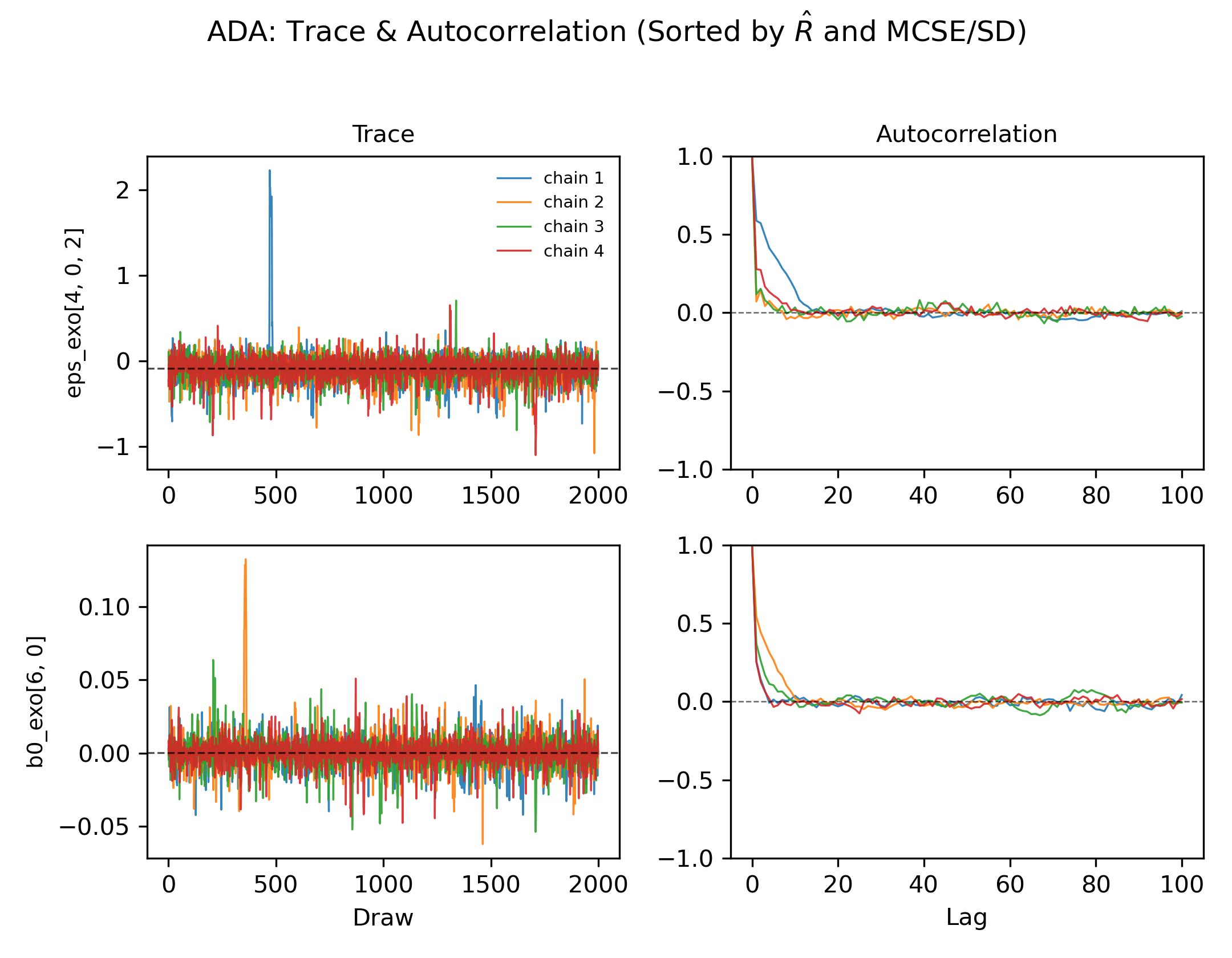}
  \end{minipage}\hfill
  \begin{minipage}{0.47\linewidth}
    \centering
    \includegraphics[width=\linewidth]{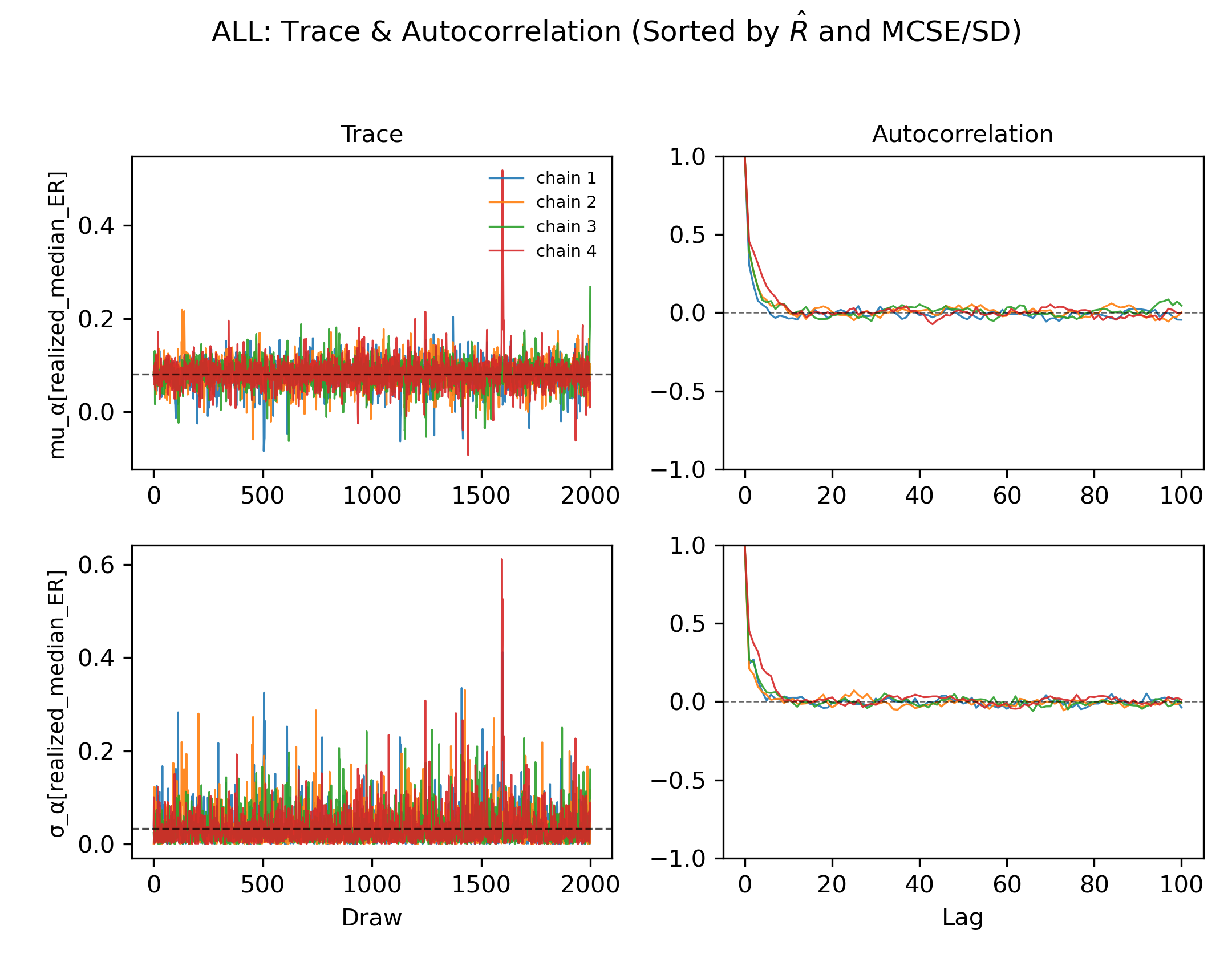}
  \end{minipage}

  \medskip

  \begin{minipage}{0.47\linewidth}
    \centering
    \includegraphics[width=\linewidth]{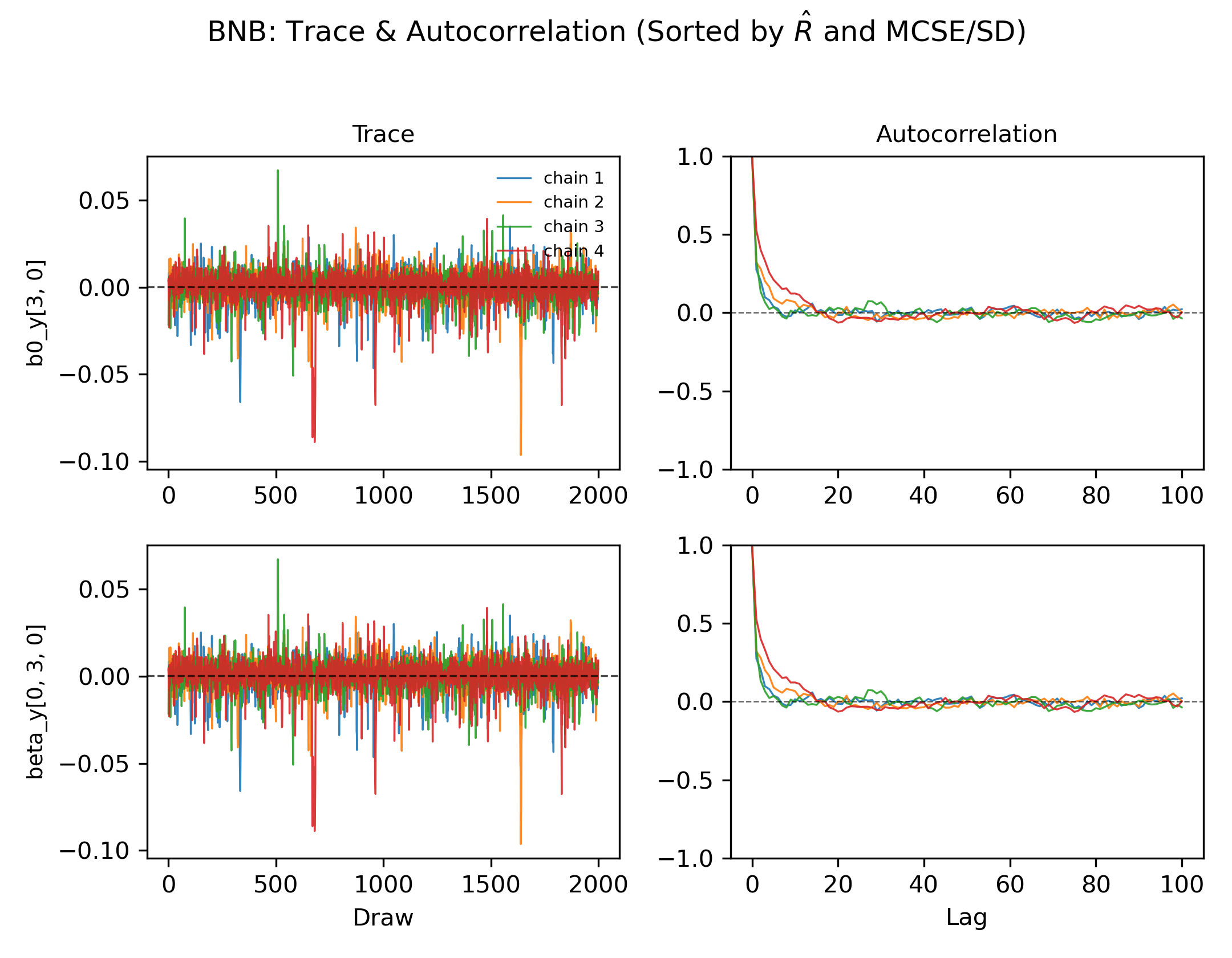}
  \end{minipage}\hfill
  \begin{minipage}{0.47\linewidth}
    \centering
    \includegraphics[width=\linewidth]{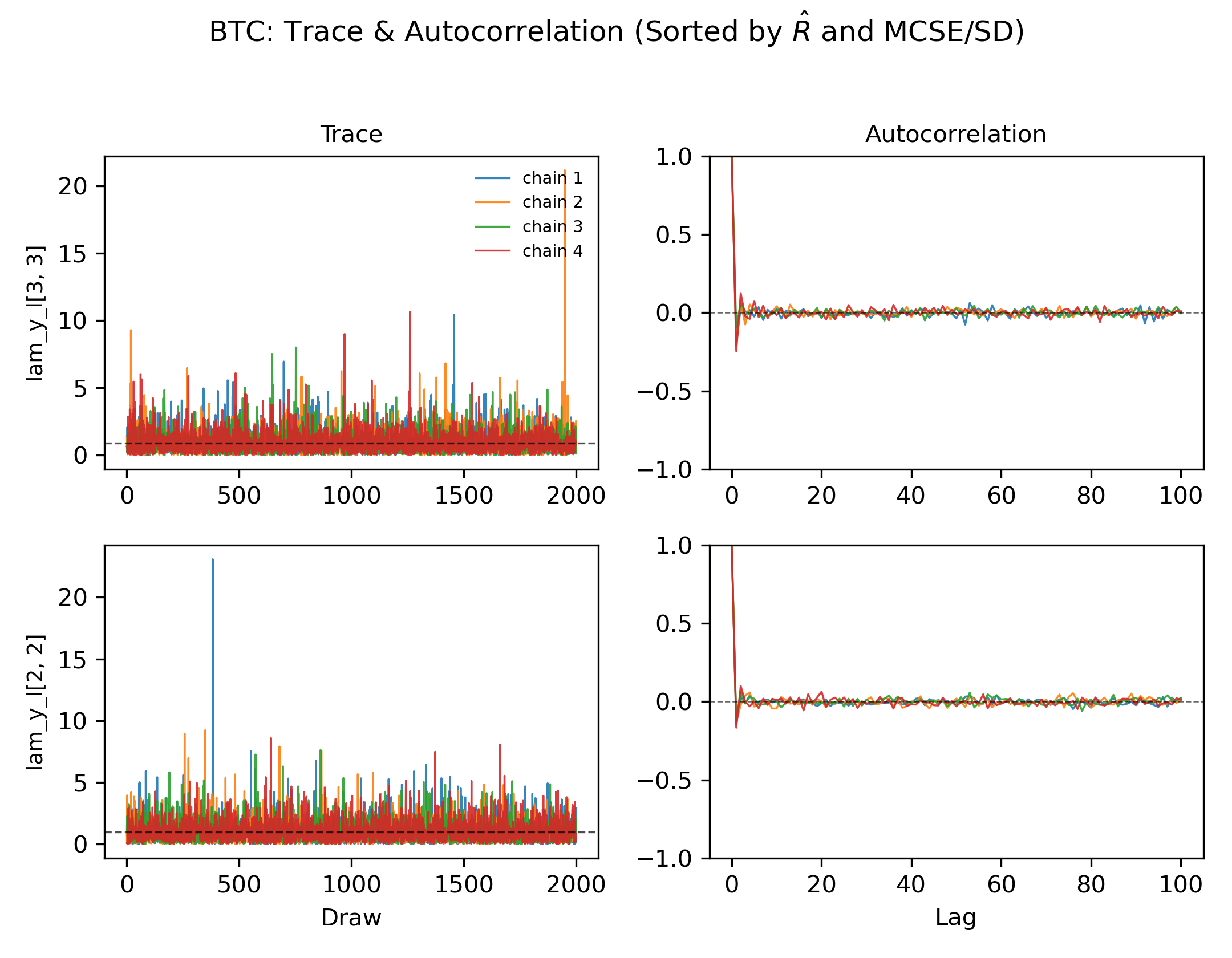}
  \end{minipage}
  \caption{Trace and autocorrelation diagnostics for ADA, ALL, BNB, and BTC baskets.}
  \label{fig:B_trace1}
\end{figure}

\begin{figure}[H]
  \centering
  \begin{minipage}{0.47\linewidth}
    \centering
    \includegraphics[width=\linewidth]{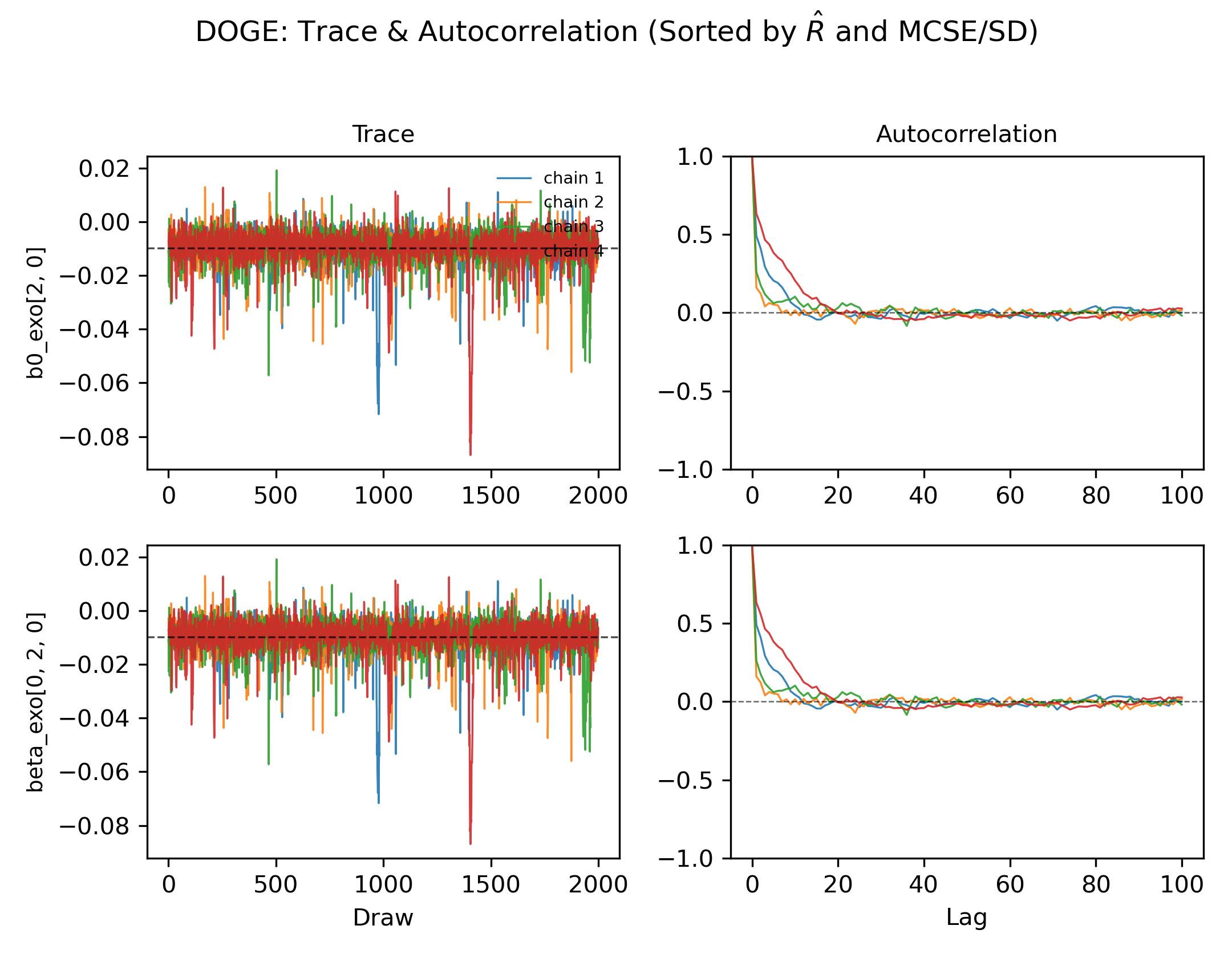}
  \end{minipage}\hfill
  \begin{minipage}{0.47\linewidth}
    \centering
    \includegraphics[width=\linewidth]{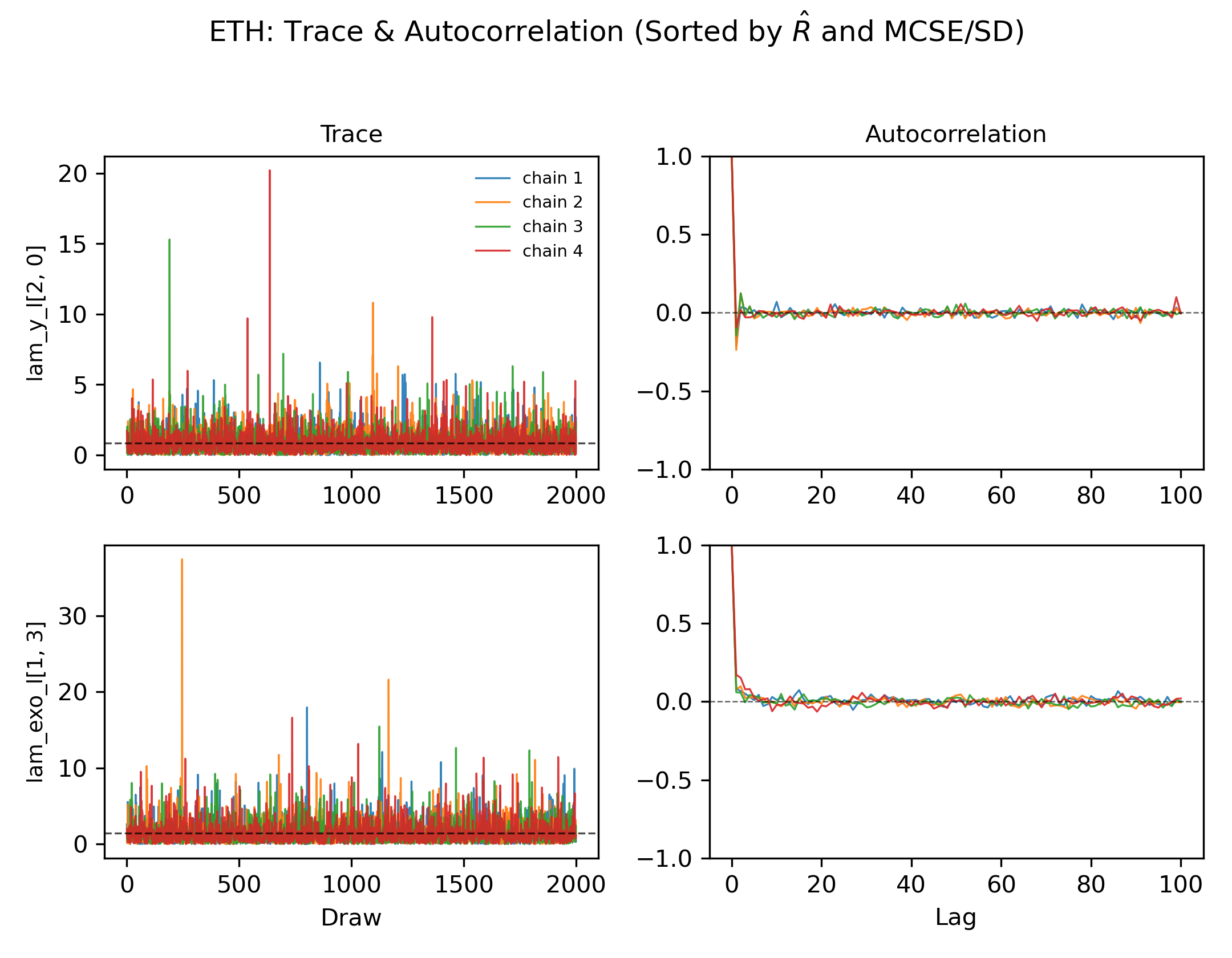}
  \end{minipage}

  \medskip

  \begin{minipage}{0.47\linewidth}
    \centering
    \includegraphics[width=\linewidth]{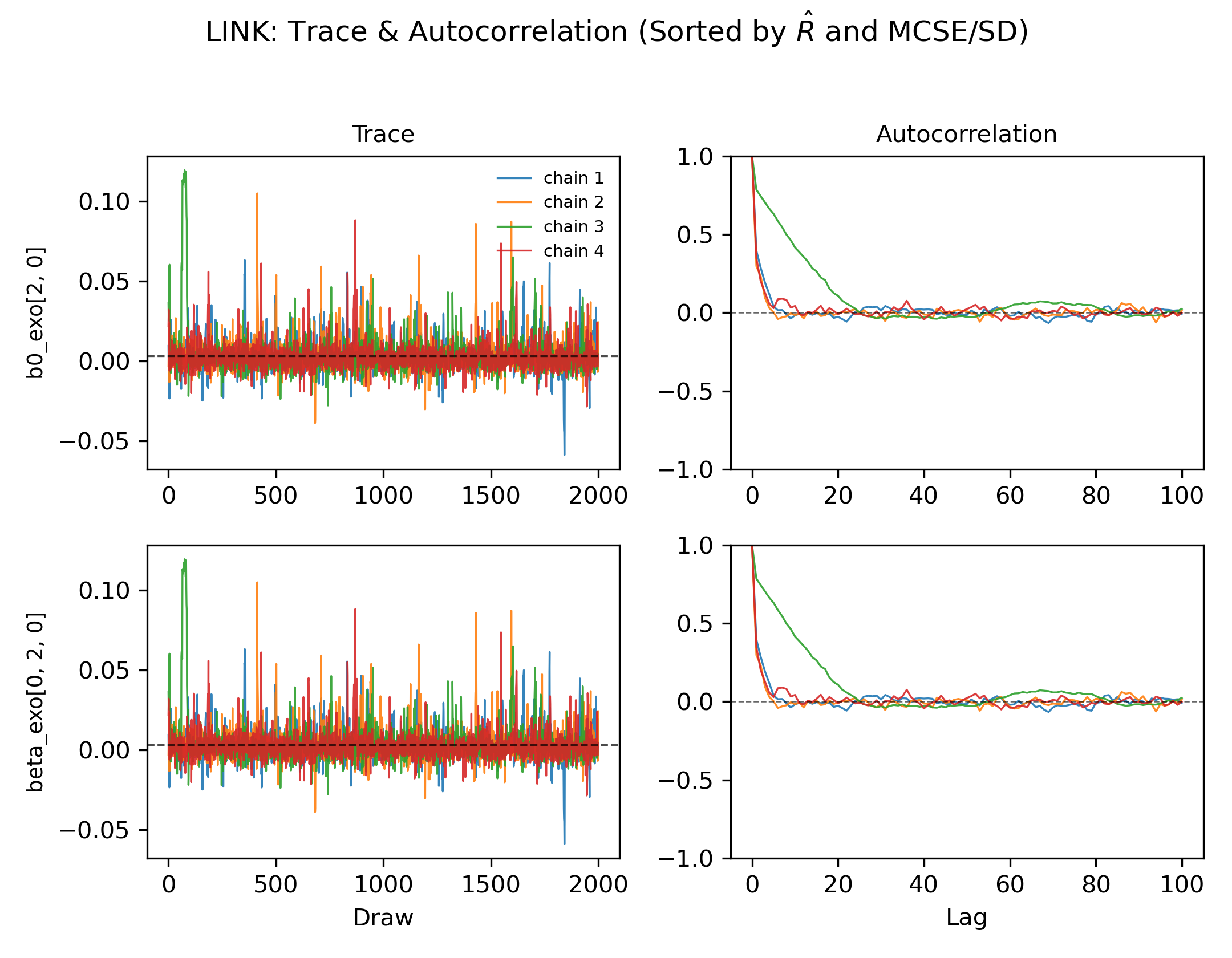}
  \end{minipage}\hfill
  \begin{minipage}{0.47\linewidth}
    \centering
    \includegraphics[width=\linewidth]{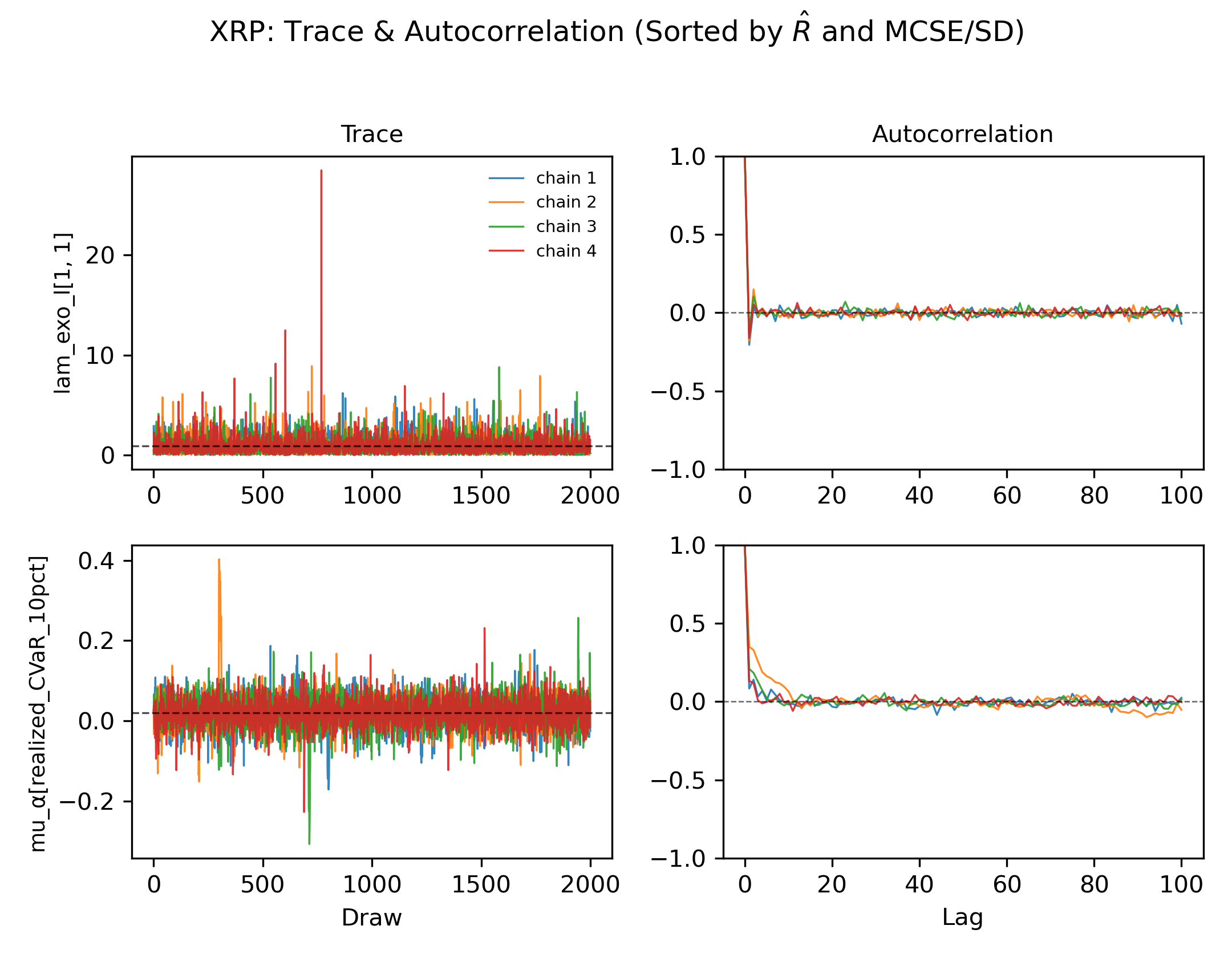}
  \end{minipage}
  \caption{Trace and autocorrelation diagnostics for DOGE, ETH, LINK, and XRP baskets.}
  \label{fig:B_trace2}
\end{figure}

Figures~\ref{fig:B_trace1} and \ref{fig:B_trace2} present trace and autocorrelation plots for the parameters with the highest $\hat{R}$ and MCSE/SD ratios in each basket. All chains exhibit stable mixing with rapidly decaying autocorrelation.

\subsubsection{Energy diagnostics}
\begin{figure}[H]
  \centering
  \includegraphics[width=0.95\linewidth]{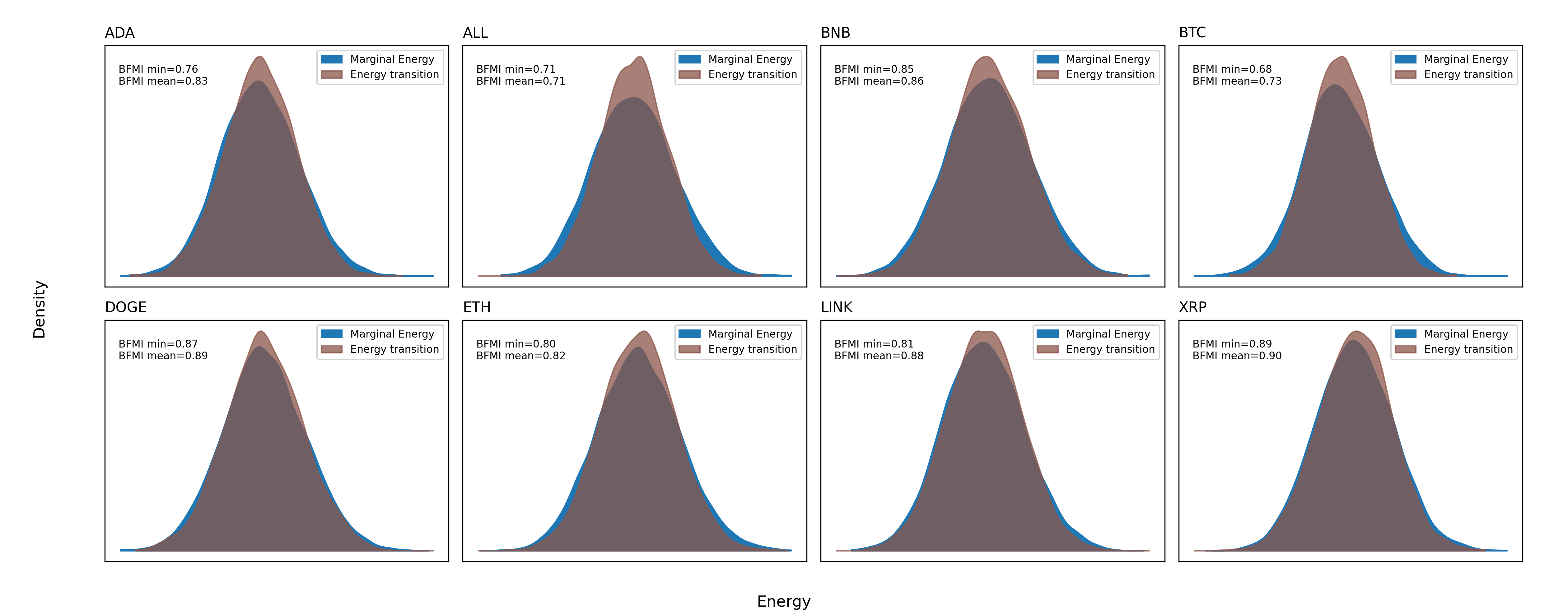}
  \caption{Energy diagnostics for all baskets.}
  \label{fig:B_energy}
\end{figure}

Figure~\ref{fig:B_energy} displays the energy diagnostics across all baskets. With minimum BFMI of 0.68, the overlapping energy transition and marginal energy distributions indicate the model did not encounter geometric pathologies.

\end{appendices}

\bibliographystyle{apalike}
\bibliography{references}

\end{document}